\documentclass[twocolumn,trackchanges,times,twocolappendix]{aastex631}

\newcommand*{\ac}{$A$(C)}
\newcommand*{\ali}{$A$(Li)}
\newcommand*{\teff}{$T_{\rm eff}$}
\newcommand*{\logg}{$\log~g$}
\newcommand*{\feh}{[Fe/H]}

\newcommand*{\vt}{$\xi_{\rm t}$}
\newcommand*{\kms}{km s$^{-1}$}

\newcommand*{\msun}{$M_\odot$}

\newcommand*{\z}{$|Z|$}

\newcommand*{\cemps}{CEMP-$s$}

\newcommand*{\cempno}{CEMP-$\rm no$}

\newcommand*{\gaia}{Gaia}

\newcommand{\RomanNumeralCaps}[1] 
    {\MakeUppercase{\romannumeral #1}} 

\usepackage{amsmath}
\usepackage{url}
\usepackage{natbib}
\usepackage{tabularx}
\usepackage{color}
\usepackage{graphicx}
\usepackage{epstopdf}
\usepackage{gensymb}
\usepackage{hyperref}
\usepackage{longtable}
\usepackage{subfigure}
\usepackage[T1]{fontenc}
\DeclareUnicodeCharacter{2212}{-}

\begin{document}


\shorttitle{Search For Extremely Metal-Poor Stars}
\shortauthors{Jeong et al.}

\title{SEARCH FOR EXTREMELY METAL-POOR STARS WITH GEMINI-N/GRACES I.
CHEMICAL-ABUNDANCE ANALYSIS}
\author{Miji Jeong}
\affiliation{Department of Astronomy, Space Science, and Geology, Chungnam National University, Daejeon 34134, Republic of Korea}
\author{Young Sun Lee}
\altaffiliation{Email: youngsun@cnu.ac.kr, Guest professor at University of Notre Dame}
\affiliation{Department of Astronomy and Space Science, Chungnam National University, Daejeon 34134, Republic of Korea}
\affiliation{Department of Physics and Astronomy and JINA Center for the Evolution of the Elements, University of Notre Dame, IN 46556, USA}
\author{Timothy C. Beers}
\affiliation{Department of Physics and Astronomy and JINA Center for the Evolution of the Elements, University of Notre Dame, IN 46556, USA}
\author{Vinicius M. Placco}
\affiliation{NSF's NOIRLab, 950 N. Cherry Ave., Tucson, AZ 85719, USA}
\author{Young Kwang Kim}
\affiliation{Department of Astronomy and Space Science, Chungnam National University, Daejeon 34134, Republic of Korea}
\author{Jae-Rim Koo}
\affiliation{Department of Earth Science Education, Kongju National University, 56 Gongjudaehak-ro, Gongju-si, Chungcheongnam-do 32588, Republic of Korea}
\author{Ho-Gyu Lee}
\affiliation{Korea Astronomy and Space Science Institute, Daejeon 34055, Republic of Korea}
\author{Soung-Chul Yang}
\affiliation{Korea Astronomy and Space Science Institute, Daejeon 34055, Republic of Korea}

\begin{abstract}

We present stellar parameters and abundances of 13 elements for 18 very metal-poor (VMP; [Fe/H] $<$ --2.0) stars, selected as extremely metal-poor (EMP; [Fe/H] $<$ --3.0) candidates from SDSS and LAMOST survey. 
High-resolution spectroscopic observations were performed
using GEMINI-N/GRACES. We find ten EMP stars among our candidates, and we newly identify three carbon-enhanced metal-poor (CEMP) stars with [Ba/Fe] $<$ 0. Although chemical abundances of our VMP/EMP stars generally follow the overall trend of other Galactic halo stars, there are a few exceptions. One Na-rich star ([Na/Fe] = +1.14) with low [Mg/Fe] suggests a possible chemical connection with second-generation stars in a globular cluster. The progenitor of an extremely Na-poor star ([Na/Fe] = --1.02) with an enhancement of K- and Ni-abundance ratios may have undergone a distinct nucleosynthesis episode, associated with core-collapse supernovae (CCSNe) having a high explosion energy. We have also found a Mg-rich star ([Mg/Fe] = +0.73) with slightly enhanced Na and extremely low [Ba/Fe], indicating that its origin is not associated with neutron-capture events. On the other hand, the origin of the lowest Mg abundance ([Mg/Fe] = --0.61) star could be explained by accretion from a dwarf galaxy, or formation in a gas cloud largely polluted by SNe Ia. We have also explored the progenitor masses of our EMP stars by comparing their chemical-abundance patterns with those predicted by Population III SNe models, and find a mass range of 10 -- 26 \msun, suggesting that such stars were primarily responsible for the chemical enrichment of the early Milky Way.
\end{abstract}


\section{Introduction} \label{sec:intro}

Population III (Pop III) stars are believed to be responsible
for the chemical enrichment of the early Universe, influencing
the formation of subsequent generation of stars \citep{bromm2003}.
The characterization of their physical properties is indispensable
to draw a complete picture of the origin of the chemical elements and the chemical-evolution
history of the early Milky Way (MW). Cosmological simulations predict that the Pop III
stars were predominantly very massive, with a characteristic mass of
$\sim$ 100 \msun\ \citep[e.g.,][]{bromm2004}. More recent sophisticated
high-resolution simulations equipped with detailed physical
processes are able to produce lower-mass stars with a few tens
of \msun\ \citep{stacy2013,hirano2014,susa2014,stacy2016}. Given that such stars no long exist,
owing to their high masses, we are not able to directly observe
and study them. At present the most practical observational probe
of the physical properties of the first-generation stars relies
on detailed elemental abundances of old, and
very metal-poor (VMP; [Fe/H]\footnote[7]{We follow the conventional nomenclature,
[X/Y] = log($N_{\rm X}/N_{\rm Y})_*$ -- log($N_{\rm X}/N_{\rm Y})_{\odot}$,
where $N_{\rm X}$ and $N_{\rm Y}$ are the number density of elements X and Y,
respectively, and  $\log\epsilon$(X) = $A$(X) = log($N_{\rm X}/N_{\rm H}$)+12,
where $N_{\rm X}$ and $N_{\rm H}$ are the
number densities of a given element X and hydrogen, respectively.} $<$ --2.0)
stars.

Various chemical elements in the atmospheres of VMP stars enable investigation of the
underlying physical processes originating in the different nucleosynthetic pathways
that produced them \citep[e.g.,][]{beers2005,norris2013,frebel2015,yoon2016}, as well as
the chemical yields from the supernovae (SNe) of their progenitors, including Pop III
SNe \citep[e.g.,][]{heger2010,ishigaki2014,nordlander2019}.
In turn, this allows us to provide important
constraints on not only the nucleosynthesis of Pop III stars by comparing
with theoretical model predictions \citep{nomoto2013}, but also on the chemical-evolution history
of the MW by examining the elemental-abundance trends \citep{frebel2015}.

Among the VMP stars, extremely metal-poor (EMP; [Fe/H] $<$ --3.0)
stars are the most suitable objects for the
studies mentioned above. Previous investigations have revealed that,
while EMP stars generally exhibit similar abundance
trends for most elements at similar metallicities, they
also show over- or under-abundances of C, N, O, the $\alpha$-elements,
and light elements such as Na and Al \citep[e.g.,][]{caffau2013,norris2013,frebel2015,bonifacio2018}.
The diversity of the chemical-abundance ratios among the EMP stars suggests that
a range of core-collapse supernovae (CCSNe) with various progenitor masses
and explosion energies have contributed to the stochastic
chemical-enrichment history of the MW.

One remarkable feature found from studies of low-metallicity stars
is that the fraction of so-called carbon-enhanced metal-poor (CEMP; originally
specified as metal-poor stars with [C/Fe] $>$ +1.0, more recently by [C/Fe] $>$ +0.7)
stars dramatically increases with decreasing metallicity
\citep{rossi1999,lucatello2006,lee2013,lee2014,lee2017,lee2019,yong2013,placco2014,yoon2018,arentsen2022}.
CEMP stars account for about 20\% of all stars with [Fe/H] $<$ --2.0, over 30\% at [Fe/H] $<$ --3.0,
and approach 100\% for [Fe/H] $< -4.0$.  The clear implication is that prodigious amounts of
carbon was produced in the early history of the MW.

CEMP stars can be divided into four major
categories: CEMP-$s$, CEMP-$r$, CEMP-$r/s$, and CEMP-no, according to
the level of enhancement of their neutron-capture elements \citep{beers2005}.
CEMP-$s$ stars exhibit over-abundances of $s$-process elements such
as Ba. CEMP-$r$ stars are strongly enhanced with $r$-process elements
such as Eu, and CEMP-$r/s$ objects are mildly enhanced with both $r$-process
and $s$-process elements. CEMP-no stars lack enhancements of neutron-capture elements.
Recent studies (e.g., \citealt{hampel2016,cowan2021}) have reported
that the production of the CEMP-$r/s$ stars is associated with an intermediate neutron-capture
process (the ``$i$-process''), first suggested by \citet{cowan1977}. The diversity of
the nature of CEMP stars implies that the formation of the various sub-classes is closely
linked to different specific astrophysical sites.

CEMP-no stars are the dominant fraction of stars with
[Fe/H] $<$ --3.0 \citep[e.g.,][]{aoki2007,yoon2016}.
Because of their low-metallicity nature and low abundances
of neutron-capture elements, they are
regarded as the likely direct descendants of Pop III
stars \citep{christlieb2002,frebel2005,caffau2011,keller2014,placco2016,yoon2016,hartwig2018,starkenburg2018,placco2021b}.
The C and O enrichment of CEMP-no stars are expected to have a profound
impact on the formation of low-mass stars, since these species play a major
role as efficient gas coolants, so that low-mass stars can form
in even extremely low-metallicity environments \citep{bromm2003,norris2013,frebel2015}.

Upon recognizing the importance of the EMP stars,
several large-scale surveys have been undertaken over
the past few decades to discover such iron-poor objects. In the early stages of this effort,
a large number of metal-poor stars were
identified by objective-prism based searches such as the HK survey \citep{beers1985,beers1992} and
Hamburg/ESO survey \citep[HES;][]{wisotzki1996,frebel2006,christlieb2008}.
Later on, it was led by large spectroscopic surveys such as the legacy Sloan Digital Sky
Survey (SDSS; \citealt{york2000}), the Sloan Extension for Galactic Understanding and
Exploration (SEGUE; \citealt{yanny2009,rockosi2022}),
and the Large sky Area Multi-Object Fiber Spectroscopic Telescope (LAMOST; \citealt{cui2012}) survey,
which are equipped with the capability for multi-object observation using hundreds to thousands of
fibers in their focal planes. Recently, narrow-band photometric surveys such
as SkyMapper \citep{keller2007}, Pristine \citep{starkenburg2017}, the Southern Photometric
Local Universe Survey (S-PLUS; \citealt{mendes2019,placco2022}),
and the Javalambre-Photometric Local Universe Survey (J-PLUS; \citealt{cenarro2019}) are
dramatically increasing the number of EMP candidates.

Despite the extensive searches for EMP stars in the last few decades,
thus far only several hundred such stars have been confirmed
by high-resolution spectroscopic analysis from which their detailed elemental abundances
have been derived \citep[e.g.,][]{ryan1996,norris2001,aoki2005,aoki2013,cayrel2004,yong2013,matsuno2017,aguado2019,yong2021,li2022}.
This is mainly because of their rarity and the difficulty of obtaining high-quality, high-resolution spectra.
Due to the stochastic nature of the chemical enrichment of the
early MW, and considering the importance of EMP stars to constrain the properties of
the first-generation stars and early Galactic chemical evolution, the confirmation, analysis, and
interpretation of their varied chemical-abundance patterns of more
EMP stars are clearly required.

In this study, we report on 18 newly identified VMP stars, of which
10 are EMP stars and 3 are CEMP stars.
We derive chemical-abundance ratios for 13 elements in these objects,
and discuss the overall abundance trends and possible origin of the chemically
peculiar objects. In addition, we use stellar explosion models to predict the progenitor masses of
our EMP stars in order to constrain the mass distribution of Pop III stars.

This paper is organized as follows. The selection of the EMP candidates and
their observation are covered in Section \ref{sec:Sample}.
The derivation of stellar parameters and elemental abundances
is presented in Sections \ref{sec:sparam} and \ref{sec:abun}, respectively.
In Section \ref{sec:abun_comp}, we discuss the derived abundance
trends of our EMP candidates by comparing with other Galactic field stars in
previous studies, and estimate the progenitor masses of our
confirmed EMP stars. We close with a summary in Section \ref{sec:summary}.

\begin{deluxetable*}{lcccccrcr}[!t]
\tabletypesize{\scriptsize}
\tablecaption{Observation Details of Our Program Stars}
\tablehead{\colhead{~~~~~~~~Object} & \colhead{Short} & \colhead{Date} & \colhead{RA} & \colhead{Dec} & \colhead{$g$} & \colhead{Exposure Time} & \colhead{S/N} & \colhead{$V_{\rm H}$} \\
\colhead{~~~~~~~~~~ID} & \colhead{ID} & \colhead{(UT)} & \colhead{} & \colhead{} & \colhead{(mag)} & \colhead{(sec)} & \colhead{(pixel$^{-1}$)} & \colhead{(\kms)}}
\startdata
\multicolumn{9}{c}{2016A (GN-2016A-Q-17)}\\
\hline
SDSS J075824.42+433643.4	& J0758	&	2016-04-07	& 07 58 24.42 & +43 36 43.4 &	16.3 	 &	 1500$\times$3	 &	 38	 &    +67.2~~	 \\
SDSS J092503.50+434718.4	& J0925	&	2016-04-05	& 09 25 03.50 & +43 47 18.4 &	16.5 	 &	 1150$\times$3	 &	 31	 &	 +28.5~~	 \\
SDSS J131116.58+001237.7	& J1311	&	2016-04-06	& 13 11 16.58 & +00 12 37.7 &	16.3 	 &	 1150$\times$3	 &	 34  &   --8.3~~	 \\
SDSS J131708.66+664356.8	& J1317	&	2016-04-08	& 13 17 08.66 & +66 43 56.8 &	15.9 	 &	 1500$\times$3	 &	 44	 &	 --185.8~~	 \\
SDSS J152202.10+305526.3	& J1522	&	2016-04-05	& 15 22 02.10 & +30 55 26.3 &	16.5	 &	 1400$\times$3	 &	 38	 &	 --353.9~~	 \\
\hline
\multicolumn{9}{c}{2018B (GN-2018B-Q-122)}\\
\hline
LAMOST J001032.66+055759.1	& J0010	&	2018-09-03	& 00 10 32.66 & +05 57 59.1 &	14.6	 &	 ~900$\times$2	 &	 77	 &	 --151.3~~	 \\
LAMOST J010235.03+105245.5	& J0102	&	2018-09-06	& 01 02 35.03 & +10 52 45.5 &	15.5	 &	 1400$\times$3	 &	 47	 &	 --130.9~~	 \\
LAMOST J015857.38+382834.7	& J0158	&	2018-09-05	& 01 58 57.38 & +38 28 34.7 &	15.2 	 &	 1200$\times$3	 &	 68	 &	 --44.54~~	 \\
BD+44 493	                & J0226	&	2018-09-05	& 02 26 49.73 & +44 57 46.5 &	 9.1\tablenotemark{\scriptsize a} &	 ~~20$\times$1\tablenotemark{\scriptsize b}	 &	 75	 &	 --149.3~~	 \\
LAMOST J035724.49+324304.3	& J0357	&	2018-09-06	& 03 57 24.49 & +32 43 04.3 &	13.8	 &	 ~600$\times$1	 &	 62	 &	 +114.5~~	 \\
LAMOST J042245.27+180824.3	& J0422	&	2018-09-05	& 04 22 45.27 & +18 08 24.3 &	15.8	 &	 1750$\times$3	 &	 76	 &	 +76.7~~	 \\
LAMOST J165056.88+480240.6	& J1650	&	2018-09-03	& 16 50 56.88 & +48 02 40.6 &	13.2 	 &	 ~600$\times$1	 &	 43	 &	 --26.1~~	 \\
LAMOST J170529.80+085559.2	& J1705	&	2018-09-03	& 17 05 29.80 & +08 55 59.2 &	14.3 	 &	 1200$\times$2	 &	 76	 &	 +43.5~~	 \\
SDSS J224145.05+292426.1	& J2241	&	2018-09-03	& 22 41 45.05 & +29 24 26.1 &	14.9	 &	 1200$\times$2	 &	 49	 &	 --218.1~~	 \\
LAMOST J224245.51+272024.5	& J2242	&	2018-09-03	& 22 42 45.51 & +27 20 24.5 &	13.5	 &	 ~600$\times$1	 &	 60	 &	 --378.8~~	 \\
LAMOST J234117.38+273557.7	& J2341	&	2018-09-03	& 23 41 17.38 & +27 35 57.7 &	15.1	 &	 1200$\times$3	 &	 77	 &	 --182.3~~	 \\
\hline
\multicolumn{9}{c}{2019B (GN-2019B-Q-115, Q-219, and Q-310)}\\
\hline
LAMOST J071349.17+550029.6  & J0713 &   2020-01-20  & 07 13 49.17 & +55 00 29.6 &   14.1    &   ~900$\times$3   &   93  &   --56.8~~    \\
LAMOST J081413.16+330557.5  & J0814 &   2020-01-21  & 08 14 13.16 & +33 05 57.5 &   15.3    &   1800$\times$3   &   57  &    +67.2~~    \\
LAMOST J090852.87+311941.2  & J0908 &   2020-01-19  & 09 08 52.87 & +31 19 41.2 &   15.4    &   1600$\times$3   &   58  &    +144.8~~    \\
LAMOST J103745.92+253134.2  & J1037 &   2020-01-20  & 10 37 45.92 & +25 31 34.2 &   15.3    &   1500$\times$3   &   46  &    +28.7~~    \\
\enddata
\tablecomments{The S/N per pixel is the average value around 5500 \AA. $V_{\rm H}$ is
the heliocentric radial velocity. The short-named
objects J0226 and J1522 are the reference stars studied by \citet{ito2013} and \citet{matsuno2017},
respectively.}
\tablenotetext{a}{$V$ magnitude.}
\tablenotetext{b}{Even though we obtained multiple exposures, we used the spectrum obtained with
20 second exposure in our analysis to validate the abundance analysis of our program stars,
which have the spectrum with the similar S/N to this object.}
\label{tab1}
\end{deluxetable*}

\section{Target Selection and High-resolution Observations} \label{sec:Sample}
\subsection{Target Selection}

We selected EMP candidates from the low-resolution ($R \sim$ 1800) spectroscopic surveys of SDSS
and LAMOST for follow-up observations with high-resolution spectroscopy.
Using an updated version of the SEGUE Stellar Parameter
Pipeline (SSPP; \citealt{allende2008,lee2008a,lee2008b,lee2011,
smolinski2011,lee2013}), which now has the capability of deriving [N/Fe] \citep{kim2022} and
[Na/Fe] \citep{koo2022} (in addition to [C/Fe] and [Mg/Fe]), we analyzed the stellar spectra obtained by the
legacy SDSS survey, SEGUE, the Baryon Oscillation Spectroscopic Survey (BOSS; \citealt{dawson2013}), and
the extended Baryon Oscillation Spectroscopic Survey (eBOSS; \citealt{blanton2017}),
and determined stellar atmospheric parameters such as effective
temperature (\teff), surface gravity (\logg), and metallicity (\feh).

Similarly, we utilized the SSPP to estimate the stellar parameters and abundances from
the LAMOST stellar spectra, made feasible due to their
similar spectral coverage (3700\,{\AA} -- 9000\,{\AA}) and resolution ($R \sim$ 1800)
to those of the SDSS. We refer interested readers to \citet{lee2015} for
details of this application.

After obtaining the stellar parameters from stars in both surveys, we selected as EMP candidates
the objects meeting the following criteria: $g$ $<$ 16, [Fe/H] $<$ --2.8, and 4000\,K $<$ \teff\ $<$ 7000\,K.
The relaxed cut on metallicity was adopted because the \ion{Ca}{2} K line, which plays an
important role in determining the metallicity for VMP stars could be blended with
interstellar calcium in low-resolution spectra, causing over-estimation of the metallicity.
We eliminated stars that had already been observed with high-dispersion
spectrographs, and carried out a visual inspection of the low-resolution spectra
to make sure that their estimated metallicities did not arise from defects in the spectrum.

\subsection{High-resolution Follow-up Observations}\label{sec:highR}

We carried out high-resolution spectroscopic
observations for twenty stars (18 EMP candidates and two reference stars),
making use of Gemini Remote Access to the CFHT ESPaDOnS Spectrograph \citep[GRACES;][]{chene2014}
on the 8 m Gemini-North telescope during the 2016A (GN-2016A-Q-17), 2018B (GN-2018B-Q-122),
and 2019B (GN-2019B-Q-115, Q-219, and Q-310) semesters.
We used the 2-fiber mode (sky + target) of the GRACES \'Echelle spectrograph, which
yields a maximum resolving power of $R \sim$ 45,000 in the spectral range
of 4,000\,{\AA} -- 10,000\,{\AA}. This mode provides for better handling of sky subtraction.
One limitation of the GRACES approach is significant reduction of blue photons,
owing to the 270 m-long fiber cable, which guides light from the focal plane
of the Gemini telescope to the ESPaDOnS spectrograph. This produces
low signal-to-noise ratio (S/N) for wavelengths shorter than 4700\,{\AA}, in which
numerous metallic lines are present, precluding abundance measurements for many atomic species.

The 2D cross-dispersed \'Echelle spectra were reduced with standard calibration
images (bias, flat-field, and arc lamp), using
the DRAGRACES\footnote[8]{\url{http://drforum.gemini.edu/topic/graces-pipeline-dragraces/}}
pipeline \citep{chene2021}, which is a reduction pipeline written in IDL to extract the
wavelength-calibrated 1D spectrum for the science target and sky.
After subtracting the sky background, we co-added spectra of each
\'Echelle order by a signal-weighted average for multiple exposures to
boost the S/N, and obtained one continuous spectrum for a given
object by stitching together adjacent orders following normalization of each order.
The overlapping wavelength regions of each order were averaged by weighting the signal.
We used this co-added spectrum for the abundance analysis.

We measured the radial velocity of each star through
cross-correlation of a synthetic template spectrum with the co-added observed spectrum
in the region of 5160\,{\AA} -- 5190\,{\AA}, where the \ion{Mg}{1}$b$ triplet lines
are located. The synthetic spectrum was generated by
considering the evolutionary stage of each target, assuming
[Fe/H] = --3.0 for our EMP candidates.

Heliocentric corrections were made with the \texttt{astropy} package. Table \ref{tab1} lists observation details,
heliocentric radial velocities, and the average S/N around 5500\,{\AA} of the co-added spectra
for our program stars. The two reference stars (denoted as J0226 and J1522 in the second column)
were observed to validate our approach of abundance analysis (see Section \ref{sec:abun}
for additional details).

\subsection{Distance and Photometry of Our Targets}

To provide better constraints on the determination of \teff\ and \logg,
we made use of the distance and photometric information
of our EMP candidates. We primarily adopted their parallaxes
from the \gaia\ Early Data Release 3 \citep[EDR3;][]{gaia2016,gaia2021} for the distance estimates.
We applied the systematic offset of --0.017 mas reported by \citet{lindegren2021}.
In cases where the parallax of a star is not available in the \gaia\ EDR3, or its
parallax uncertainty is larger than 25\%, we adopted its photometric distance derived
by the methodology of \citet{beers2000,beers2012}.

The $V$ and $K_{\rm s}$ magnitudes of our program stars used to derive
photometry-based \teff\ estimates were obtained from
the AAVSO Photometric All-Sky Survey \citep[APASS;][]{henden2016} and
Two Micron All Sky Survey \citep[2MASS;][]{skrutskie2006}, respectively.
The extinction value $E(B-V)$ was estimated from the dust map provided by \citet{sfd2011},
applying the relations reported by \citet{schlegel1998} to
correct the interstellar extinction for each bandpass.
The reddening obtained from the dust map is the upper limit
along the line-of-sight to a star, but we need to calculate the appropriate $E(B-V)$ taking
into account the star's distance. It is known that a difference of 0.01 mag in $E(B-V)$
leads to a 50\,K shift in the temperature estimate \citep{casagrande2010}.
We computed the $E(B-V)$ value using the reddening fraction
equation proposed by \citet{anthony1994} in the same way as performed by \citet{ito2013},
and corrected the reddening for each target. Table \ref{tab2} lists the \gaia\ EDR3 ID, distance,
$V$, $K_{\rm s}$, computed reddening, and absolute $V$ magnitude of the observed EMP candidates.

\begin{deluxetable*}{cccccccccr}
\tabletypesize{\scriptsize}
\tablecaption{Distance and Photometric Information}
\tablehead{\colhead{Short} & \colhead{\textit{Gaia} EDR3 ID~~~~~} & \colhead{Distance} & \colhead{$V$} & \colhead{Error} & \colhead{$K_{\rm s}$} & \colhead{Error} & \colhead{$V-K_s$} & \colhead{$E(B-V)$} & \colhead{$M_{\rm V}$}\\
\colhead{ID} & \colhead{} & \colhead{(kpc)} & \colhead{} & \colhead{} & \colhead{} & \colhead{} & \colhead{} & \colhead{} & \colhead{}}
\startdata
J0010 & 2742456847516899328  & 16.94\tablenotemark{\scriptsize a}    & 13.93 & 0.04 & 10.82 & 0.02 & 3.11 & 0.03 & --2.33 \\
J0102 & 2583190938965264256  & 0.93\tablenotemark{\scriptsize b}    & 15.31 & 0.11 & 13.75 & 0.05 & 1.56 & 0.04 & ~~5.24 \\
J0158 & 343053198441295616   & 7.10\tablenotemark{\scriptsize a}    & 14.95 & 0.02 & 12.78 & 0.02 & 2.17 & 0.05 & ~~0.53 \\
J0226 & 341511064663637376   & 0.20\tablenotemark{\scriptsize b}    & ~9.11 & 0.02 & ~7.20 & 0.01 & 1.91 & 0.09 & ~~2.45 \\
J0357 & 170370808491343360   & 0.72\tablenotemark{\scriptsize b}    & 13.37 & 0.10 & 11.09 & 0.02 & 2.28 & 0.26 & ~~3.39 \\
J0422 & 47471831142473216    & 5.29\tablenotemark{\scriptsize a}    & 15.08 & 0.03 & 11.79 & 0.02 & 3.29 & 0.42 & ~~0.11 \\
J0713 & 988031250483121792   & 10.64\tablenotemark{\scriptsize b}   & 13.40 & 0.09 & 10.42 & 0.02 & 2.98 & 0.08 & --2.00 \\
J0758 & 923068053360267136   & 1.60\tablenotemark{\scriptsize b}    & 16.20 & 0.25 & 14.88 & 0.10 & 1.32 & 0.04 & ~~5.06 \\
J0814 & 902428295962842496   & 3.18\tablenotemark{\scriptsize b}    & 15.04 & 0.03 & 13.38 & 0.03 & 1.67 & 0.05 & ~~2.65 \\
J0908 & 700085583420082176   & 18.18\tablenotemark{\scriptsize a}   & 14.92 & 0.05 & 12.42 & 0.03 & 2.50 & 0.02 & --1.46 \\
J0925 & 817672820091831552   & 6.46\tablenotemark{\scriptsize a}    & 16.12 & 0.09 & 14.43 & 0.08 & 1.70 & 0.02 & ~~2.02 \\
J1037 & 724816554864554368   & 1.19\tablenotemark{\scriptsize b}    & 15.08 & 0.05 & 13.85 & 0.04 & 1.23 & 0.02 & ~~4.63 \\
J1311 & 3687718706290945920  & 2.75\tablenotemark{\scriptsize b}    & 16.28 & 0.07 & 14.99 & 0.14 & 1.29 & 0.03 & ~~3.97 \\
J1317 & 1678584136808089344  & 2.13\tablenotemark{\scriptsize b}    & 15.85 & 0.06 & 14.74 & 0.11 & 1.11 & 0.01 & ~~4.15 \\
J1522 & 1276882477044162688  & 4.97\tablenotemark{\scriptsize a}    & 16.46 & 0.15 & 14.80 & \nodata & 1.66 & 0.02 & ~~2.91 \\
J1650 & 1408719281332527616  & 0.84\tablenotemark{\scriptsize b}    & 13.14 & 0.01 & 12.01 & 0.02 & 1.13 & 0.02 & ~~3.44 \\
J1705 & 4443271696395120128  & 0.75\tablenotemark{\scriptsize b}    & 14.12 & 0.03 & 12.66 & 0.03 & 1.44 & 0.11 & ~~4.39 \\
J2241 & 1887491346088043008  & 1.82\tablenotemark{\scriptsize b}    & 15.06 & 0.09 & 13.73 & 0.05 & 1.32 & 0.06 & ~~3.55 \\
J2242 & 1884009398920040192  & 6.42\tablenotemark{\scriptsize b}    & 13.12 & 0.06 & 10.68 & 0.02 & 2.44 & 0.06 & --1.10 \\
J2341 & 2865251577418971392  & 15.59\tablenotemark{\scriptsize a}    & 14.49 & 0.09 & 11.79 & 0.02 & 2.70 & 0.08 & --1.75 \\
\enddata
\tablecomments{The $V$ and $K_{\rm s}$ magnitudes come from APASS and 2MASS, respectively. The
reddening value was rescaled according to the distance of a star (see text for detail). $M_{\rm
V}$ is the absolute magnitude in the $V$ band.}
\tablenotetext{a}{Based on spectroscopic parallax.}
\tablenotetext{b}{Based on $Gaia$ EDR3 parallax.}
\label{tab2}
\end{deluxetable*}

\section{Stellar Atmospheric Parameters} \label{sec:sparam}

The determination of stellar parameters is of central importance for
deriving abundance estimates of chemical elements. In this section, we describe how
we derived \teff, \logg, \feh, and microturbulence velocity \vt.

\subsection{Initial Guess of Stellar Parameters}\label{sec:initial}

Any high-resolution spectroscopic analysis to determine the stellar
parameters requires a model atmosphere as a staring point.
We obtained initial stellar-atmospheric parameters, which are needed to
generate a model atmosphere, through fitting an observed spectrum of
our EMP target to a synthetic template. We refer the interested reader
to \citet{kim2016} for a more detailed description of this approach.

\begin{figure} 
\epsscale{1.15}
\plotone{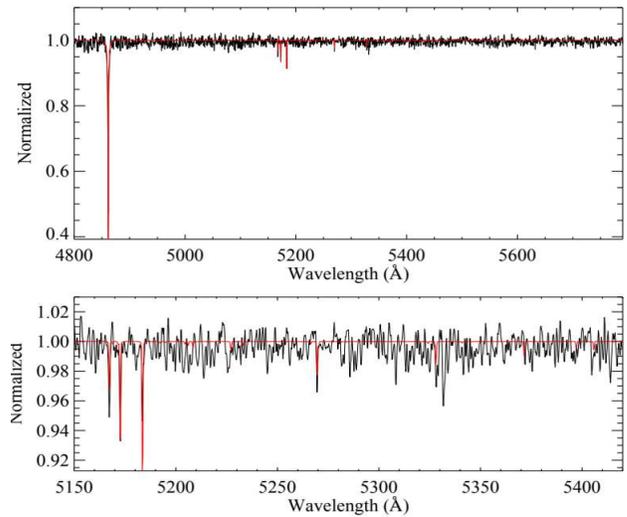}
\caption{An example (J0226, one of the reference stars) of our spectral-matching technique.
The black line is the observed spectrum, whereas
the red is the best-matching synthetic spectrum. The bottom panel is a close-up view
of the \ion{Mg}{1}$b$ triplet region and a few iron lines.}
\label{fig1}
\end{figure}

In this procedure, we used the spectral
range of 4800\,{\AA} -- 5500\,{\AA}, in which several metallic lines and the
temperature-sensitive H$\beta$ line are present.
In addition, we degraded the original spectrum to $R$ = 10,000 for fast and
efficient spectral-template matching to derive stellar parameters.
The smoothing of the spectrum also has the benefit of increasing the spectral S/N.
Figure \ref{fig1} shows an example of our spectral-matching technique for one of our targets (J0226).
In this figure, the black and red lines represent the observed spectrum and best-fit
synthetic spectrum, respectively. The strongest feature is the H$\beta$ line.
The bottom panel of the figure is a close-up view of the \ion{Mg}{1}$b$ triplet lines
and a few iron lines. We performed this spectral fitting on our entire sample of stars, and obtained initial estimates of
stellar parameters, which were used as starting points in the process of determining more accurate estimates,
as described in Section \ref{sec:routine}.

\subsection{Measurement of Equivalent Widths}

To determine accurate stellar parameters and abundances of various chemical elements,
we need to measure the equivalent widths (EWs) of Fe lines, as well as other metallic lines that are
detectable in a spectrum. For this, we collected the information for various atomic lines
from several literature sources \citep{aoki2013,aoki2018,spite2018,placco2020,rasmussen2020}.
Then, the EWs were measured via fitting a Gaussian profile, using the  Image Reduction
and Analysis Facility \citep[IRAF;][]{tody1986,tody1993} task \texttt{splot}.
We did not measure EWs for blended lines, but only for well-isolated lines. The line information of Li, C,
and Ba were produced with the \texttt{linemake}\footnote[9]{\url{https://github.com/vmplacco/linemake}}
code \citep{placco2021a}, and their abundances were determined through spectral synthesis rather
than the EW analysis. Table \ref{a_1} in the Appendix lists the line information used for the abundance analysis.

\begin{figure} [!t]
\epsscale{1}
\plotone{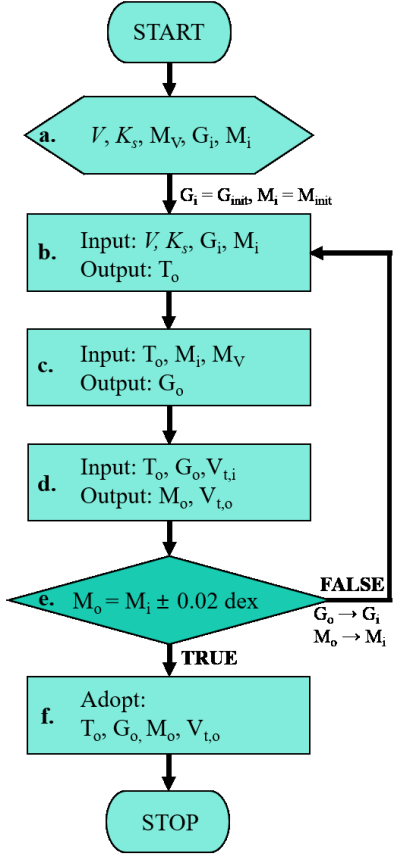}
\caption{Flow chart of our iterative procedure to determine \teff, \logg, \feh, and \vt.
For convenience, the effective temperature, surface gravity, metallicity, and microturbulence velocity are expressed
as T, G, M, and V$_{\rm t}$, respectively. $M_{\rm V}$ is the absolute $V$ magnitude.}
\label{fig2}
\end{figure}

\subsection{An Iterative Procedure for Determining Stellar Parameters}\label{sec:routine}

Because our targets are low-metallicity stars with weak metallic absorption lines, and
the GRACES spectra have relatively lower S/N in the wavelength region
shorter than 4700\,{\AA}, which includes many iron lines that
play a crucial role in constraining the stellar parameters, we did not follow the
traditional ionization-equilibrium technique to derive
\teff, \logg, and \vt. Instead, we have devised an iterative procedure
to determine the stellar parameters, as illustrated in Figure \ref{fig2}.
It begins with the information on $V$, $K_{\rm s}$, $M_{\rm V}$, and the
initial stellar parameters obtained by the spectral-matching procedure described above.
Note that for brevity we express effective temperature as T, surface gravity as G, metallicity as M,
and microturbulence velocity (\vt) as V$_{\rm t}$. A detailed
description of the procedure is as follows:

\begin{description}
  \item[Step a] Preparation of required information. We gather information on $V$, $K_{\rm s}$, $M_{\rm V}$, G$_{\rm init}$, and M$_{\rm init}$ of each star, where $M_{\rm V}$ is the absolute $V$ magnitude, and G$_{\rm init}$ and M$_{\rm init}$ are the gravity and the metallicity, respectively, estimated from the spectral matching. The adopted $V$, $K_{\rm s}$, and M$_{V}$ for our program stars are listed in Table \ref{tab2}.

  \item[Step b] Estimation of effective temperature. We input $V$, $K_{\rm s}$, G$_{\rm i}$ (G$_{\rm init}$),
      and M$_{\rm i}$ (M$_{\rm init}$), and estimate three \teff\ values for each star by following the procedure described in Section \ref{sec:teff}. The gravity G$_{\rm i}$ is used to separate giants from dwarfs. We obtain a bi-weight average \teff\ (T$_{\rm o}$) from the three estimates.

  \item[Step c] Estimation of surface gravity. Following the prescription in Section \ref{sec:logg}, we estimate \logg\ (G$_{\rm o}$), based on isochrone fitting. We assume a stellar age of 13 Gyr and [$\alpha$/Fe] = +0.3,  appropriate for our EMP candidates. In this step, the inputs are T$_{\rm o}$, M$_{\rm i}$, and $M_{\rm V}$.

  \item[Step d] Estimation of metallicity and microturbulence velocity. While fixing T$_{\rm o}$ and G$_{\rm o}$ determined in \textbf{Step b} and \textbf{Step c}, we estimate M$_{\rm o}$ and V$_{\rm t,o}$ by the prescription addressed in Section \ref{sec:feh}.

  \item[Step e] Convergence check. We check if M$_{\rm i}$ and M$_{\rm o}$ agree with each other within $\pm$0.02 dex. If the convergence criterion is satisfied, the routine goes to \textbf{Step f}, and if not, it goes back to \textbf{Step b}, and G$_{\rm o}$ and M$_{\rm o}$ are used as G$_{\rm i}$ and M$_{\rm i}$ to repeat until it converges to the tolerance level of 0.02 dex in the metallicity difference between the input and output.

  \item[Step f] Determination of adopted stellar parameters. If the convergence criterion is met in \textbf{Step e}, the derived T$_{\rm o}$, G$_{\rm o}$, M$_{\rm o}$, and V$_{\rm t,o}$ are taken as the adopted stellar parameters.
\end{description}

\subsubsection{Effective Temperature}\label{sec:teff}
We employed three different methods to derive accurate and precise effective temperature estimates.
The underlying principle of the methods is the InfraRed Flux Method (IRFM) method, which
is based on the extinction-corrected $V-K_{\rm s}$ color-\teff\ relation. We adopted
the color-temperature relations provided by \citet{alonso1999b}, \citet{gonzales2009},
and \citet{casagrande2010}. These studies reported different relations over a
range of the surface gravities of stars, and by following their prescription, we made
use of Equations 8 and 9 in Table 2 of \citet{alonso1999b} for giants, Equation 10
of \citet{gonzales2009} for giants and dwarfs, and Equation 3 of \citet{casagrande2010}
for subgiants and dwarfs. All of these equations have some dependency
on the metallicity, with valid ranges of \feh\ and $V-K_{\rm s}$.

We require $V-K_{\rm s}$, \logg, and \feh\ information to work with the above equations.
The criteria for subdividing the luminosity class of a star are \logg\ $\leq$ 3.0 for a
giant, 3.0 $<$ \logg\ $\leq$ 3.75 for a subgiant, and \logg\ $>$ 3.75 for a dwarf. As
Equations 8 and 9 of \citet{alonso1999b} use the same color range for a giant, we derived
the temperature from both equations
and took an average. If the $V-K_{\rm s}$ color is out of the range specified
in a color-temperature relation, the effective temperature was determined by using
the color range closest to the observed one. Through this process, we obtained two
estimates of \teff\ for each object.

In addition, we introduced the method by \citet{frebel2013} to correct the
spectroscopically determined \teff\ from the spectral fitting. They reported
a procedure of adjusting the systematic offset between the photometric and spectroscopic
based \teff. This provides a third \teff\ estimate. We ultimately calculated
a bi-weight \teff\ estimate from the three \teff\ determinations for each star. We
then follow the iterative process laid out in Figure \ref{fig2} until convergence occurs,
and a final adopted \teff\ is obtained. Note that for the first trial of the \teff\ estimate
from the color-temperature relations, we adopted \logg\ and \feh\ derived by the spectral-fitting method.
We used the final adopted \teff\ for the abundance analysis. The standard error of the three temperatures
is reported as the uncertainty in \teff\ of each star.

\begin{deluxetable}{ccccc}
\tabletypesize{\scriptsize}
\tablecaption{Determined Atmospheric Parameters}
\tablehead{\colhead{Short} & \colhead{\teff} & \colhead{\logg} & \colhead{[Fe/H]} & \colhead{\vt} \\
           \colhead{ID}    & \colhead{(K)}   & \colhead{} & \colhead{}  & \colhead{(km s$^{-1}$)}}
\startdata
J0010 & 4309$\pm$81  & $0.34~^{+0.23}_{-0.26}$  & --2.48$\pm$0.11 & 2.3 \\
J0102 & 5974$\pm$35  & $4.55~^{+0.02}_{-0.02}$  & --3.09$\pm$0.06 & 0.7 \\
J0158 & 5165$\pm$44  & $2.48~^{+0.15}_{-0.14}$  & --3.04$\pm$0.05 & 1.7 \\
J0226 & 5461$\pm$98  & $3.00~^{+0.29}_{-0.23}$  & --3.78$\pm$0.05 & 1.9 \\
J0357 & 5631$\pm$27  & $3.49~^{+0.00}_{-0.04}$  & --2.75$\pm$0.05 & 1.2 \\
J0422 & 5027$\pm$48  & $2.07~^{+0.17}_{-0.15}$  & --3.33$\pm$0.06 & 2.3 \\
J0713 & 4380$\pm$91  & $0.22~^{+0.19}_{-0.20}$  & --3.15$\pm$0.08 & 2.3 \\
J0758 & 6453$\pm$254 & $4.33~^{+0.14}_{-0.20}$  & --2.96$\pm$0.13 & 1.0 \\
J0814 & 5853$\pm$237 & $3.52~^{+0.16}_{-0.22}$  & --3.39$\pm$0.05 & 1.2 \\
J0908 & 4713$\pm$145 & $1.21~^{+0.40}_{-0.32}$  & --3.67$\pm$0.06 & 2.3 \\
J0925 & 5631$\pm$111 & $3.39~^{+0.09}_{-0.14}$  & --3.53$\pm$0.06 & 1.8 \\
J1037 & 6569$\pm$423 & $4.37~^{+0.16}_{-0.28}$  & --2.50$\pm$0.05 & 1.3 \\
J1311 & 6510$\pm$237 & $4.23~^{+0.19}_{-0.30}$  & --2.74$\pm$0.04 & 1.0 \\
J1317 & 6810$\pm$336 & $4.25~^{+0.11}_{-0.17}$  & --2.37$\pm$0.05 & 1.3 \\
J1522 & 5698$\pm$27  & $3.43~^{+0.01}_{-0.04}$  & --3.70$\pm$0.06 & 1.5 \\
J1650 & 6800$\pm$335 & $3.95~^{+0.02}_{-0.05}$  & --2.17$\pm$0.05 & 1.3 \\
J1705 & 6580$\pm$238 & $4.29~^{+0.12}_{-0.16}$  & --2.64$\pm$0.05 & 1.3 \\
J2241 & 6608$\pm$152 & $3.94~^{+0.05}_{-0.02}$  & --2.71$\pm$0.13 & 0.7 \\
J2242 & 4857$\pm$144 & $1.46~^{+0.52}_{-0.39}$  & --3.40$\pm$0.08 & 2.3 \\
J2341 & 4628$\pm$46  & $0.96~^{+0.17}_{-0.15}$  & --3.17$\pm$0.07 & 2.3 \\
\enddata
\tablecomments{The objects named J0226 and J1522 are
the reference stars studied by \citet{ito2013} and \citet{matsuno2017}, respectively.}
\label{tab3}
\end{deluxetable}

\subsubsection{Surface Gravity}\label{sec:logg}

Iron abundances estimated from Fe I lines are known to
suffer from large non-local thermodynamic equilibrium (NLTE)
effects \citep[e.g.,][]{lind2012,amarsi2016},
and the number of available Fe II lines is very limited in the GRACES spectra
for our EMP candidates. This makes it difficult to determine the surface
gravity by the traditional approach of the ionization equilibrium (forcing
a balance between abundances estimated from the neutral atomic lines and singly ionized lines).

Instead, we determine the surface-gravity estimate through fitting isochrones.
For this technique to work, we first generated a hundred isochrones with
metallicities resampled from a normal distribution with an estimated
[Fe/H] and a conservative error of 0.2 dex. In this process, we employed
Yonsei-Yale \citep[{$Y^{2}$};][]{kim2002,demarque2004} isochrone,
and assumed a stellar age of 13 Gyr and [$\alpha$/Fe] = +0.3. Since
the minimum value of \feh\ is --3.75 in the $Y^{2}$ isochrone,
any metallicity of our target lower than this limit was forced to
be \feh\ = --3.75.

In the $M_{\rm V}$-\teff\ plane based on the generated isochrones,
we searched for a gravity that most closely reflected
the observables ($M_{\rm V}$ and \teff) of our target.
We took a median value from one hundred such gravity estimates.
The errors were taken at 34\% to the left and right from the median in the \logg\
distribution. When the metallicity tolerance is satisfied, as illustrated in Figure \ref{fig2},
the final estimate of \logg\ is adopted.

\subsubsection{Metallicity and Microturbulence Velocity}\label{sec:feh}

The metallicity was determined by minimizing the slope of log(EWs) of \ion{Fe}{1} lines
as a function of excitation potential. In this process, a total of 129 \ion{Fe}{1} lines were used,
and a starting model atmosphere was generated with \teff, \logg, and \feh\ estimated in
\textbf{Step b}, \textbf{Step c}, and the spectral fitting, respectively. The \vt\ value
was assumed according to the evolutionary stage (\vt\ = 0.5 \kms\ for dwarfs, 1.0 \kms\
for turnoff stars, and 2.0 \kms\ for giants). We adopted Kurucz model
atmospheres \citep{castelli2003}\footnote[10]{\url{https://www.user.oats.inaf.it/castelli/grids.html}},
using newly calculated opacity distribution functions with
updated opacities and abundances \citep{castelli2004}.
A desired model atmosphere was created by interpolating the existing
grid at given stellar parameters, using the ATLAS9 code \citep{castelli1997}.

Note that, because \ion{Fe}{2} lines are less subject to NLTE
effects \citep[e.g.,][]{lind2012,amarsi2016}, it is desirable to use
their abundance. However, because detectable \ion{Fe}{2} lines are very limited
for our EMP targets, we used the abundance derived from the \ion{Fe}{1} lines.
The uncertainty of [Fe/H] is given by the standard error of mean of the \ion{Fe}{1} abundances.

The microturbulence velocity (\vt) of a star was estimated by seeking
a flat slope of log (EWs) for \ion{Fe}{1} lines as a function of the reduced
equivalent width, log (EWs/$\lambda$). We used the line-analysis
program MOOG (\citealt{sneden1973,sobeck2011}). As shown in
Figure \ref{fig2}, if the metallicity M$_{\rm o}$ agrees with M$_{\rm i}$ within a difference of $\pm$0.02 dex,
then T$_{\rm o}$, G$_{\rm o}$, M$_{\rm o}$, and V$_{\rm t,o}$ are adopted as the final stellar parameters.
When the absolute difference between M$_{\rm i}$ and M$_{\rm o}$ is larger than 0.02 dex,
M$_{\rm i}$ is updated by M$_{\rm o}$, and the routine run again until the metallicity
tolerance is satisfied. Note that we do not update the \vt\ value when repeating the routine.

Table \ref{tab3} lists the derived stellar parameters by the iterative procedure.
According to the table, we have confirmed 8 VMP
and 10 EMP stars, except two objects (J0226 and J1522), which were already
analyzed by \citet{ito2013} and \citet{matsuno2017}, respectively. We observed these
objects as benchmark stars, and re-analyzed them to validate our approach to
determining the stellar parameters and chemical abundances.
In Section \ref{sec:comp_lit}, we present a detailed
comparison between our estimates and those of the previous studies.

\begin{figure}[t]
\epsscale{1.15}
\plotone{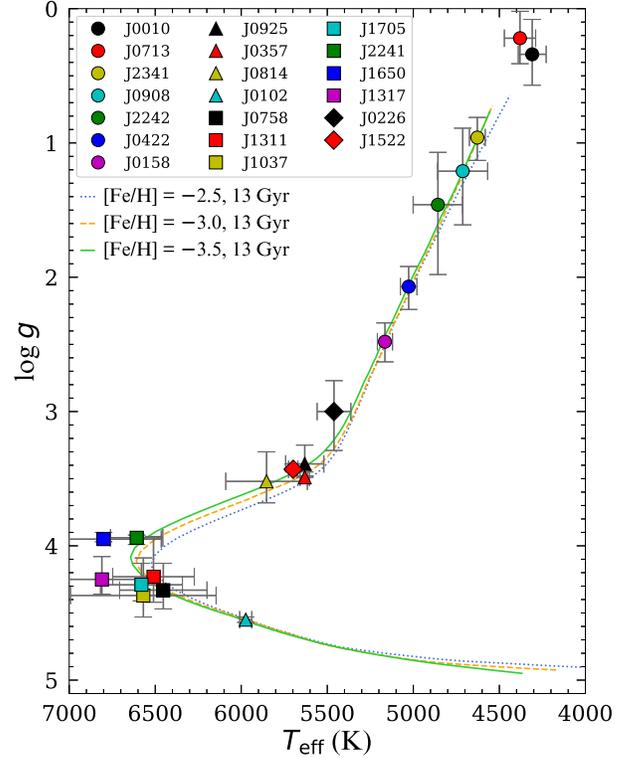}
\caption{Adopted surface gravity, as a function of effective temperature, for
our program stars. They are grouped into three different symbols according to \teff. The square
symbols represent turnoff stars, while the circle symbols represent
cool giants. The dwarfs and subgiants are denoted by triangles.
The two comparison stars are symbolized by diamonds (black for J0226 and red for
J1522). The three isochrones shown, with age 13~Gyr and [$\alpha$/Fe] = +0.3,
represent [Fe/H] = --2.5 (blue dotted),
[Fe/H] = --3.0 (orange dashed), and [Fe/H] = --3.5 (green solid).}
\label{fig3}
\end{figure}

Figure \ref{fig3} shows the position of our program stars
in the \logg\/-\teff\ plane, together with isochrones with age 13 Gyr and
[$\alpha$/Fe] = +0.3, for \feh\ = --2.5, --3.0, and --3.5,
represented by blue-dotted, orange-dashed, and green-solid lines, respectively.
The turnoff stars are indicated with squares, while the giants are indicated circles.
The dwarfs and subgiants are denoted by triangles. The two benchmark stars are
represented by diamonds (black for J0226 and red for J1522).
It is clear to see that our program stars populate a wide range of luminosity
classes and spectral types.

\section{Elemental Abundances}\label{sec:abun}

To derive the abundances of individual elements, we carried out a one-dimensional (1D),
local thermodynamic equilibrium (LTE) abundance analysis and spectral synthesis
using MOOG. We adopted the solar atmospheric abundances in \citet{asplund2009} to determine
the abundance ratios relative to the Sun. Whenever difficulties in measuring the
EWs of a line arose, we also performed spectral synthesis. The atomic line data were
compiled from several literature sources \citep{aoki2013,aoki2018,spite2018,placco2020,rasmussen2020}; and
the line information is listed in Table \ref{a_1} in the Appendix. In this section,
we address how we derived the chemical abundances.

\subsection{Chemical Abundances by Equivalent Width Analysis}\label{sec:ewa}

We derived the abundances for the odd-$Z$ elements (Na, K, and Sc), $\alpha$-elements (Mg and Ca),
and iron-peak elements (Cr, Mn, Ni, and Zn) by measuring the EWs of their neutral or singly ionized lines.
The Na abundance was determined using the \ion{Na} {1} doublet located at 5889\,{\AA} and
5895\,{\AA}, which are the only available sodium lines in low-metallicity stars.

Only one line at 7699\,{\AA} was used to measure the K abundance, while
five \ion{Sc}{2} lines were used to measure [Sc/Fe]. We used four \ion{Mg}{1} and
12 \ion{Ca}{1} lines to derive Mg and Ca abundance ratios, respectively.

Concerning the abundances of the iron-peak elements, we utilized 10 \ion{Cr}{1},
3 \ion{Mn}{1}, 15 \ion{Ni}{1}, and 2 \ion{Zn}{1} lines to determine [Cr/Fe], [Mn/Fe], [Ni/Fe], and [Zn/Fe], respectively. Depending on the effective temperature, metallicity, and S/N of a spectrum, the number of measured EWs for each element
differs from star to star.

\subsection{Chemical Abundances by Spectral Synthesis}\label{sec:ss}

\begin{figure}[t]
\epsscale{1.15}
\plotone{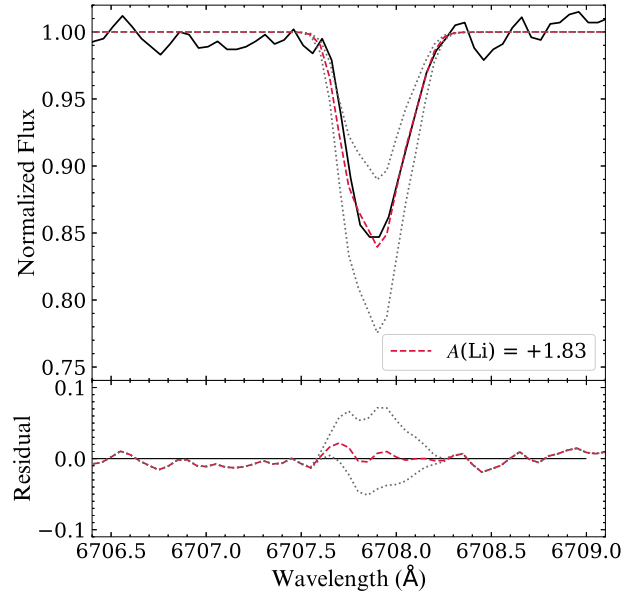}
\caption{Top panel: Example of Li spectral synthesis for one  of our program stars (J0357).
The black-solid line is the observed spectrum, and the red-dashed line is the best-matching
synthetic spectrum. The two dotted lines are the upper and lower limits
of the spectral fits, which deviate by $\pm$0.2 dex. Bottom panel: Residual
plot of the top panel.}
\label{fig4}
\end{figure}

The abundance ratios for Li, C, and Ba were determined by spectral synthesis, using
atomic line data from \texttt{linemake}. For the Li
abundance ratio, we used the \ion{Li}{1} resonance doublet line around 6707\,{\AA}.
Figure \ref{fig4} provides an example of the Li spectral synthesis (J0357).
The black-solid line is the observed spectrum;
the red-dashed line is the best-matching synthetic spectrum. The two dotted lines present
the upper and lower limits of the spectral fits by deviating by $\pm$0.2 dex in $A$(Li) from
the best-matching one. We conservatively considered these limits as the error of
the determined [Li/Fe].

\begin{figure}[t]
\epsscale{1.15}
\plotone{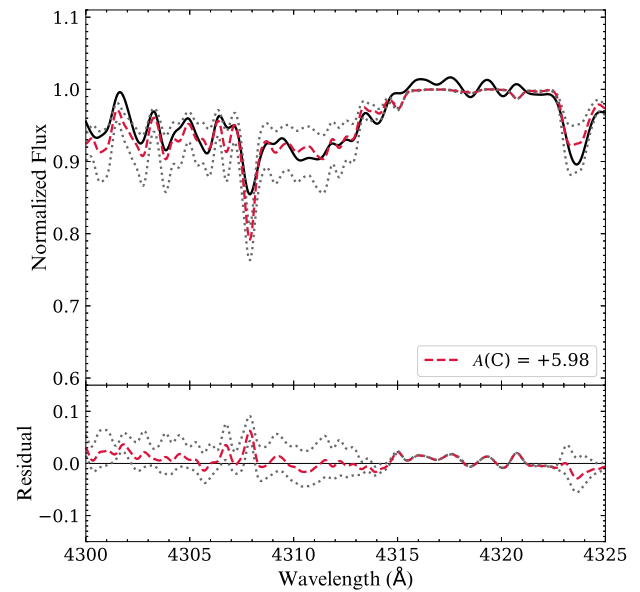}
\caption{Same as in Figure \ref{fig4}, but for C and J0226.}
\label{fig5}
\end{figure}

\begin{figure*}[t]
\epsscale{1.15}
\plotone{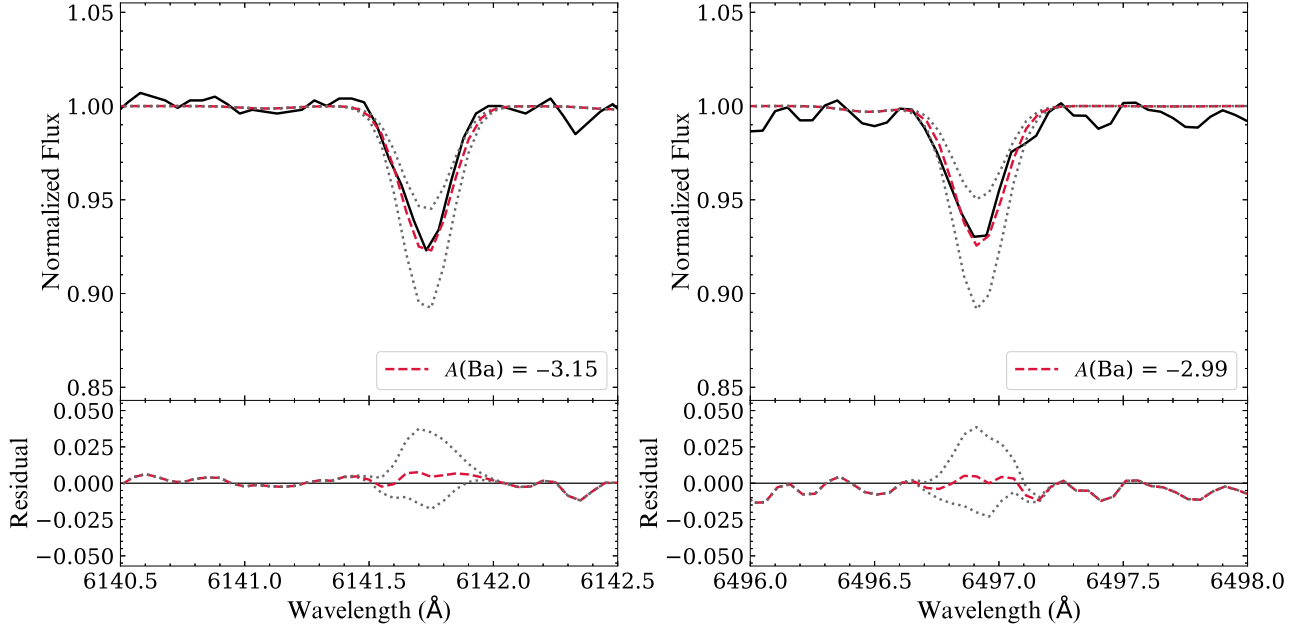}
\caption{Same as in Figure \ref{fig4}, but for Ba and J0713.}
\label{fig6}
\end{figure*}

\begin{deluxetable*}{lcrcccccccccr}
\tabletypesize{\scriptsize}
\renewcommand{\tabcolsep}{2pt}
\tablecaption{[X/Fe] Abundance Ratios for Program Stars}
\tablehead{\colhead{ID} & \colhead{\ion{Li}{1}} & \colhead{C} & \colhead{\ion{Na}{1}} & \colhead{\ion{Mg}{1}} & \colhead{\ion{K}{1}} & \colhead{\ion{Ca}{1}} & \colhead{\ion{Sc}{2}} & \colhead{\ion{Cr}{1}} & \colhead{\ion{Mn}{1}} & \colhead{\ion{Ni}{1}} & \colhead{\ion{Zn}{1}} & \colhead{\ion{Ba}{2}}}
\startdata
J0010 & \nodata & 0.37 & 0.24$\pm$0.07 & 0.73$\pm$0.04 & 0.13$\pm$0.09 & 0.19$\pm$0.05 & --0.33$\pm$0.11 & --0.10$\pm$0.05 & 0.09$\pm$0.06 & 0.13$\pm$0.05 & 0.00$\pm$0.12 & --1.72$\pm$0.06 \\
J0102 & 3.79    & 0.75 & --0.60$\pm$0.12 & 0.21$\pm$0.12 & \nodata & \nodata & \nodata & --0.04$\pm$0.17 & \nodata & \nodata & \nodata & \nodata \\
J0158 & 2.78    & 0.11 & 1.14$\pm$0.08 & 0.06$\pm$0.06 & \nodata & 0.41$\pm$0.04 & \nodata & --0.09$\pm$0.06 & --0.01$\pm$0.10 & 0.30$\pm$0.06 & 0.61$\pm$0.10 & $<$--1.60$\pm$0.20 \\
J0226 & \nodata & 1.33 & 0.29$\pm$0.04 & 0.59$\pm$0.05 & \nodata & \nodata & \nodata & \nodata & \nodata & \nodata & \nodata & \nodata \\
J0357 & 3.53    & 0.40 & 0.13$\pm$0.07 & 0.34$\pm$0.04 & \nodata & 0.33$\pm$0.03 & 0.50$\pm$0.10 & --0.16$\pm$0.05 & 0.01$\pm$0.09 & 0.37$\pm$0.07 & 0.81$\pm$0.10 & --0.70$\pm$0.04 \\
J0422 & \nodata & $<$0.11 & --0.46$\pm$0.08 & 0.18$\pm$0.04 & \nodata & 0.21$\pm$0.05 & \nodata & --0.36$\pm$0.08 & \nodata & 0.23$\pm$0.08 & \nodata & --1.15$\pm$0.20 \\
J0713 & \nodata & 0.08 & 0.37$\pm$0.08 & 0.31$\pm$0.06 & 0.18$\pm$0.10 & 0.00$\pm$0.05 & --0.26$\pm$0.10 & --0.33$\pm$0.03 & --0.02$\pm$0.10 & --0.01$\pm$0.04 & 0.31$\pm$0.13 & --2.08$\pm$0.04 \\
J0758 & \nodata & $<$1.20 & --0.75$\pm$0.24 & 0.06$\pm$0.17 & \nodata & \nodata & \nodata & --0.11$\pm$0.24 & \nodata & \nodata & \nodata & \nodata \\
J0814 & 4.10    & $<$1.10 & --0.05$\pm$0.05 & 0.33$\pm$0.04 & \nodata & 0.61$\pm$0.07 & \nodata & \nodata & \nodata & \nodata & \nodata & \nodata \\
J0908 & \nodata & $<$0.01 & --0.19$\pm$0.08 & 0.17$\pm$0.07 & 0.96$\pm$0.11 & 0.35$\pm$0.04 & \nodata & --0.47$\pm$0.08 & \nodata & 0.35$\pm$0.08 & \nodata & $<$--1.40$\pm$0.20 \\
J0925 & 3.80    & 1.40 & --0.28$\pm$0.05 & 0.35$\pm$0.05 & \nodata & \nodata & \nodata & \nodata & \nodata & \nodata & \nodata & \nodata \\
J1037 & 3.66    & 1.10 & --0.41$\pm$0.06 & 0.30$\pm$0.03 & \nodata & 0.12$\pm$0.03 & \nodata & --0.24$\pm$0.09 & \nodata & 0.34$\pm$0.09 & \nodata & \nodata \\
J1311 & \nodata & $<$1.10 & --0.12$\pm$0.03 & --0.22$\pm$0.03 & \nodata & \nodata & \nodata & --0.35$\pm$0.02 & \nodata & \nodata & \nodata & $<$--0.20$\pm$0.20 \\
J1317 & 3.73    & $<$0.60 & --0.08$\pm$0.08 & 0.30$\pm$0.06 & \nodata & 0.24$\pm$0.05 & \nodata & --0.09$\pm$0.11 & \nodata & 0.69$\pm$0.06 & \nodata & \nodata \\
J1522 & 4.30    & $<$1.03 & \nodata & 0.42$\pm$0.07 & \nodata & \nodata & \nodata & \nodata & \nodata & \nodata & \nodata & \nodata \\
J1650 & 3.47    & $<$0.30 & --0.60$\pm$0.10 & 0.13$\pm$0.07 & 0.69$\pm$0.14 & 0.35$\pm$0.04 & \nodata & --0.22$\pm$0.10 & \nodata & 0.73$\pm$0.10 & \nodata & $<$--0.64$\pm$0.20 \\
J1705 & 3.55    & $<$0.30 & --0.44$\pm$0.06 & 0.25$\pm$0.03 & \nodata & 0.30$\pm$0.02 & \nodata & --0.07$\pm$0.08 & \nodata & \nodata & \nodata & $<$--0.50$\pm$0.20 \\
J2241 & \nodata & $<$1.30 & --0.60$\pm$0.15 & --0.61$\pm$0.15 & \nodata & \nodata & \nodata & \nodata & \nodata & \nodata & \nodata & $<$--0.50$\pm$0.20 \\
J2242 & \nodata & 0.05 & --1.02$\pm$0.26 & 0.24$\pm$0.04 & 1.48$\pm$0.36 & 0.71$\pm$0.14 & --0.11$\pm$0.36 & 0.01$\pm$0.26 & \nodata & 0.89$\pm$0.36 & \nodata & \nodata \\
J2341 & \nodata & 0.03 & 0.74$\pm$0.10 & 0.36$\pm$0.04 & 0.59$\pm$0.12 & 0.25$\pm$0.04 & 0.09$\pm$0.07 & --0.38$\pm$0.04 & \nodata & 0.07$\pm$0.09 & \nodata & --1.61$\pm$0.06 \\
\enddata
\tablecomments{The errors of Li and C are conservatively assumed to be 0.2 dex; see text for the errors estimated for other elements.
The $<$ symbol indicates the upper limit estimate. Note that the [C/Fe] value is corrected for the evolutionary
stage following the prescription of \citet{placco2014}. The applied correction is listed in Table \ref{a_2} in the Appendix.}
\label{tab4}
\end{deluxetable*}

\begin{figure} [b] 
\epsscale{1.15}
\plotone{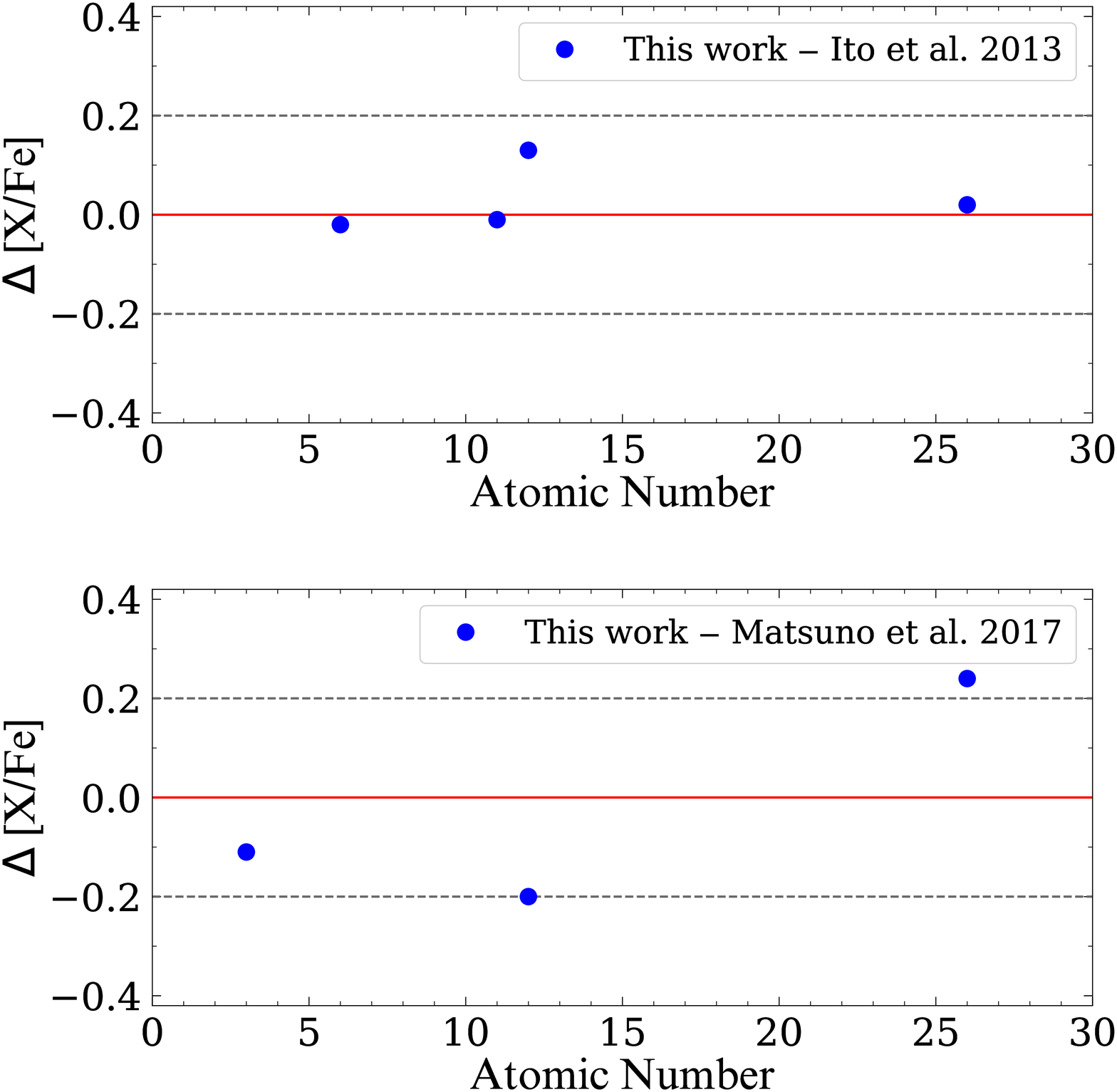}
\caption{Differences in the chemical abundances between our study and the work by \citet{ito2013} (top panel)
and \citet{matsuno2017} (bottom panel). The red-solid line shows the zero point, while the gray-dashed
lines indicate deviations of $\pm$0.2 dex.}
\label{comp}
\end{figure}

We estimated the C-abundance ratio by spectral synthesis of the CH
$G$-band around 4310\,{\AA}. While estimating [C/Fe], we adopted
$^{12}$C/$^{13}$C of 89 (\logg $>$ 3.75), 30 (3.0 $<$ \logg\ $\leq$ 3.75),
or 20 (\logg\ $\leq$ 3.0) according to the luminosity class of our
program stars \citep{asplund2009}, and we used spectra degraded to $R$ = 10,000
in order to increase the S/N around this feature.
Figure \ref{fig5} exhibits an example of the spectral synthesis for
the carbon-abundance determination (J0226). As before, the observed spectrum
is represented by the black line, while the best-matching synthetic spectrum is represented
by the red-dashed line. The upper and lower limits of the spectral fits are denoted by the
two dotted lines, which are located at $\pm$0.2 dex away from the determined value.
This limit is considered as the uncertainty of the estimated [C/Fe]. We corrected the
determined [C/Fe] following the prescription of
\citet{placco2014} to restore the natal carbon on the surface of stars that have been
altered due to evolutionary effects (the depletion of C as stars climb the giant branch).
Table \ref{a_2} in the Appendix summarizes the measured [C/Fe] and its corrected value for each
object.

We were able to measure the abundance of one of the neutron-capture
elements (Ba), using two \ion{Ba}{2} lines at 6141\,{\AA} and 6497\,{\AA}, through spectral synthesis.
Isotopic splitting using the values from \citet{sneden2008} and hyperfine structure
were considered when the line information is generated with \texttt{linemake}. An example is displayed in
Figure \ref{fig6}; the lines are the same
as in Figure \ref{fig4}. The error estimate comes from the
standard error of the two estimates, or is conservatively set to 0.2 dex for
the objects with only one line available.

\subsection{Errors of Derived Abundances}\label{sec:error}

When computing the error in the derived abundance for
individual elements, we have considered both the random errors arising from
the line-to-line scatter and the systematic error caused by the errors of
the adopted stellar parameters. The random error represents the variation of the individual lines
for a given element, calculated by $\sigma/\sqrt{N}$, where
$N$ is the number of lines and $\sigma$ is the standard deviation
of the derived abundance. This has the advantage of including
the uncertainty from the oscillator strength values of the lines considered.
In case for an element where the number of detectable
lines is less than three, we took $\sigma$ from the Fe I lines and
computed its standard error by $\sigma$(\ion{Fe}{1})/$\sqrt{N}$.

To derive the systematic error due to stellar atmospheric-parameter errors,
we perturbed the stellar parameters by $\pm$100\,K for \teff, $\pm$0.2 dex for \logg,
and $\pm$0.2 \kms\ for \vt\ one at a time, and estimated the systematic error for
each case. The systematic error for each parameter is an average
of the two values derived by perturbing two ways ($\pm$). The final reported error on
the abundance of each element is the quadratic sum of the random error and the systematic
error. Table \ref{tab4} summarizes the estimated abundances for our program stars;
Table \ref{a_3} in the Appendix provides more detailed abundances and their associated errors measured for our VMP stars.

\subsection{Abundance Comparison with Previous Studies}\label{sec:comp_lit}

Among our sample of stars, two objects (J0226 and J1522) were previously
studied by \citet{ito2013} and \citet{matsuno2017}, respectively.
The two studies observed these stars with Subaru/HDS to obtain
high-resolution ($R \sim$ 60,000) spectra, and carried out a detailed
abundance analysis. We observed these stars to validate the strategy
of our stellar parameter and abundance determination by comparing our estimates
with their values.

We intentionally observed the bright ($V \sim$ 9.1) object J0226
multiple times with different exposure times to obtain spectra with a range of S/N.
This provides an opportunity to check how the S/N of a spectrum affects the
derivation of the stellar parameters and abundances. By comparing
the literature stellar parameters and chemical abundances with the ones derived from
our spectrum that has similar S/N to the rest of our program stars,
we can appreciate if the stellar parameters and abundances are reliable
at S/N ($\sim$ 60), which is the mean S/N of the relatively bright ($g \leq$ 15.9)
objects among our program stars. Similarly, the relatively faint object J1522 can be used to
validate the derived abundance for the faint ($g > $ 15.9) objects with S/N $\sim$ 40.

We obtained \teff\ = 5461\,K, \logg\ = 3.0, [Fe/H] = --3.78 for the object J0226
from a GRACES spectrum with S/N = 75. Comparison with the stellar
parameters estimated by \citet{ito2013} reveals that our estimates are 30\,K
higher in \teff, 0.4 dex lower in \logg, and 0.01 dex higher in [Fe/H], indicating
good agreement. In addition, we compared the elemental differences in common between
their study and our work, as shown in the top panel of Figure \ref{comp}. The
gray-dashed lines indicate the abundance deviation of $\pm$0.2 dex. The plot clearly
indicates a good match within $\pm$0.2 dex, validating our abundance analysis.

The stellar parameters derived from our GRACES spectrum for J1522
are \teff\ = 5698\,K, \logg\ = 3.43, and [Fe/H] = --3.7. This object
is relatively faint, and has a low S/N of 38. Comparing
with the values reported by \citet{matsuno2017}, our estimates
are about 200\,K higher for \teff, 0.3 dex lower for \logg, and 0.25 dex more
metal-rich. We ascribe the [Fe/H] difference primarily to the temperature difference.
We note that they adopted \teff\ = 5505\,K determined by fitting the H$\beta$ profile
for their abundance analysis, instead of \teff\ derived from $V-K_{\rm s}$, which
is 5813\,K, and much closer to our estimate. Nonetheless, in the bottom panel of Figure \ref{comp},
we see that Li and Mg abundance agree within $\pm$0.2 dex, again validating our abundance
determinations even at lower S/N. Note that in the figure, we did not include the carbon-abundance ratio
because both theirs and ours are the upper limit estimate.

\begin{figure*} [!t]
\epsscale{1.15}
\plotone{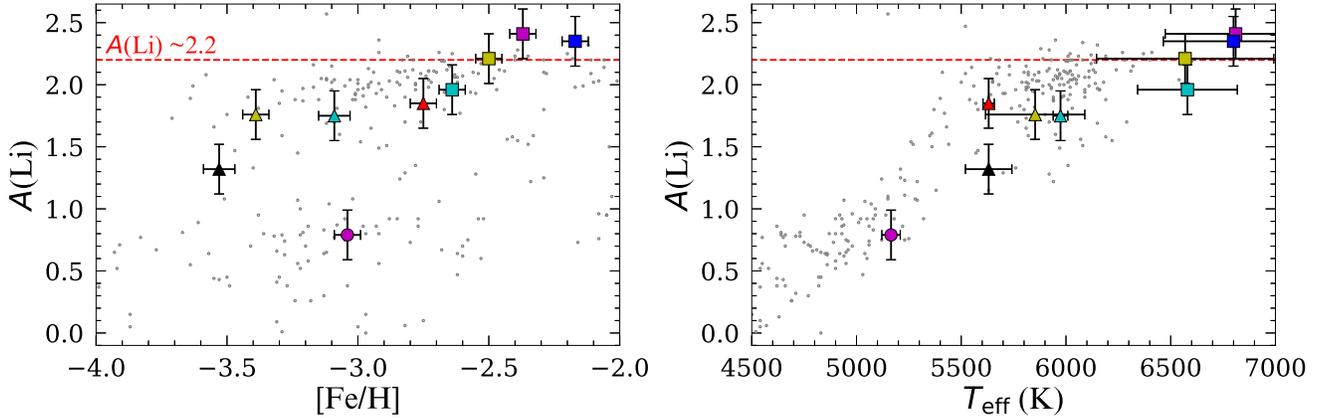}
\caption{\ali\ distribution, as a function of \feh\ (left panel)
and a function of \teff\ (right panel). The colored symbols are our program stars,
while the gray circles are gathered from \citet{roederer2014}.
The red-dashed line indicates the Spite plateau at $A$(Li) = 2.2.}
\label{ali}
\end{figure*}

\section{Discussion}\label{sec:abun_comp}

In this section, we compare the chemical abundances of our
program stars with those from other previously studied Galactic halo stars.
We find that most of our program stars follow the general trends of the previous data
for VMP stars.  However, there are a few exceptional stars with peculiar abundances.
We first discuss the overall abundance trends of individual elements, and
then focus on the chemically peculiar objects. These stars will be
valuable to provide constraints on rare astrophysical production sites or nucleosynthetic channels.
Note that in the following discussion that we only consider 1D LTE abundances,
to be consistent with other literature abundances, which are not generally
corrected for either 3D or NLTE effects. We do not include the two benchmark
stars in the following discussion.

\subsection{Overall Abundance Trends}

\subsubsection{Lithium}\label{sec:Li}

Lithium is one of the most important elements, as its abundance can be used
to constrain Big Bang nucleosynthesis. Figure \ref{ali} exhibits the behavior
of the absolute Li abundances, $A$(Li), as a function of [Fe/H] (left panel) and \teff\ (right panel).
The colored symbols are our program stars, while the gray circles are adopted
from \citet{roederer2014}, which are not corrected for NLTE effects.
The stellar sequence located at \ali\ $\sim$ 2.2
for the region of --2.9 $<$ [Fe/H] is known as the Spite plateau \citep{spite1982}.
Although the location of the Spite plateau varies over \ali\ = 2.10 -- 2.35 from study to
study \citep[e.g.,][]{ryan1999,ryan2000,bonifacio2010,sbordone2010,simpson2021}, we indicate
it as the red-dashed line at $A$(Li) = 2.2, which is the most commonly accepted value.
The primordial Li predicted by the Big Bang nucleosynthesis is in the range \ali\ = 2.67 --
2.74 \citep{spergel2007,cyburt2016,coc2017};  the difference between the
primordial Li and the Spite plateau is known as the longstanding cosmological lithium problem.

Inspection of the figure reveals that our sample of stars follows the general
trend of the other literature abundances, but there are some subtle differences as well.
The object J0158 (magenta circle) with \ali\ $<$ 1.0 is a giant.
When stars leave the main sequence (MS), their surface Li material is reduced
by dilution caused by the first dredge-up (FDU). Observationally,
we can see the difference in Li abundances between the
turnoff (TO) stars and giants, and then the Li abundance for giants does not
change much across a large range of [Fe/H] after the FDU to the red giant
branch (RGB) bump \citep[e.g.,][]{lind2009}. The Li abundance further decreases as a
star ascends above the RGB bump (\logg\ $<$ 2.0), due to poorly understood
extra mixing. This characteristic abundance pattern due
to the stellar evolution is clearly seen in Figure \ref{ali}.

Among our program stars, the four stars hotter than 6500\,K have $A$(Li) values
closely scattered around the Spite plateau, as can be seen in the right panel. These
objects can be very useful to check the possibility of the extension of the
Li plateau to the hotter temperature region, where relatively few data exist at present.
Three stars (black, red, and yellow triangles) are subgiants, which are
close to the giant phase (see Figure \ref{fig3}). Their Li abundance is slightly
lower than the overall trend, probably due to the increasing convection zone.
The object J0102 (cyan triangle) is a warm dwarf with [Fe/H] $<$ --3.0.
Its expected convective zone is not very deep; hence its surface Li should be
preserved without much destruction since its birth. However, its \ali\ is
slightly lower than the Spite plateau. One possible cause of its lower $A$(Li)
compared to the Spite plateau and other literature sample is the temperature scale we have adopted.
It is known that a temperature difference of 100\,K results in the
change of 0.08 dex of $A$(Li). Consequently, it may be possible that our temperature
scale is slightly lower for dwarf stars than the ones used in the literature.
An analysis of a larger number of stars in a uniform and
consistent manner is clearly critical when discussing the observed Li trend.
A follow-up study of our TO and dwarf stars can provide
useful constraints on the primordial lithium problem, considering
their evolutionary stage and metallicity.

One more interesting aspect is that our sample of stars also gives a clue
to a behavior known as the ``double sequence'' in the Spite plateau noted
by \citet{melendez2010}, whereby the stars with [Fe/H] $<$ --2.5 possess slightly lower $A$(Li)
than the ones with [Fe/H] $>$ --2.5. The similar behavior
was reported in other studies \citep[e.g.,][]{ryan1999,ryan2000,sbordone2010},
and we notice this characteristic in the figure as well.

\begin{figure*}
\epsscale{1.1}
\plotone{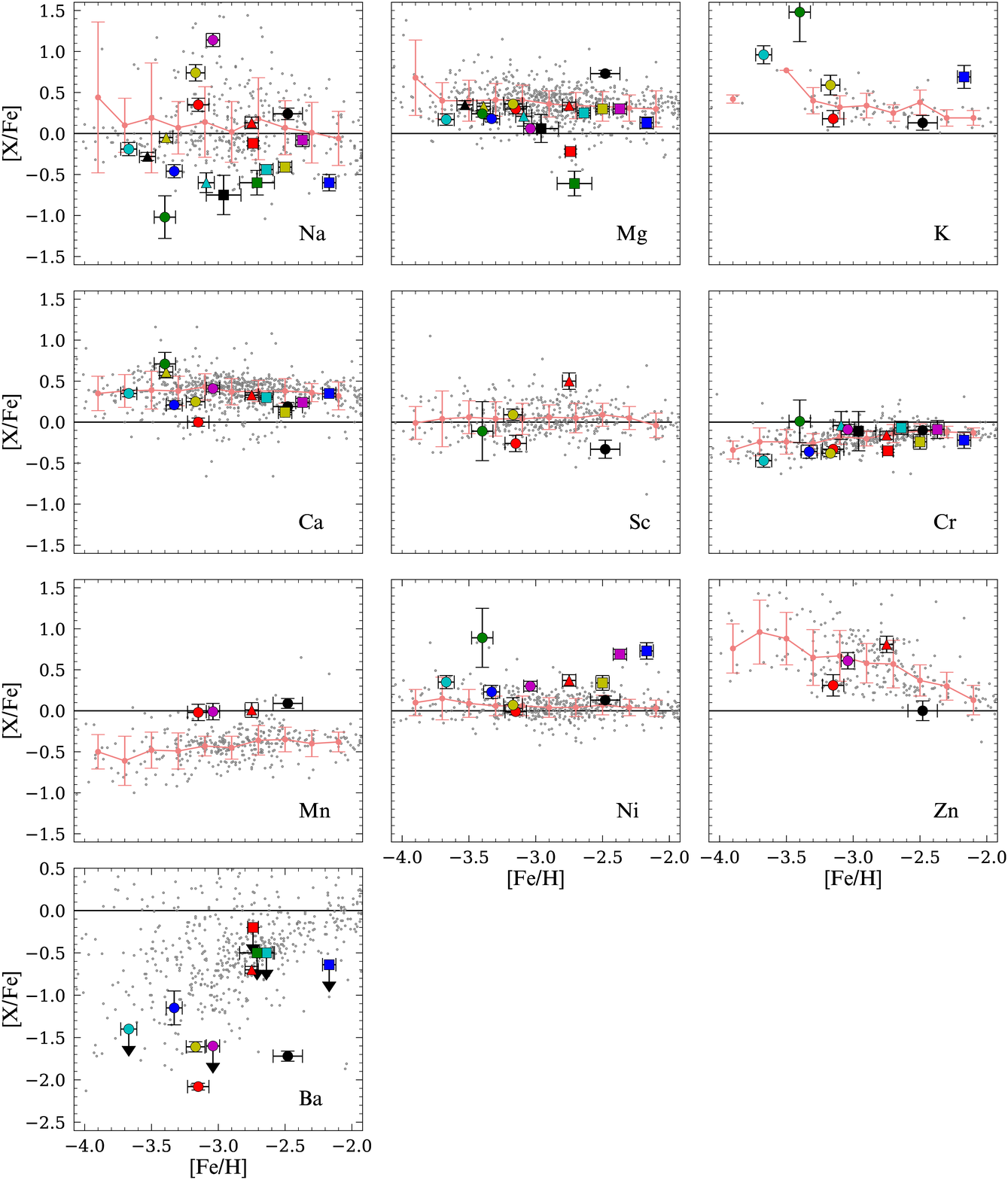}
\caption{[X/Fe], as a function of [Fe/H], for the odd-$Z$ (Na, K, and Sc),
$\alpha$- (Mg and Ca), iron-peak (Cr, Mn, Ni, and Zn), and neutron-capture (Ba) elements.
Our program stars are represented by colored symbols, while
the objects compiled from various literature sources \citep{venn2004,aoki2013,yong2013,roederer2014} are shown with
gray circles. The pink-solid line and error bars represent the 3$\sigma$-clipped mean
trend and standard deviation of the literature sample for a bin size of 0.2 dex in [Fe/H], respectively.
Our estimates are not included in these calculations. The black-dashed line
denotes the solar abundance ratio. The colors and symbols for our program stars are the same as in
Figure \ref{fig3}. The down arrow indicates  an upper limit.}
\label{other}
\end{figure*}

\subsubsection{Odd-$Z$ Elements: Na, K, and Sc}\label{sec:odd}

Figure \ref{other} compares our derived abundances with other previously studied
Galactic VMP stars for the odd-$Z$ (Na, K, and Sc), $\alpha$- (Mg and Ca),
iron-peak (Cr, Mn, Ni, and Zn), and neutron-capture (Ba) elements. Our program stars
are represented by the large colored symbols, while the objects compiled
from various literature sources are indicated by gray circles \citep{venn2004,aoki2013,yong2013,roederer2014}.
Note that, even though the NLTE corrected values for some elements are available in some
studies, we used LTE values for this comparison.
The colors and symbols for our program stars are the same as in Figure \ref{fig3}.
The pink-solid line and error bar represent the $3\sigma$-clipped mean
trend and standard deviation for the literature sample using a bin size of 0.2 dex in [Fe/H], respectively.
Our estimates are not included in the calculations. The
black-solid line denotes the solar abundance ratio. Some
selection biases may be underlying in the individual samples
from the literature, according to their science goals. Nevertheless, Figure \ref{other}
can convey useful information on understanding the early chemical evolution
of the MW, as well as for identifying the chemically peculiar objects.

Among the odd-$Z$ elements, we were able to derive Na, K, and Sc abundance
ratios. The Na-abundance ratios of our program stars in Figure \ref{other}
show that many of them are consistently lower than the main trend (the pink line),
even though some of them are in the  range of other VMP stars.
Sodium is formed during C-burning in massive stars, as well as during the hydrogen
burning via the Ne-Na cycle \citep[e.g.,][]{romano2010}. It is dispersed into the
interstellar medium (ISM) by CCSNe as well as by mass loss from asymptotic giant
branch (AGB) stars. Hence, the objects with [Na/Fe] $<$ --0.5 may be
formed in isolated gas clouds, which were not well-mixed or chemically enriched
by a few CCSNe or AGB stars. Two objects stand out in the figure (J0158 and J2242),
with [Na/Fe] $>$ +1.0 and [Na/Fe] $<$ --1.0, respectively. We discuss these
objects in more detail in Section \ref{sec:na}.

Unlike Na, K and Sc are the product of explosive Si-burning and/or O-burning in
the CCSN stage \citep{woosley1995}. They are also created by the $\nu$-process
in CCSNe \citep{kobayashi2011b}. Hence, they are good probes for tracing the
nucleosynthesis process in explosive events. We measured K abundances for six stars.
Figure \ref{other} indicates that the K-abundance trend increases with decreasing [Fe/H]
for [Fe/H] $<$ --3.2; one of our program stars (J0908; cyan circle) is on this increasing trend.
There are two objects (J1650 and J2242) at [Fe/H] $\sim$ --2.2 and --3.4, respectively,
which possess much higher [K/Fe] than the other objects at the same metallicity, as further discussed in Section \ref{sec:na}.

The Sc-abundance ratio was derived for five objects, and two of their abundances
exhibit somewhat larger scatter than the other Galactic field stars,
as can be seen in Figure \ref{other}. Studies \citep[e.g.,][]{chieffi2002} show that
its production depends on the mass of the CCSN progenitors. If stars have formed in gas clouds
which were enriched by a few SNe from stars with different masses, we would expect to observe a large scatter
in the Sc abundances, which may be the case for our program stars. It is known that
Galactic chemical-evolution models do not reproduce the evolution of the observed K and Sc abundances;
the model predictions are consistently lower than the observations for these two elements \citep[e.g.,][]{kobayashi2020}.

\subsubsection{Alpha Elements: Mg and Ca}\label{sec:alpha}

The so-called $\alpha$-elements are mostly produced by the hydrostatic and explosive
nucleosynthesis process in massive stars, and they are ejected into the
interstellar space by CCSNe explosions. Specifically, Mg is produced through
a hydrostatic nucleosynthetic process during the C-burning of a massive star. Calcium is mainly
created during the O-burning of CCSNe. Although the $\alpha$-elements are mainly formed
by CCSNe, some amount of Ca is also generated by SNe Ia \citep{iwamoto1999}. Thus, the
Mg abundance, which has a single production channel, is frequently used
as an important tracer to examine the contribution of CCSNe. In addition,
thanks to the strong Mg absorption lines, its abundance is readily
measured from the spectra of VMP stars, and it has proven to be a powerful
tool to track the star-formation history.

Among the $\alpha$-elements, we were able to derive the abundances of Mg
for all our EMP candidates and Ca for 13 objects from the GRACES spectra.
In Figure \ref{other}, we can observe that most of our targets exhibit a similar
[Mg/Fe] trend at [Mg/Fe] $\sim$ +0.3, with small scatter relative to other VMP stars,
but a few objects (J0010, J1311, and J2241) distinguish themselves from the rest.
We discuss these objects in Section \ref{sec:mg}. The Ca abundance for most of our program stars
exhibit a very small dispersion, following the trends of other VMP stars.
The behavior of the observed $\alpha$-element abundances indicates
that CCSNe had a dominant role in the chemical enrichment for our program stars.

\subsubsection{Iron-peak Elements: Cr, Mn, Ni and Zn}\label{sec:iron-peak}

Iron-peak elements are produced during Si-burning, and also can be synthesized in
both SNe Ia and CCSNe \citep{kobayashi2009,kobayashi2020}. Since the innermost
region of C+O white dwarfs is sufficiently hot to burn Si, iron-peak elements can also be
synthesized \citep{hoyle1960,arnett1971} by this pathway.
CCSNe contribute to explosive nucleosynthesis to form iron-peak elements
in two distinct regions: incomplete and complete Si-burning
regions \citep[e.g.,][]{hashimoto1989,woosley1995,thielemann1996,umeda2002}.
Chromium and Mn are synthesized in the incomplete Si-burning region in the ejecta
of CCSNe, while Ni and Zn are produced in the complete Si-burning
region of the deeper inner portion of the ejecta during the CCSN explosion.

We are able to determine the abundance of Cr, Mn, Ni, and Zn among the iron-peak
elements for some of our program stars. Inspection of Figure \ref{other}
reveals that the overall trend of Cr, Mn, and Ni of the VMP/EMP stars exhibits
a relatively small dispersion, but the Zn abundance shows a rather large
scatter among the iron-group elements.

Similar to other Galactic VMP stars, the Cr abundances of our
targets exhibit ordinary behavior, consistent with other literature results.
The very small dispersion over a range
of metallicities suggests that their formation is closely related \citep[e.g.,][]{reggiani2017}.
We also notice the declining trend with decreasing
[Fe/H], suggesting chemical evolution driven by CCSNe in the early epochs of the MW.

Figure \ref{other} indicates the most of the Galactic field stars
exhibit a decreasing trend of the Mn abundance with decreasing metallicity, similar to  the Cr abundance,
supporting the claim of early chemical enrichment from CCSNe.
We have measured the Mn abundance for four stars. It can be seen in
Figure \ref{other} that their abundances are near the solar value, independent
of their metallicity, and appear elevated compared to the main locus. However,
the absorption lines of Mn are often too weak in VMP/EMP stars,
causing systematic uncertainty. This might be the case for our
targets, and three of the four objects rely on one line,
resulting in uncertain measurements. An abundance analysis based on higher-quality
spectra are required to confirm the Mn enhancements in our four program stars.

The Ni abundances of the previously studied VMP stars are generally close
to the solar level, as are some of our program stars. Notably, we
observe three objects (J1317, J1650, and J2242) that have [Ni/Fe] $>$ +0.5.
We further discuss the objects J1650 and J2242 in Section \ref{sec:na}.
The abundances for elements other than Ni in J1317 appear normal, so it may
have been enriched by SNe Ia, considering its relatively high metallicity.
However, the Ni abundance is derived from one absorption
line, and its uncertainty is rather large; its chemical
peculiarity is desirable to confirm with additional lines.

In Figure \ref{other}, the average Zn-abundance pattern tends to increase with decreasing metallicity.
This trend has been argued to be caused by problems detecting the Zn lines
at low metallicity (see \citealt{yong2021} for a more detailed discussion).
This is particularly problematic for EMP stars. These factors
may explain the larger dispersion of Zn compared to other iron-peak elements.
Notwithstanding, the Zn-abundance trend can be used to understand the physics of CCSNe.
Zinc is generated in the deepest region of
hypernovae \citep[HNe;][]{umeda2002}, and a more significant explosion
energy leads to higher [Zn/Fe] ratios \citep{nomoto2013}. Consequently,
HNe may be responsible for higher values of [Zn/Fe] at the lowest metallicity.

\subsubsection{Neutron-capture Element: Ba}\label{sec:ba}

Heavier elements beyond the iron peak are created by capturing neutrons and their
subsequent $\beta$-decay. At least two processes -- the slow ($s$-) and rapid ($r$-) neutron-capture
processes -- are thought to be responsible for the synthesis of these elements.
An intermediate neutron-caption process (the $i$-process; \citealt{cowan1977}) may also be involved.
The slow neutron-capture process occurs in an environment
where the neutron flux is much lower than the rapid neutron-capture
process. The main $r$-process elements are created during violent events
such as CCSNe, neutron star mergers, gamma ray burst, etc. \citep{nishimura2015,drout2017,cote2019,siegel2019},
whereas the main $s$-process elements are produced during the AGB
phase of low-mass stars \citep{suda2004,herwig2005,komiya2007,masseron2010,lugaro2012}.
A number of sites have been suggested for the $i$-process,
including AGB stars \citep{hampel2016,cowan2021,choplin2022} and rapidly accreting white dwarfs \citep{denissenkov2017,denissenkov2019,cote2018}.

In the GRACES spectra of our program stars, the only measurable lines for
neutron-capture elements are two \ion{Ba}{2} lines. We took an average
of the abundances derived from these two lines. The bottom panel of
Figure \ref{other} displays [Ba/Fe] as a function of [Fe/H]
for our sample (colored symbols) and other field stars.
We clearly observe a very large scatter especially in the low-metallicity region,
and all our program stars have [Ba/Fe] $<$ 0. The large spread
of the Ba abundance at low metallicity is a well-known pattern \citep{ryan1996,aoki2005,roederer2013}.

Because the favored astrophysical site for the operation
of the $s$-process is AGB stars, stars with the lowest metallicity, and oldest ages,  do not
have sufficient time to be polluted by progenitors during their thermally pulsing AGB phase.
Instead, in the early Universe, Ba could be produced by the $r$-process
in massive stars \citep{travaglio1999,cescutti2006}, or by
fast rotating, low-metallicity stars \citep{frischknecht2016,choplin2018}.
Thus, we expect the Ba abundance for our EMP stars was probably produced by
the main $r$-process. In both cases, the main neutron source is the
$^{22}$Ne($\alpha$,$n$)$^{25}$Mg reaction. In addition, due to inefficient
mixing of the ISM in the early Galaxy, stars born in giant molecular clouds polluted
by single SN may have unusually high Ba. However, we do not see evidence
for this in our sample, as all our stars exhibit low Ba abundances ([Ba/Fe] $<$ 0.0).

Although we are not able to measure Sr
abundances for our program stars, other studies \citep[e.g.,][]{spite2014,cowan2021,mataspinto2021}
reported a large scatter of [Sr/Ba] among the VMP stars, indicating that a Ba-poor star
can still be Sr-rich. This leads to invoking other processes,
such as a non-standard $s$-process \citep[e.g.,][]{frischknecht2016}
and the $i$-process \citep{cowan1977,hampel2016}. These
mechanisms are believed to be relatively dominant only in Ba-deficient stars.
Additional follow-up studies to determine the Sr abundance ratio for our program stars
will be valuable to confirm these characteristics.

Recently, \citet{li2022} reported two different behaviors of Ba abundance ratios
for their VMP giant sample. The stars with [Fe/H] $<$ --3.0 exhibit
much lower [Ba/Fe] than the ones with [Fe/H] $>$ --3.0. Although we do not observe
the pattern from our giants, the inclusion of the TO stars reveals a similar behavior.
Furthermore, the plot for [Ba/Fe] indicates that most of stars with
[Ba/Fe] $<$ --1.0 have [Fe/H] $<$ --3.0, and are giants. \citet{li2022}
also found such a feature in their sample. The variety of Ba abundances for
VMP stars suggests stochastic pollution from neutron-capture
elements in the chemical evolution of the early MW.

One intriguing object is J0713, which has the lowest [Ba/Fe] among our sample.
Given that there is nothing unusual in other abundances, this object may
be born in a natal cloud having no association with a neutron-capture event.

\begin{figure}
\centering
\epsscale{1.2}
\plotone{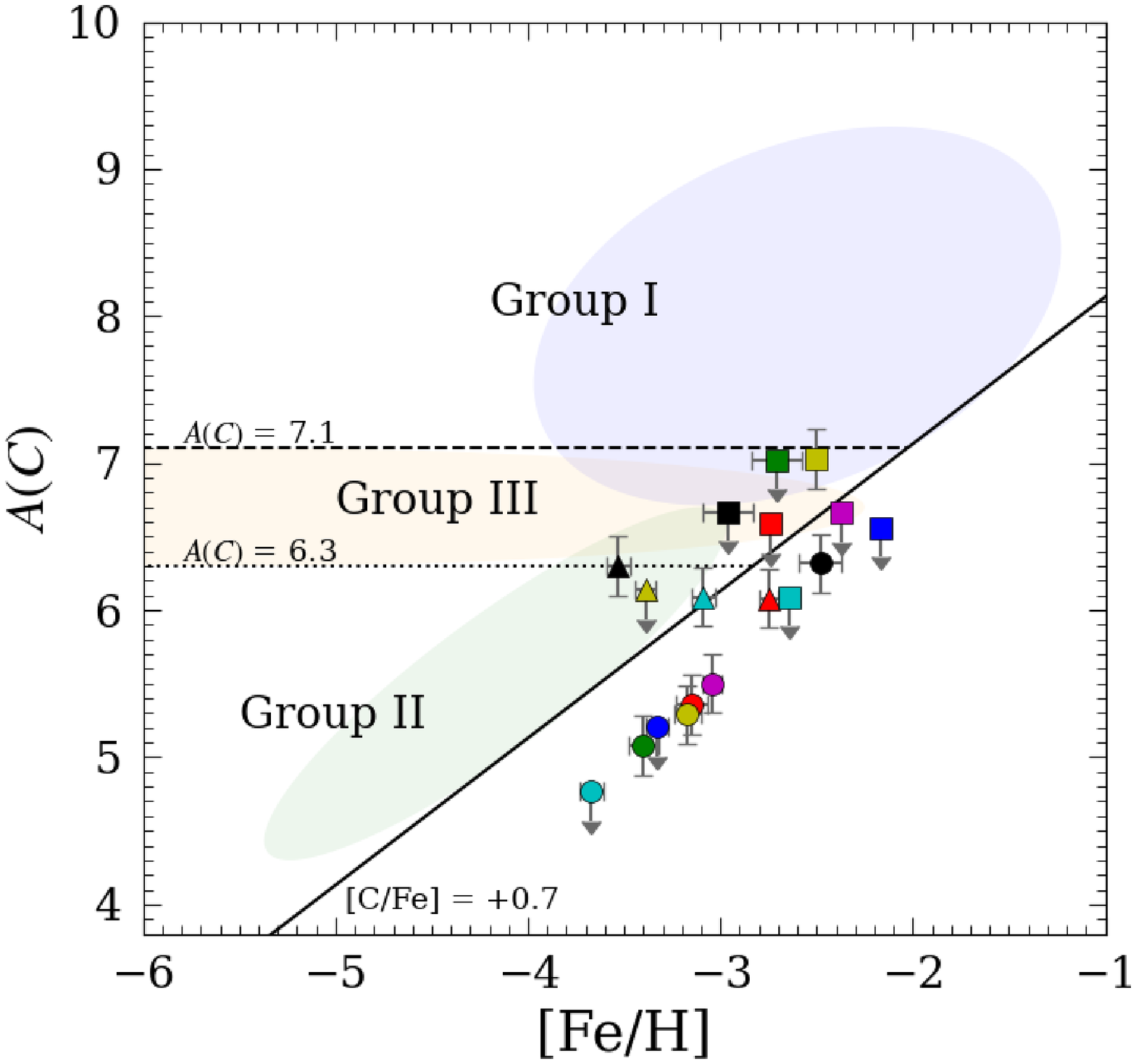}
\caption{Yoon-Beers diagram of $A$(C) as a function of [Fe/H]. Light purple, green, and yellow circles indicate morphological Groups I, II, and III, respectively, as described by \citet{yoon2016,yoon2019}. The black solid, dashed, and
dotted lines denote [C/Fe] = +0.7, $A$(C) = 7.1, and $A$(C) = 6.3, respectively.
Note that the \ac\ values are corrected for the evolutionary effects following
the prescription of \citet{placco2014}. Their corrections are
listed in Table \ref{a_2} in the Appendix. The color and symbol are the same
as in Figure \ref{fig3}. The down arrow indicates the
upper limit estimate.}
\label{cemp}
\end{figure}

\subsection{Carbon-Enhanced Metal-Poor (CEMP) Stars}\label{sec:cemp}

Based on the [C/Fe] estimates of our program stars, three objects (excluding
upper limit estimates) can be classified as CEMP stars ([C/Fe] $>$ +0.7),
resulting in a CEMP fraction of 17\% (3/18),
which is not far from the literature value of $\sim$ 20\% for
[Fe/H] $<$ --2.0 \citep{lee2013,placco2014}. As they exhibit low Ba-abundance
ratios ([Ba/Fe] $<$ 0.0), they are \cempno\ stars.

Previous studies reported that some CEMP (especially CEMP-no) stars are often
enhanced with Na, Mg, Al, and Si \citep{norris2013,aoki2018,bonifacio2018}.
These characteristics can be useful to identify the mechanisms for the production of the
CEMP stars. However, this is not the case for our CEMP stars, as they exhibit
normal or relatively low abundance ratios of Na and Mg. In fact,
only one potential CEMP object (J0758) with an upper limit
of [C/Fe] = 1.2 exhibits a very low sodium ratio, [Na/Fe] = --0.75.
The stars with peculiar abundances are mostly non-CEMP stars in our
sample.

We have examined where our target stars are located in the \ac-[Fe/H] diagram,
the so-called Yoon-Beers Diagram, as shown in Figure \ref{cemp}. In the figure, the light purple, green, and yellow
circles indicate the morphological regions for Group I, II, and III stars, respectively, as described
by \citet{yoon2016,yoon2019}. The black solid, dashed, and dotted lines denote [C/Fe] = +0.7, $A$(C) = 7.1,
and $A$(C) = 6.3, respectively. The legend for colors and symbols is the same as in
Figure \ref{fig3}. The down arrow indicates an upper limit. Note that the \ac\ values are corrected for
the evolutionary effects following the prescription of \citet{placco2014}, and their uncorrected
values and the amount of the correction are listed in Table \ref{a_2} in the Appendix.

Inspection of Figure \ref{cemp} reveals that, among our confirmed three CEMP stars,
taking into account the low-metallicity ([Fe/H] $<$ --3.0)
and $A$(C) level, two objects (black and cyan triangles) may belong to Group II,
and one star (J1037, yellow square) with [Fe/H] $>$ --3.0 occupies
Group I. According to the study by \citet{yoon2016}, stars
in Group I are dominated by CEMP-$s$ or $r/s$ stars, which exhibit
enhancements of $s$- or $r$-process elements (and may in fact be CEMP-$i$ stars), while Groups II and III contain
mostly CEMP-no stars, which exhibit low abundance of neutron-capture elements.
Considering its classification as CEMP-no star (due to their low Ba abundances), J1037 is
likely to be in Group III rather than Group I. A recent study by \citet{norris2019} also
reported that about 14\% of the Group I CEMP objects belong to CEMP-no category.
Since most of \cemps\ stars show radial-velocity variation \citep[e.g.,][]{starkenburg2014,placco2015b,hansen2016a,hansen2016b,jorissen2016},
indicative of their binarity, radial-velocity monitoring of these stars can
further confirm their assigned CEMP sub-class.

As mentioned above, CEMP-no stars can be further sub-divided into two groups:
one (Group II) with a good correlation between \ac\ and [Fe/H], and
the other (Group III) without clear correlation between them \citep{yoon2016}.
The fact that CEMP-no objects exhibit excesses in carbon
with low neutron-capture elements implies that
the sources of their chemical patterns are unlikely
to be due to mass transfer from a binary companion, as in the
\cemps\ stars. Rather, they have likely been enriched through distinct nucleosynthesis channels.
There are two channels of the formation of the \cempno\ stars that have been widely considered.
One is pollution from faint SNe associated with Pop III stars.
This type of SN experiences mixing and fallback
\citep{umeda2003,tominaga2007,tominaga2014,heger2010,nomoto2013,ezzeddine2019}, and
ejects less iron due to its small explosion energy.
Thus, only the outer layers, which have
copious amounts of lighter elements, including carbon, are ejected,
whereas the inner part, which includes a large amount of Fe
falls back onto the neutron star or black hole, increasing
the [C/Fe] ratio,

Another mechanism is a so-called spinstar. A rapidly rotating massive
ultra metal-poor ([Fe/H] $<$ --4.0) star can have large
amounts of carbon at its surface (due to efficient mixing with carbon
production deeper in the star), and the surface material
is blown by a stellar wind to pollute the
ISM \citep{meynet2006,hirschi2007,frischknecht2012,maeder2015}.
Additional formation mechanisms for CEMP-no stars are discussed in detail by \citet{norris2013}.

These two models cannot account completely for the observed chemical patterns of
the CEMP-no stars. Nevertheless, one can infer from the different
level of \ac\ and the distinct behaviors in the
$A$(Na)-$A$(C) and $A$(Mg)-$A$(C) spaces that the two Group II and Group III sub-groups
may be associated with different formation mechanisms \citep{yoon2016}.
However, a much larger sample of CEMP-no stars (especially Group III stars)
with accurate elemental-abundance estimates is required to better distinguish
between these two channels. In this aspect, the CEMP-no stars identified
through our work certainly help increase their sample size.

\begin{figure*} [t]
\includegraphics[width=\textwidth]{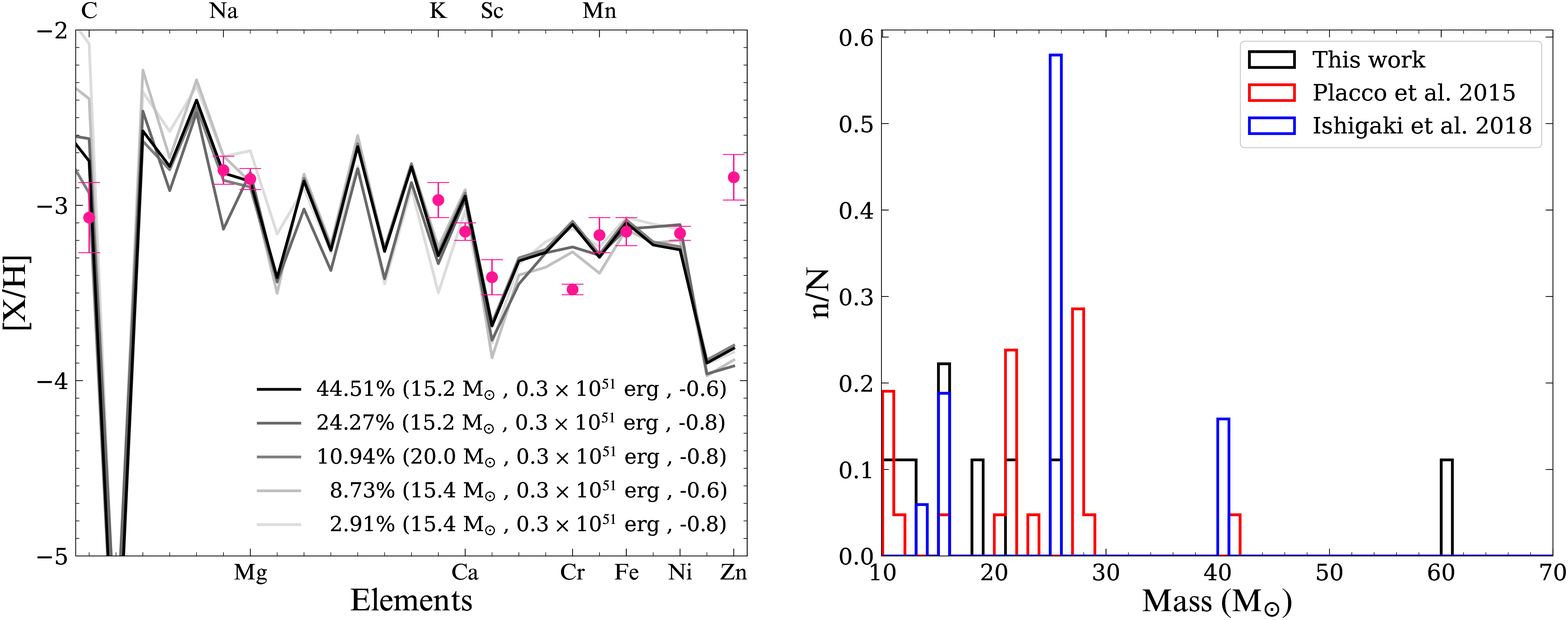}
\caption{Left~panel: Example of the abundance-pattern matching for J0713. Magenta symbols are
the measured abundances. The solid lines represent the five best-fit models with different masses and mixing
efficiencies. In this particular example, the explosion energy is the same for all five models.
The percentage in the legend indicates the occurrence rate of the model among 10,000 predicted models.
Right~panel: Histogram (black) of the predicted progenitor masses of our EMP stars.
The red and blue histograms are the mass distributions of the UMP stars predicted
by \citet{placco2015a} and the EMP stars derived by \citet{ishigaki2018},
respectively. Each histogram is normalized by the total number of stars to
compare the mass distribution.}
\label{mass}
\end{figure*}

\subsection{Chemically Peculiar Stars}\label{sec:cpo}

\subsubsection{Sodium-Peculiar Stars}\label{sec:na}

In Figure \ref{other}, we identified one object (J0158) with [Na/Fe] = +1.14.
This high Na-abundance ratio is a typical property of the second
population (P2) in globular clusters (GCs). It is known
that materials synthesized in stars of the primordial population (P1) of GCs,
chemically enriched ISM with light elements such as Na and Al, so that the P2 stars exhibit
distinct chemical characteristics from the P1 objects, establishing anti-correlations
between Na-O and Al-Mg among stars in GCs \citep[e.g.,][]{gratton2004,
martell2011,carretta2012,pancino2017}. This object thus may have originated in a GC.

Another piece of evidence for the chemical signature of a GC P2 in J0158  is its
low Mg-abundance ratio, [Mg/Fe] = +0.06, yielding a very
high [Na/Mg] ratio of +1.08. Its normal carbon content ([C/Fe] = +0.11) also
points to P2, as P2 stars mostly exhibit normal to low values of carbon.
Even though a further detailed abundance analysis is required, if this object is
indeed a P2 star from a GC, such a cluster may once have
belonged to a dwarf galaxy, because its metallicity ([Fe/H] = --3.04) is quite low
compared to GCs in the MW. In line with this, \citet{trincado2017} argued
that their low Mg-abundance stars with P2 chemical abundances could originate
from outside of the MW. Additional kinematic study will help confirm
whether or not it has been accreted from a dwarf satellite galaxy. We plan to carry out
a thorough kinematics analysis of our program stars in a forthcoming paper.

The object J2242 is extremely Na poor ([Na/Fe] = --1.02); J1650 also
has a relatively low [Na/Fe] = --0.6. However, both these stars are
enhanced in K and Ni. As K is produced by CCSNe, and Ni is formed
in the inner region of the explosion, it is plausible
to infer that the gas clouds  which formed these objects
may have polluted by CCSNe with high explosion energies.
Especially, considering its metallicity being [Fe/H] = --3.4,
the progenitor of J2242 was unlikely to have been enriched by AGBs.
Because the chemical abundances in the early MW were likely established by a
limited number of chemical-enrichment events, this particular
object may have undergone a peculiar nucleosynthesis episode. Its
enhancement of K and Ni abundances supports the distinct nucleosynthesis hypothesis.
However, recall that K and Ni for this star
were only estimated from single absorption lines, thus additional high-resolution
spectroscopic study of this object is necessary to confirm its peculiarity.

It is worthwhile mentioning that, in the case of globular clusters for
NGC 2419 \citep{cohen2012,mucciarelli2012} and NGC 2808 \citep{mucciarelli2015},
it has been reported that some of their member stars exhibit strong
anti-correlations between their K and Mg abundance ratios. \citet{kemp2018} also
reported that a large number of their metal-poor stars selected from LAMOST
are K rich, with relatively low Mg-abundance ratios, and concluded that an
anomalous nucleosynthesis event might be associated with the progenitors of the stars.
Even though the metallicities of our two K-rich objects are much lower than that
of the stars from \citet{kemp2018} ([Fe/H\ $>$ --1.5),
because they are not strongly enhanced in Mg, they could belong to the same category.

\subsubsection{Magnesium-Peculiar Stars}\label{sec:mg}

The object J0010 (black circle) in Figure \ref{other}
has [Mg/Fe] = +0.73, much higher than other halo stars near its metallicity ([Fe/H] = --2.48).
This Mg-rich star is slightly enhanced with Na, while [Ca/Fe] looks normal (+0.19),
resulting in an elevated [Mg/Ca] ratio (+0.54). The existence of the
high-[Mg/Ca] objects has been suggested in numerous studies \citep{norris2002,andrievsky2007,frebel2008,aoki2018}.
Its carbon enhancement is mild ([C/Fe] = +0.37). This star also stands out as a very
low-Ba object with respect to other halo stars
at [Fe/H] = --2.5; its [Ba/Fe] of --1.72 is over 1 dex lower than other objects at the same metallicity.
Judging from its extremely low [Ba/Fe] ratio, this object may be not associated with events that produced large amounts of neutron-capture elements, but more likely with CCSNe, since most of its Ca and iron-peak show low or normal abundances,
with the exception of Mn, whose abundance is uncertain due to weak \ion{Mn}{1} lines.

J2241, indicated by the Mg-abundance plot of Figure \ref{other},
has [Mg/Fe] = --0.61, which is much lower than the other VMP stars near its metallicity ([Fe/H] = --2.71).
Interestingly, this star's Na-abundance ratio is also somewhat deficient
with [Na/Fe] = --0.6. J1311 (red square) also has a relatively
low [Mg/Fe], but nothing abnormal among the other abundances.

Mg-poor halo stars have also been reported in other studies \citep[e.g.,][]{ivans2003,aoki2014}.
According to \citet{ivans2003}, the origin of these objects
can be explained by larger pollution from SNe Ia compared to other
halo objects with similar metallicities. Magnesium is primarily produced by massive
stars, while Ca is created by both SNe Ia and CCSNe, resulting
in a deficiency of Mg relative to Ca. Unfortunately, we do not
have measured Ca abundances for the two low-[Mg/Fe] stars to confirm this scenario.
\citet{ivans2003} also reported very low [Na/Fe] and [Ba/Fe] ratios
for their Mg-poor stars. One of our Mg-poor objects also exhibits this signature.

Stars with low $\alpha$-abundance ratios are sometimes explained by
enhancement of their Fe \citep{cayrel2004,yong2013,jacobson2015}.
In this case, the abundance ratios of other elements are expected to be relatively
lower as well. It will be worthwhile to carry out higher S/N, high-resolution follow-up
observation for these low-[Mg/Fe] objects to see if other elements behave
in this manner.

Another plausible explanation for the Mg-poor stars is that they came
from classical or ultra-faint dwarf galaxies.
These systems have had very low star-formation rates, and the
contribution of SNe Ia started occurring at much lower
metallicities \citep[e.g.,][]{shetrone2003,tolstoy2009}.
In order to test this, a kinematic analysis of the Mg-deficient stars
is presently being pursued.

\begin{deluxetable}{ccc} [!t]
\label{tab5}
\tablecaption{Predicted Progenitor Masses and SN Explosion Energies of Our EMP Stars}
\tablehead{\colhead{Short ID} & \colhead{Mass (\msun)}& \colhead{Energy (10$^{51}$ erg)}}
\startdata
J0102 & 10.9 & ~~0.9 \\ 
J0158 & 21.0 & ~~0.3 \\ 
J0422 & 60.0 & ~~3.0 \\ 
J0713 & 15.2 & ~~0.3 \\ 
J0758 & 11.9 & ~~0.9 \\ 
J0814 & 25.5 & ~~0.9 \\ 
J0908 & 12.8 & ~~0.3 \\ 
J2242 & 18.6 & ~10.0 \\ 
J2341 & 15.2 & ~~0.3 \\ 
\enddata
\tablecomments{Note that we did not attempt to determine the progenitor mass of
the two reference stars (J0226 and J1522) and one program star (J0925),
because only a few elements are available to constrain their progenitor mass.}
\end{deluxetable}

\subsection{Progenitor Masses of EMP Stars}\label{sec:mass}

Extremely metal-poor stars are regarded as fossil probes for understanding
the chemical evolution of the early MW, because
they preserve the chemical information of their natal gas clouds,
permitting constraints on their
predecessors, presumably massive Pop III stars. We
explore the characteristics (especially mass and explosion energy)
of the progenitors of our EMP stars by comparing their abundance patterns with
theoretical predictions of Pop III SN models by \citet{heger2010}.

Their SN models consist of a grid of 16,800 combinations,
which have a range of the explosion energy of 0.3 -- 10 $\times$ $10^{51}$ erg
and progenitor masses between of 10 -- 100 \msun. There exist 120 initial
masses, and the grid includes the mixing
efficiencies from no mixing to nearly complete mixing. Using
this grid, one can retrieve the progenitor properties of an EMP star by finding a
best-matching SN chemical yield with the observed abundance patterns.
We have made use of the \texttt{starfit} online
tool\footnote[11]{\url{http://starfit.org}} to carry out this
exercise. We only considered EMP stars among our program
stars, and assumed that our EMP stars were formed out of the gas polluted
by a single Pop III SN. If the measured abundance was derived from only one line,
we treated it as upper limit, and we attempted to fit with all available abundances.
In this exercise, we did not include the two reference stars (J0226 and J1522) and
one program star (J0925), because the abundances of only a few elements are
available for them to constrain their progenitor mass.

To confidently recover progenitor masses for our EMP stars, we generated
10,000 different abundance patterns by resampling each abundance from a normal
error distribution for each object, obtaining distributions of masses and explosion
energies of 10,000 possibilities.
The left panel of Figure \ref{mass} shows an example
of the best-matching chemical yield for J0713. The magenta symbols represent the
observed abundances. Each solid line represents the theoretically predicted abundance patterns produced by
a combination of different masses, explosion energies, and mixing efficiencies,
as indicated in the legend at the top of the panel. In this particular example,
all five best-fit models have the explosion energy of 0.3 $\times$ $10^{51}$ erg, and the best-fit model
has a mass of 15.2 \msun\ with mixing efficiency of --0.6, which accounts for 44.5\%
among 10,000 predicted models. The other four top models are followed with a range of masses, 15 -- 20 \msun\
and mixing efficiency of --0.6 or --0.8. We chose the most frequently
occurring model, and adopted its mass as the progenitor mass of our EMP star.

The right panel of Figure \ref{mass} displays a histogram of the predicted
progenitor masses of our EMP stars. The histogram implies that except one object (60 \msun),
our EMP objects have a progenitor mass of less than 26 \msun. Table \ref{tab5} lists the most probable
masses and explosion energies for our EMP stars.

\citet{placco2015a} used the same SNe models by \citet{heger2010} to determine the progenitor
masses of 21 UMP stars, and found that most of their progenitors have the mass range 20 -- 28 \msun\ and
explosion energies 0.3 -- 0.9 $\times$ $10^{51}$ erg (see also \citealt{placco2016} for the progenitor
masses of additional UMP stars). The red histogram of the right panel of Figure \ref{mass} is
their mass estimates of the UMP
progenitors, which are mostly less than 40 \msun. The mass range of our EMP stars is somewhat less
than that of \citet{placco2015a}, but generally in good agreement.
The majority of our EMP stars have their progenitor SN explosion energy between 0.3 and 1.0 $\times$ $10^{51}$ erg,
as can be read off from Table \ref{tab5}, again consistent with that of \citet{placco2015a}.
Placco et al. pointed out that the estimated mass
and explosion energy are very sensitive to the present of carbon
and nitrogen abundance. The relatively lower masses of our EMP progenitors
compared to theirs may thus be due to the absence of the measured N abundance for
our EMP stars.

\citet{ishigaki2018} also carried out a similar work to derive the mass function of the
first stars, using about 200 EMP stars, but with different SN models.
They used mixing and fallback SNe models, which have the combination
of five initial masses (13, 15, 25, 40, and 100 \msun)
and explosion energies of 0.5, 1.0, 10, 30, and 60 $\times$ $10^{51}$ erg to represent
low-, normal-, and high-energy explosions. They found that the progenitor
masses of their sample are mostly less than 40 \msun, as can be seen
in the blue histogram of the right panel of Figure \ref{mass}. Our predicted
masses also agree well with theirs. They also suggested that the C, N, and O abundances
are sensitive to the progenitor mass, and that the Na, Mg, and Al abundance ratios are
also useful tracers of the progenitor mass when ignoring the impact of stellar rotation.

To sum up, the (relatively) low-mass range of
the progenitors of our sample and other studies suggests that
stars  with $M <$ 40 \msun were likely primarily responsible
for the chemical enrichment of the early MW.

\section{Summary and Future Work} \label{sec:summary}

We have carried out high-resolution spectroscopic follow-up observations using GEMINI-N/GRACES
for 20 stars (including two reference stars), selected as EMP candidates from
SDSS and LAMOST medium-resolution spectra. We have presented stellar parameters
and abundance estimates for Li, C, Na, Mg, K, Ca, Sc,
Cr, Mn, Fe, Ni, Zn, and Ba, which are derived from a 1D LTE abundance analysis.
The chemical abundances of our EMP candidates are compared with those of
two benchmark stars from previous studies, and they show good agreement,
validating our measurements of the program star chemical abundances.

Based on our chemical abundances, we have found that all our candidates
are VMP stars, and include 10 objects that are EMP stars. In addition, three
CEMP stars are newly identified, and their low-Ba abundances ([Ba/Fe] $<$ 0.0)
indicate that they are all \cempno\ objects. As a result, our work nicely
contributes to increasing the sample size of the VMP/EMP stars as well as CEMP-no objects.

The Li abundance of our warm dwarf is slightly lower
than the Spite Plateau, possibly due to the different temperature
scale adopted. On the other hand, the $A$(Li) values
of our TO stars are distributed around the Spite Plateau. Consequently,
they can be used to constrain the Spite plateau by
follow-up studies, and especially, the TO stars may contribute to extending the
Spite plateau to the warmer temperature region.

Comparison with other Galactic halo VMP stars reveals that the chemical abundances of
our VMP objects generally follow similar abundance trends as a function of [Fe/H].
However, there exist a few objects that stand out from the majority of the VMP stars. We have
identified one star (J0158) with [Na/Fe] = +1.14. Its low [Fe/H], [Mg/Fe],
and [C/Fe] values imply that this object is closely connected to a second-generation
star from a GC that once belonged to a dwarf galaxy. The object J2242
has an extremely low-sodium abundance ratio of [Na/Fe] = --1.02.
This object also exhibits enhancement of K and Ni abundances. Taking into account
its metallicity of [Fe/H] = --3.4, this star may have formed in a gas cloud that
was chemically enriched by CCSNe with high explosion energies.

We also found a Mg-rich star ([Mg/Fe] = +0.73)
that is also slightly enhanced with Na, while its [Ca/Fe] is normal. These
abundance characteristics, together with the extremely low [Ba/Fe], suggest
that this VMP star is not likely to be associated with CCSNe that produce a
large amount of neutron-capture elements. The star J2241 exhibits the lowest
[Mg/Fe] (--0.61). The origin of this Mg-poor star
can be explained by larger pollution by SNe Ia, or it could be accreted from
a dwarf galaxy that experienced low star-formation efficiency.

We have also explored the progenitor characteristics (mass and explosion energy)
of our EMP stars by comparing their chemical-abundance patterns with those
predicted by Pop III SN models. Except one object (60 \msun), our estimate masses of the EMP stars
are in the range of 10 -- 26 \msun, which is in good agreement with the previous studies
by \citet{placco2015a} and \citet{ishigaki2018}. This suggests that the chemical evolution
of the early MW was driven primarily by stars with masses $M <$ 40 \msun.

Since there are some uncertainties in the derived abundance
for some elements due to weak metallic lines, low-S/N spectra, or a small number
of measured lines, we plan to carry out high-resolution, high-S/N spectroscopic
observations for the chemically peculiar objects to determine more reliable
measurements of their elemental abundances. We also plan to supplement the
available chemical information for our program stars with a chemodynamical analysis in an upcoming paper.

\software{\texttt{Astropy} \citep{astropy1,astropy2,astropy3}, \texttt{matplotlib}
\citep{matplotlib}, \texttt{NumPy} \citep{numpy}, \texttt{SciPy} \citep{scipy}.}

\begin{acknowledgments}
We thank Christopher Sneden for his suggestions and advice on
the spectral synthesis of CH $G$ band.
Y.S.L. acknowledges support from the National Research Foundation (NRF) of
Korea grant funded by the Ministry of Science and ICT (NRF-2021R1A2C1008679).
Y.S.L. also gratefully acknowledges partial support for his visit to the University
of Notre Dame from OISE-1927130: The International Research Network for Nuclear Astrophysics (IReNA),
awarded by the US National Science Foundation. Y.K.K. acknowledges support from Basic Science
Research Program through the NRF of Korea funded by the Ministry of
Education (NRF-2021R1A6A3A01086446). T.C.B. acknowledges partial support for
this work from grant PHY 14-30152; Physics Frontier Center/JINA Center for the Evolution
of the Elements (JINA-CEE), awarded by the U.S. National Science Foundation. The work of V.M.P. is
supported by NOIRLab, which is managed by the Association of Universities for Research in
Astronomy (AURA) under a cooperative agreement with the National Science Foundation.
This work was supported by K-GMT Science Program (PID: GN-2016A-Q-17), GN-2018B-Q-122,
GN-2019B-Q-115, GN-2019B-Q-219, and GN-2019B-Q-310) of Korea Astronomy and Space Science Institute (KASI).

This research is based on observations obtained through the Gemini Remote Access to CFHT ESPaDOnS
Spectrograph (GRACES). ESPaDOnS is located at the Canada-France-Hawaii Telescope (CFHT), which is
operated by the National Research Council of Canada, the Institut National des Sciences de l’Univers
of the Centre National de la Recherche Scientifique of France, and the University of Hawai’i.
ESPaDOnS is a collaborative project funded by France (CNRS, MENESR, OMP, LATT), Canada (NSERC),
CFHT and ESA. ESPaDOnS was remotely controlled from the international Gemini Observatory,
a program of NSF’s NOIRLab, which is managed by the Association of Universities for Research in Astronomy (AURA) under
a cooperative agreement with the National Science Foundation on behalf of the Gemini Observatory
partnership: the National Science Foundation (United States), National Research Council (Canada),
Agencia Nacional de Investigaci\'{o}n y Desarrollo (Chile), Ministerio de Ciencia,
Tecnolog\'{i}a e Innovaci\'{o}n (Argentina), Minist\'{e}rio da Ci\^{e}ncia, Tecnologia,
Inova\c{c}\~{o}es e Comunica\c{c}\~{o}es (Brazil), and Korea Astronomy and
Space Science Institute (Republic of Korea).
\end{acknowledgments}


 \label{reference}

\appendix
Table \ref{a_1} lists the line information used to derived
chemical abundances. The carbon abundance ratios and their corrected
value according to the evolutionary stage are listed in Table \ref{a_2}.
Table \ref{a_3} provides detailed abundance measurements of other elements and their associated errors.
\startlongtable
\begin{deluxetable}{cccc}
\label{a_1}
\tabletypesize{\scriptsize}
\tablecaption{Line Lists}
\tablehead{\colhead{Species} & \colhead{Wavelength (\AA)} & \colhead{EP (eV)} & \colhead{log $gf$}}
\startdata
~~3.0 & 6707.70 & 0.00 & ~~0.17 \\
11.0 & 5895.92 & 0.00  & --0.19 \\
11.0 & 5889.95 & 0.00  & ~~0.11 \\
12.0 & 5528.41 & 4.34  & --0.50 \\
12.0 & 5183.60 & 2.72  & --0.24 \\
12.0 & 4702.99 & 4.33  & --0.38 \\
12.0 & 5172.68 & 2.71  & --0.45 \\
19.0 & 7664.90 & 0.00  & ~~0.14 \\
19.0 & 7698.96 & 0.00  & --0.17 \\
20.0 & 5262.24 & 2.52  & --0.47 \\
20.0 & 5265.56 & 2.52  & --0.26 \\
20.0 & 5349.47 & 2.71  & --0.31 \\
20.0 & 5588.76 & 2.52  & ~~0.21 \\
20.0 & 6493.79 & 2.52  & --0.11 \\
20.0 & 6462.57 & 2.52  & ~~0.26 \\
20.0 & 6102.72 & 1.88  & --0.79 \\
20.0 & 6122.22 & 1.89  & --0.32 \\
20.0 & 6162.17 & 1.90  & --0.09 \\
20.0 & 6439.07 & 2.52  & ~~0.47 \\
20.0 & 5598.49 & 2.52  & --0.09 \\
20.0 & 5594.47 & 2.52  & ~~0.10 \\
21.1 & 5318.37 & 1.36  & --2.01 \\
21.1 & 5031.01 & 1.36  & --0.40 \\
21.1 & 5657.91 & 1.51  & --0.60 \\
21.1 & 5526.79 & 1.77  & ~~0.02 \\
21.1 & 5239.81 & 1.46  & --0.77 \\
22.0 & 4840.87 & 0.90  & --0.45 \\
22.0 & 4991.07 & 0.84  & ~~0.44 \\
22.0 & 4981.73 & 0.84  & ~~0.56 \\
22.0 & 5192.97 & 0.02  & --0.95 \\
22.0 & 5210.39 & 0.05  & --0.83 \\
22.0 & 5064.65 & 0.05  & --0.94 \\
22.0 & 5007.21 & 0.82  & ~~0.17 \\
22.0 & 5024.84 & 0.82  & --0.55 \\
22.0 & 5009.65 & 0.02  & --2.20 \\
22.0 & 5039.96 & 0.02  & --1.13 \\
22.0 & 5014.19 & 0.00  & --1.22 \\
22.0 & 5014.28 & 0.81  & ~~0.11 \\
22.0 & 5038.40 & 1.43  & ~~0.07 \\
22.0 & 5016.16 & 0.85  & --0.52 \\
22.0 & 5036.46 & 1.44  & ~~0.19 \\
22.0 & 5035.90 & 1.46  & ~~0.26 \\
22.0 & 5020.02 & 0.84  & --0.36 \\
22.0 & 5173.74 & 0.00  & --1.06 \\
22.0 & 4999.50 & 0.83  & ~~0.31 \\
24.0 & 5296.69  & 0.98 & --1.36 \\
24.0 & 5409.77  & 1.03 & --0.67 \\
24.0 & 4789.34  & 2.54 & --0.33 \\
24.0 & 5206.04  & 0.94 & ~~0.02 \\
24.0 & 5348.31  & 1.00 & --1.21 \\
24.0 & 5345.80  & 1.00 & --0.95 \\
24.0 & 5247.56  & 0.96 & --1.64 \\
24.0 & 5208.42  & 0.94 & ~~0.16 \\
24.0 & 5300.74  & 0.98 & --2.00 \\
24.0 & 5298.28  & 0.98 & --1.14 \\
25.0 & 4783.43  & 2.30 & ~~0.04 \\
25.0 & 4754.05  & 2.28 & --0.09 \\
25.0 & 4823.53  & 2.32 & ~~0.14 \\
26.0 & 5379.57  & 3.69 & --1.51 \\
26.0 & 5383.37  & 4.31 & ~~0.65 \\
26.0 & 5027.23  & 3.64 & --1.89 \\
26.0 & 5389.48  & 4.42 & --0.41 \\
26.0 & 5322.04  & 2.28 & --2.80 \\
26.0 & 5397.13  & 0.92 & --1.98 \\
26.0 & 5405.78  & 0.99 & --1.85 \\
26.0 & 5288.53  & 3.68 & --1.51 \\
26.0 & 5410.91  & 4.47 & ~~0.40 \\
26.0 & 5415.20  & 4.39 & ~~0.64 \\
26.0 & 4920.50  & 2.83 & ~~0.07 \\
26.0 & 5393.17  & 3.24 & --0.91 \\
26.0 & 5371.49  & 0.96 & --1.64 \\
26.0 & 5367.47  & 4.42 & ~~0.44 \\
26.0 & 5307.36  & 1.61 & --2.91 \\
26.0 & 5365.40  & 3.56 & --1.02 \\
26.0 & 5424.07  & 4.32 & ~~0.52 \\
26.0 & 5364.87  & 4.45 & ~~0.23 \\
26.0 & 4800.65  & 4.12 & --1.03 \\
26.0 & 5302.30  & 3.28 & --0.72 \\
26.0 & 5339.93  & 3.27 & --0.72 \\
26.0 & 4924.77  & 2.28 & --2.11 \\
26.0 & 5332.90  & 1.55 & --2.78 \\
26.0 & 5328.53  & 1.56 & --1.85 \\
26.0 & 5328.04  & 0.92 & --1.47 \\
26.0 & 5324.18  & 3.21 & --0.10 \\
26.0 & 5369.96  & 4.37 & ~~0.54 \\
26.0 & 5429.70  & 0.96 & --1.88 \\
26.0 & 5446.92  & 0.99 & --1.91 \\
26.0 & 5283.62  & 3.24 & --0.52 \\
26.0 & 6137.69  & 2.59 & --1.35 \\
26.0 & 4736.77  & 3.21 & --0.75 \\
26.0 & 6191.56  & 2.43 & --1.42 \\
26.0 & 6230.72  & 2.56 & --1.28 \\
26.0 & 6252.56  & 2.4  & --1.69 \\
26.0 & 6297.79  & 2.22 & --2.64 \\
26.0 & 6393.60  & 2.43 & --1.58 \\
26.0 & 6400.00  & 3.6  & --0.29 \\
26.0 & 6430.85  & 2.18 & --1.95 \\
26.0 & 4733.59  & 1.49 & --2.99 \\
26.0 & 6494.98  & 2.40 & --1.24 \\
26.0 & 6677.99  & 2.69 & --1.42 \\
26.0 & 4710.28  & 3.02 & --1.61 \\
26.0 & 4871.32  & 2.87 & --0.36 \\
26.0 & 4707.27  & 3.24 & --1.08 \\
26.0 & 6136.62  & 2.45 & --1.41 \\
26.0 & 4872.14  & 2.88 & --0.57 \\
26.0 & 6065.48  & 2.61 & --1.41 \\
26.0 & 4890.76  & 2.88 & --0.39 \\
26.0 & 5455.61  & 1.01 & --2.09 \\
26.0 & 5497.52  & 1.01 & --2.83 \\
26.0 & 5501.46  & 0.96 & --3.05 \\
26.0 & 5506.78  & 0.99 & --2.79 \\
26.0 & 4918.99  & 2.85 & --0.34 \\
26.0 & 5572.84  & 3.40 & --0.28 \\
26.0 & 5586.76  & 3.37 & --0.14 \\
26.0 & 5434.52  & 1.01 & --2.13 \\
26.0 & 4789.65  & 3.53 & --0.96 \\
26.0 & 5615.64  & 3.33 & ~~0.05 \\
26.0 & 5624.54  & 3.42 & --0.76 \\
26.0 & 4903.31  & 2.88 & --0.93 \\
26.0 & 5662.52  & 4.18 & --0.57 \\
26.0 & 5686.53  & 4.55 & --0.45 \\
26.0 & 5753.12  & 4.26 & --0.69 \\
26.0 & 4891.49  & 2.85 & --0.11 \\
26.0 & 4786.81  & 3.00 & --1.61 \\
26.0 & 5281.79  & 3.04 & --0.83 \\
26.0 & 5028.13  & 3.56 & --1.12 \\
26.0 & 5150.84  & 0.99 & --3.04 \\
26.0 & 5110.41  & 0.00 & --3.76 \\
26.0 & 5123.72  & 1.01 & --3.06 \\
26.0 & 5125.12  & 4.22 & --0.14 \\
26.0 & 5127.36  & 0.92 & --3.25 \\
26.0 & 5001.87  & 3.88 & ~~0.05 \\
26.0 & 5131.47  & 2.22 & --2.52 \\
26.0 & 5098.70  & 2.18 & --2.03 \\
26.0 & 5133.69  & 4.18 & ~~0.14 \\
26.0 & 5141.74  & 2.42 & --2.24 \\
26.0 & 5142.93  & 0.96 & --3.08 \\
26.0 & 5269.54  & 0.86 & --1.33 \\
26.0 & 5151.91  & 1.01 & --3.32 \\
26.0 & 5162.27  & 4.18 & ~~0.02 \\
26.0 & 5166.28  & 0.00 & --4.12 \\
26.0 & 5137.38  & 4.18 & --0.40 \\
26.0 & 5171.60  & 1.49 & --1.72 \\
26.0 & 5090.77  & 4.26 & --0.36 \\
26.0 & 5080.95  & 3.27 & --3.09 \\
26.0 & 5022.24  & 3.98 & --0.53 \\
26.0 & 5014.94  & 3.94 & --0.30 \\
26.0 & 5012.07  & 0.86 & --2.64 \\
26.0 & 5041.07  & 0.96 & --3.09 \\
26.0 & 5041.76  & 1.49 & --2.20 \\
26.0 & 5044.21  & 2.85 & --2.02 \\
26.0 & 5083.34  & 0.96 & --2.84 \\
26.0 & 5049.82  & 2.28 & --1.36 \\
26.0 & 5060.08  & 0.00 & --5.46 \\
26.0 & 5006.12  & 2.83 & --0.62 \\
26.0 & 5068.77  & 2.94 & --1.04 \\
26.0 & 5074.75  & 4.22 & --0.20 \\
26.0 & 5079.22  & 2.20 & --2.11 \\
26.0 & 5079.74  & 0.99 & --3.25 \\
26.0 & 5051.63  & 0.92 & --2.76 \\
26.0 & 4994.13  & 0.92 & --2.97 \\
26.0 & 5227.19  & 1.56 & --1.23 \\
26.0 & 4957.60  & 2.81 & ~~0.23 \\
26.0 & 5204.58  & 0.09 & --4.33 \\
26.0 & 5250.21  & 0.12 & --4.94 \\
26.0 & 5216.27  & 1.61 & --2.08 \\
26.0 & 5217.39  & 3.21 & --1.16 \\
26.0 & 5225.53  & 0.11 & --4.76 \\
26.0 & 4957.30  & 2.85 & --0.41 \\
26.0 & 5202.34  & 2.18 & --1.87 \\
26.0 & 4859.74  & 2.88 & --0.76 \\
26.0 & 4939.69  & 0.86 & --3.25 \\
26.0 & 4946.39  & 3.37 & --1.17 \\
26.0 & 5263.31  & 3.27 & --0.88 \\
26.0 & 5242.49  & 3.63 & --0.97 \\
26.0 & 5247.05  & 0.09 & --4.95 \\
26.0 & 5254.96  & 0.11 & --4.76 \\
26.0 & 5232.94  & 2.94 & --0.06 \\
26.0 & 5266.56  & 3.00 & --0.39 \\
26.0 & 5250.65  & 2.20 & --2.18 \\
26.0 & 5198.71  & 2.22 & --2.09 \\
26.0 & 5194.94  & 1.56 & --2.02 \\
26.0 & 4938.81  & 2.88 & --1.08 \\
26.0 & 4973.10  & 3.96 & --0.95 \\
26.0 & 4966.09  & 3.33 & --0.87 \\
26.0 & 5191.46  & 3.04 & --0.55 \\
26.0 & 5192.34  & 3.00 & --0.42 \\
28.0 & 5017.59  & 3.54 & --0.08 \\
28.0 & 5035.37  & 3.63 & ~~0.29 \\
28.0 & 4714.42  & 3.38 & ~~0.23 \\
28.0 & 4980.16  & 3.61 & --0.11 \\
28.0 & 5080.52  & 3.65 & ~~0.13 \\
28.0 & 5081.11  & 3.85 & ~~0.30 \\
28.0 & 5084.08  & 3.68 & ~~0.03 \\
28.0 & 5115.40  & 3.83 & --0.11 \\
28.0 & 5578.73  & 1.68 & --2.64 \\
28.0 & 5137.08  & 1.68 & --1.99 \\
28.0 & 4904.41  & 3.54 & --0.17 \\
28.0 & 5476.90  & 1.83 & --0.89 \\
28.0 & 5155.76  & 3.90 & --0.09 \\
28.0 & 4855.41  & 3.54 & ~~0.00 \\
28.0 & 5754.68  & 1.94 & --2.33 \\
30.0 & 4810.53  & 4.08 & --0.14 \\
30.0 & 4722.15  & 4.03 & --0.39 \\
56.1 & 6496.90  & 0.60 & --0.38 \\
56.1 & 4934.09  & 0.00 & --0.16 \\
\enddata
\tablecomments{EP is the excitation potential.}
\end{deluxetable}  
\startlongtable
\begin{deluxetable}{ccccc} 
\label{a_2}
\tabletypesize{\scriptsize}
\tablecaption{Carbon-Abundance Ratios and Their Corrections for Evolutionary Stage}
\tablehead{\colhead{Short ID} & \colhead{[C/Fe]$_{\rm BC}$} & \colhead{C$_{\rm cor}$} & \colhead{[C/Fe]} & \colhead{$A$(C)}}
\startdata
J0010  &  ~~--0.22  &  0.59  &  ~~0.37  &  ~~6.32 \\
J0102  &  ~~~~0.75  &  0.00  &  ~~0.75  &  ~~6.09 \\
J0158  &  ~~~~0.10  &  0.01  &  ~~0.11  &  ~~5.50 \\
J0226  &  ~~~~1.33  &  0.00  &  ~~1.33  &  ~~5.98 \\
J0357  &  ~~~~0.40  &  0.00  &  ~~0.40  &  ~~6.08 \\
J0422  &  $<$0.10   &  0.01  & $<$0.11  &  $<$5.21 \\
J0713  &  ~~--0.60  &  0.68  &  ~~0.08  &  ~~5.36 \\
J0758  &  $<$1.20   &  0.00  & $<$1.20  &  $<$6.67 \\
J0814  &  $<$1.10   &  0.00  & $<$1.10 &   $<$6.14 \\
J0908  &  $<$--0.40 &  0.41  & $<$0.01 &   $<$4.77 \\
J0925  &  ~~~~1.40  &  0.00  & ~~1.40   &  ~~6.30 \\
J1037  &  ~~~~1.10  &  0.00  & ~~1.10   &  ~~7.03 \\
J1311  &  $<$0.90   &  0.00  & $<$0.90  &  $<$6.59 \\
J1317  &  $<$0.60   &  0.00  & $<$0.60  &  $<$6.66 \\
J1522  &  $<$1.03   &  0.00  & $<$1.03  &  $<$5.76 \\
J1650  &  $<$0.30   &  0.00  & $<$0.30  &  $<$6.56 \\
J1705  &  $<$0.30   &  0.00  & $<$0.30  &  $<$6.09 \\
J2241  &  $<$1.30   &  0.00  & $<$1.30  &  $<$7.02 \\
J2242  &  ~~--0.20  &  0.25  &  ~~0.05  &  ~~5.08 \\
J2341  &  ~~--0.50  &  0.53  &  ~~0.03  &  ~~5.29 \\
\enddata
\tablecomments{C$_{\rm cor}$ indicates the amount of correction following \citet{placco2014} due to the
evolution status. [C/Fe]$_{\rm BC}$ is the value before C$_{\rm cor}$ is applied, while
[C/Fe] and $A$(C) are the values after the correction is applied. The $<$ symbol indicates an upper limit
estimate. As mentioned in Section \ref{sec:ss}, the measured error is conservatively assumed to be 0.2 dex.}
\end{deluxetable}  
\startlongtable
\begin{deluxetable}{ccccccccc}
\label{a_3}
\tablewidth{0pt}
\tabletypesize{\scriptsize}
\renewcommand{\tabcolsep}{1pt}
\tablecaption{Detailed Chemical Abundances for Each Object}
\tablehead{\colhead{Species}           & \colhead{$N$}           & \colhead{$\log\epsilon$\,(X)} & \colhead{[X/Fe]} & \colhead{SE} &
           \colhead{$\Delta$\teff} & \colhead{$\Delta$\logg} & \colhead{$\Delta$\vt}         & \colhead{TE}  \\
           \colhead{}              & \colhead{}              & \colhead{}                    & \colhead{}       & \colhead{(dex)} &
          \colhead{$\pm$100 K}    & \colhead{$\pm$0.2 dex}  & \colhead{$\pm$0.2 \kms}       & \colhead{(dex)}}
\startdata
\multicolumn{9}{c}{J0010}\\
\hline
\ion{Li}{1} & \nodata & \nodata & \nodata & \nodata & \nodata & \nodata & \nodata & \nodata \\
\ion{Na}{1} & ~~~~2 & ~~4.00 & ~~0.24 & 0.12 & 0.06 & 0.00 & 0.04 & 0.07 \\
\ion{Mg}{1} & ~~~~4 & ~~5.85 & ~~0.73 & 0.02 & 0.06 & 0.05 & 0.02 & 0.04 \\
\ion{K}{1}  & ~~~~1 & ~~2.68 & ~~0.13 & 0.17 & 0.03 & 0.02 & 0.07 & 0.09 \\
\ion{Ca}{1} & ~~12 & ~~4.05 & ~~0.19 & 0.06 & 0.06 & 0.00 & 0.05 & 0.05 \\
\ion{Sc}{2} & ~~~~4 & ~~0.34 & --0.33 & 0.06 & 0.18 & 0.09 & 0.07 & 0.11 \\
\ion{Cr}{1} & ~~~~9 & ~~3.06 & --0.10 & 0.08 & 0.05 & 0.00 & 0.02 & 0.05 \\
\ion{Mn}{1} & ~~~~3 & ~~3.04 & ~~0.09 & 0.10 & 0.02 & 0.04 & 0.02 & 0.06 \\
\ion{Fe}{1} & 121 & ~~5.02 & ~~0.00 & 0.02 & 0.18 & 0.06 & 0.10 & 0.11 \\
\ion{Ni}{1} & ~~13 & ~~3.87 & ~~0.13 & 0.03 & 0.07 & 0.01 & 0.07 & 0.05 \\
\ion{Zn}{1} & ~~~~2 & ~~2.08 & ~~0.00 & 0.12 & 0.20 & 0.04 & 0.08 & 0.12 \\
\ion{Ba}{2} & ~~~~2 & --2.02 & --1.72 & 0.06 & \nodata & \nodata & \nodata & 0.06 \\
\hline
\multicolumn{9}{c}{J0102}\\
\hline
\ion{Li}{1} & ~~~~1 & ~~1.75 & ~~3.79 & 0.20 & \nodata & \nodata & \nodata & 0.20 \\
\ion{Na}{1} & ~~~~2 & ~~2.55 & --0.60 & 0.23 & 0.02 & 0.01 & 0.01 & 0.12  \\
\ion{Mg}{1} & ~~~~2 & ~~4.72 & ~~0.21 & 0.23 & ~~0.01 & 0.07 & 0.01 & 0.12 \\
\ion{K}{1}  &\nodata& \nodata &\nodata  & \nodata  & \nodata & \nodata & \nodata & \nodata   \\
\ion{Ca}{1} &\nodata& \nodata &\nodata  & \nodata  & \nodata & \nodata & \nodata & \nodata  \\
\ion{Sc}{2} &\nodata& \nodata &\nodata  & \nodata  & \nodata & \nodata & \nodata & \nodata \\
\ion{Cr}{1} & ~~~~1 & ~~2.51 & --0.04 & 0.33 & 0.01 & 0.01 & 0.01 & 0.17 \\
\ion{Mn}{1} &\nodata& \nodata &\nodata  & \nodata  & \nodata & \nodata & \nodata & \nodata  \\
\ion{Fe}{1} & ~~13 & ~~4.41 & ~~0.00 & 0.09 & 0.08 & 0.0 & 0.02 & 0.06  \\
\ion{Ni}{1} &\nodata& \nodata &\nodata  & \nodata  & \nodata & \nodata & \nodata & \nodata\\
\ion{Zn}{1} &\nodata& \nodata &\nodata  & \nodata  & \nodata & \nodata & \nodata & \nodata \\
\ion{Ba}{2} &\nodata& \nodata &\nodata & \nodata & \nodata & \nodata & \nodata & \nodata  \\
\hline
\multicolumn{9}{c}{J0158}\\
\hline
\ion{Li}{1} & ~~~~1 & ~~0.79 & ~~2.78 & 0.20 & \nodata & \nodata & \nodata & 0.20  \\
\ion{Na}{1} & ~~~~2 & ~~4.34 & ~~1.14 & 0.13 & 0.02 & 0.06 & 0.05 & 0.08  \\
\ion{Mg}{1} & ~~~~4 & ~~4.62 & ~~0.06 & 0.11 & 0.03 & 0.02 & 0.02 & 0.06  \\
\ion{K}{1}  &\nodata& \nodata &\nodata  & \nodata  & \nodata & \nodata & \nodata & \nodata \\
\ion{Ca}{1} & ~~~~8 & ~~3.71 & ~~0.41 & 0.08 & 0.04 & 0.00 & 0.01 & 0.04  \\
\ion{Sc}{2} & \nodata & \nodata & \nodata & \nodata & \nodata & \nodata & \nodata & \nodata  \\
\ion{Cr}{1} & ~~~~3 & ~~2.51 & --0.09 & 0.11 & 0.00 & 0.00 & 0.00 & 0.06  \\
\ion{Mn}{1} & ~~~~1 & ~~2.38 & --0.01 & 0.19 & 0.02 & 0.00 & 0.02 & 0.10  \\
\ion{Fe}{1} & ~~62 & ~~4.46 & --0.00 & 0.02 & 0.10 & 0.01 & 0.02 & 0.05  \\
\ion{Ni}{1} & ~~~~3 & ~~3.48 & ~~0.30 & 0.11 & 0.02 & 0.01 & 0.02 & 0.06  \\
\ion{Zn}{1} & ~~~~1 & ~~2.13 & ~~0.61 & 0.19 & 0.06 & 0.04 & 0.02 & 0.10  \\
\ion{Ba}{2} & ~~~~1 & $<$--2.46 & $<$--1.60 & 0.20 & \nodata & \nodata & \nodata & 0.20 \\
\hline
\multicolumn{9}{c}{J0226}\\
\hline
\ion{Li}{1} &\nodata& \nodata &\nodata  & \nodata  & \nodata & \nodata & \nodata & \nodata\\
\ion{Na}{1} & ~~~~2 & ~~2.75 & ~~0.29 & 0.07 & 0.02 & 0.01 & 0.01 & 0.04  \\
\ion{Mg}{1} & ~~~~2 & ~~4.41 & ~~0.59 & 0.07 & 0.02 & 0.02 & 0.07 & 0.05  \\
\ion{K}{1}  &\nodata& \nodata &\nodata  & \nodata  & \nodata & \nodata & \nodata & \nodata\\
\ion{Ca}{1} &\nodata& \nodata &\nodata  & \nodata  & \nodata & \nodata & \nodata & \nodata  \\
\ion{Sc}{2} &\nodata& \nodata &\nodata  & \nodata  & \nodata & \nodata & \nodata & \nodata \\
\ion{Cr}{1} &\nodata& \nodata &\nodata  & \nodata  & \nodata & \nodata & \nodata & \nodata \\
\ion{Mn}{1} &\nodata& \nodata &\nodata  & \nodata  & \nodata & \nodata & \nodata & \nodata \\
\ion{Fe}{1} & ~~15 & ~~3.72 & ~~0.00 & 0.03 & 0.10 & 0.01 & 0.01 & 0.05 \\
\ion{Ni}{1} &\nodata& \nodata &\nodata  & \nodata  & \nodata & \nodata & \nodata & \nodata \\
\ion{Zn}{1} &\nodata& \nodata &\nodata  & \nodata  & \nodata & \nodata & \nodata & \nodata \\
\ion{Ba}{2} &\nodata& \nodata & \nodata & \nodata & \nodata & \nodata & \nodata & \nodata \\
\hline
\multicolumn{9}{c}{J0357}\\
\hline
\ion{Li}{1} & ~~~~1 & ~~1.83 & ~~3.53 & 0.20 & \nodata & \nodata & \nodata & 0.20  \\
\ion{Na}{1} & ~~~~2 & ~~3.62 & ~~0.13 & 0.13 & 0.01 & 0.03 & 0.03 & 0.07 \\
\ion{Mg}{1} & ~~~~4 & ~~5.19 & ~~0.34 & 0.06 & 0.01 & 0.04 & 0.01 & 0.04  \\
\ion{K}{1}  &\nodata& \nodata &\nodata  & \nodata  & \nodata & \nodata & \nodata & \nodata\\
\ion{Ca}{1} & ~~~~8 & ~~3.92 & ~~0.33 & 0.05 & 0.03 & 0.01 & 0.02 & 0.03  \\
\ion{Sc}{2} & ~~~~1 & ~~0.90 & ~~0.50 & 0.18 & 0.06 & 0.07 & 0.02 & 0.10  \\
\ion{Cr}{1} & ~~~~3 & ~~2.73 & --0.16 & 0.10 & 0.01 & 0.01 & 0.01 & 0.05  \\
\ion{Mn}{1} & ~~~~1 & ~~2.69 & ~~0.01 & 0.18 & 0.02 & 0.00 & 0.02 & 0.09 \\
\ion{Fe}{1} & ~~64 & ~~4.75 & ~~0.00 & 0.02 & 0.09 & 0.00 & 0.02 & 0.05 \\
\ion{Ni}{1} & ~~~~2 & ~~3.84 & ~~0.37 & 0.13 & 0.02 & 0.01 & 0.01 & 0.07 \\
\ion{Zn}{1} & ~~~~1 & ~~2.62 & ~~0.81 & 0.18 & 0.05 & 0.03 & 0.02 & 0.10 \\
\ion{Ba}{2} & ~~~~2 & --1.27 & --0.70 & 0.04 & \nodata & \nodata & \nodata & 0.04 \\
\hline
\multicolumn{9}{c}{J0422}\\
\hline
\ion{Li}{1} & \nodata & \nodata & \nodata & \nodata & \nodata & \nodata & \nodata & \nodata \\
\ion{Na}{1} & ~~~~2 & ~~2.45 & --0.46 & 0.15 & 0.02 & 0.00 & 0.01 & 0.08 \\
\ion{Mg}{1} & ~~~~4 & ~~4.45 & ~~0.18 & 0.06 & 0.04 & 0.00 & 0.03 & 0.04 \\
\ion{K}{1}  &\nodata& \nodata &\nodata  & \nodata  & \nodata & \nodata & \nodata & \nodata\\
\ion{Ca}{1} & ~~~~5 & ~~3.22 & ~~0.21 & 0.08 & 0.04 & 0.00 & 0.02 & 0.05 \\
\ion{Sc}{2} &\nodata& \nodata &\nodata  & \nodata  & \nodata & \nodata & \nodata & \nodata\\
\ion{Cr}{1} & ~~~~2 & ~~1.95 & --0.36 & 0.15 & 0.01 & 0.00 & 0.01 & 0.08  \\
\ion{Mn}{1} &\nodata& \nodata &\nodata  & \nodata  & \nodata & \nodata & \nodata & \nodata \\
\ion{Fe}{1} & ~~54 & ~~4.17 & ~~0.00 & 0.03 & 0.12 & 0.01 & 0.02 & 0.06  \\
\ion{Ni}{1} & ~~~~2 & ~~3.12 & ~~0.23 & 0.15 & 0.03 & 0.00 & 0.02 & 0.08 \\
\ion{Zn}{1} &\nodata& \nodata &\nodata  & \nodata  & \nodata & \nodata & \nodata & \nodata \\
\ion{Ba}{2} & ~~~~2 & --2.30 & --1.15 & 0.20 & \nodata & \nodata & \nodata & 0.20  \\
\hline
\multicolumn{9}{c}{J0713}\\
\hline
\ion{Li}{1} & \nodata & \nodata & \nodata & \nodata & \nodata & \nodata & \nodata & \nodata \\
\ion{Na}{1} & ~~~~2 & ~~3.45 & ~~0.37 & 0.13 & 0.06 & 0.01 & 0.06 & 0.08 \\
\ion{Mg}{1} & ~~~~4 & ~~4.75 & ~~0.31 & 0.10 & 0.05 & 0.02 & 0.02 & 0.06 \\
\ion{K}{1}  & ~~~~1 & ~~2.05 & ~~0.18 & 0.19 & 0.05 & 0.03 & 0.04 & 0.10 \\
\ion{Ca}{1} & ~~~~8 & ~~3.18 & ~~0.00 & 0.02 & 0.08 & 0.03 & 0.04 & 0.05 \\
\ion{Sc}{2} & ~~~~3 & --0.27 & --0.26 & 0.11 & 0.14 & 0.10 & 0.04 & 0.10 \\
\ion{Cr}{1} & ~~~~5 & ~~2.15 & --0.33 & 0.06 & 0.02 & 0.00 & 0.01 & 0.03 \\
\ion{Mn}{1} & ~~~~1 & ~~2.25 & --0.02 & 0.19 & 0.08 & 0.00 & 0.03 & 0.10 \\
\ion{Fe}{1} & ~~96 & ~~4.34 & ~~0.00 & 0.02 & 0.15 & 0.06 & 0.04 & 0.08 \\
\ion{Ni}{1} & ~~~~4 & ~~3.05 & --0.01 & 0.04 & 0.07 & 0.01 & 0.03 & 0.04 \\
\ion{Zn}{1} & ~~~~1 & ~~1.71 & ~~0.31 & 0.19 & 0.16 & 0.04 & 0.04 & 0.13 \\
\ion{Ba}{2} & ~~~~2 & --2.66 & --2.08 & 0.04 & \nodata & \nodata & \nodata & 0.04 \\
\hline
\multicolumn{9}{c}{J0758}\\
\hline
\ion{Li}{1} & \nodata & \nodata & \nodata & \nodata & \nodata & \nodata & \nodata & \nodata \\
\ion{Na}{1} & ~~~~1 & ~~2.53 & --0.75 & 0.47 & 0.02 & 0.01 & 0.01 & 0.24 \\
\ion{Mg}{1} & ~~~~2 & ~~4.70 & ~~0.06 & 0.33 & 0.02 & 0.02 & 0.03 & 0.17 \\
\ion{K}{1}  &\nodata& \nodata &\nodata  & \nodata  & \nodata & \nodata & \nodata & \nodata\\
\ion{Ca}{1} &\nodata& \nodata &\nodata  & \nodata  & \nodata & \nodata & \nodata & \nodata\\
\ion{Sc}{2} &\nodata& \nodata &\nodata  & \nodata  & \nodata & \nodata & \nodata & \nodata\\
\ion{Cr}{1} & ~~~~1 & ~~2.57 & --0.11 & 0.47 & 0.01 & 0.00 & 0.01 & 0.24 \\
\ion{Mn}{1} &\nodata& \nodata &\nodata  & \nodata  & \nodata & \nodata & \nodata & \nodata\\
\ion{Fe}{1} & ~~~~4 & ~~4.54 & ~0.00 & 0.24 & 0.08 & 0.00 & 0.00 & 0.13 \\
\ion{Ni}{1} &\nodata& \nodata &\nodata  & \nodata  & \nodata & \nodata & \nodata & \nodata\\
\ion{Zn}{1} &\nodata& \nodata &\nodata  & \nodata  & \nodata & \nodata & \nodata & \nodata\\
\ion{Ba}{2} &\nodata & \nodata & \nodata & \nodata & \nodata & \nodata & \nodata & \nodata \\
\hline
\multicolumn{9}{c}{J0814}\\
\hline
\ion{Li}{1} & ~~~~1 & ~~1.76 & ~~4.10 & 0.20 & \nodata & \nodata & \nodata & 0.20 \\
\ion{Na}{1} & ~~~~2 & ~~2.80 & --0.05 & 0.10 & 0.02 & 0.00 & 0.00 & 0.05 \\
\ion{Mg}{1} & ~~~~3 & ~~4.54 & ~~0.33 & 0.08 & 0.03 & 0.02 & 0.02 & 0.04 \\
\ion{K}{1}  &\nodata& \nodata &\nodata  & \nodata  & \nodata & \nodata & \nodata & \nodata\\
\ion{Ca}{1} & ~~~~1 & ~~3.56 & ~~0.61 & 0.14 & 0.04 & 0.01 & 0.01 & 0.07 \\
\ion{Sc}{2} &\nodata& \nodata &\nodata  & \nodata  & \nodata & \nodata & \nodata & \nodata\\
\ion{Cr}{1} &\nodata& \nodata &\nodata  & \nodata  & \nodata & \nodata & \nodata & \nodata\\
\ion{Mn}{1} &\nodata& \nodata &\nodata  & \nodata  & \nodata & \nodata & \nodata & \nodata\\
\ion{Fe}{1} & ~~13 & ~~4.11 & ~~0.00 & 0.04 & 0.10 & 0.00 & 0.02 & 0.05 \\
\ion{Ni}{1} &\nodata& \nodata &\nodata  & \nodata  & \nodata & \nodata & \nodata & \nodata\\
\ion{Zn}{1} &\nodata& \nodata &\nodata  & \nodata  & \nodata & \nodata & \nodata & \nodata\\
\ion{Ba}{2} &\nodata & \nodata & \nodata & \nodata & \nodata & \nodata & \nodata & \nodata \\
\hline
\multicolumn{9}{c}{J0908}\\
\hline
\ion{Li}{1} & \nodata & \nodata & \nodata & \nodata & \nodata & \nodata & \nodata & \nodata\\
\ion{Na}{1} & ~~~~2 & ~~2.38 & --0.19 & 0.15 & 0.02 & 0.01 & 0.02 & 0.08 \\
\ion{Mg}{1} & ~~~~3 & ~~4.10 & ~~0.17 & 0.12 & 0.06 & 0.01 & 0.04 & 0.07 \\
\ion{K}{1}  & ~~~~1 & ~~2.32 & ~~0.96 & 0.21 & 0.04 & 0.02 & 0.02 & 0.11 \\
\ion{Ca}{1} & ~~~~4 & ~~3.02 & ~~0.35 & 0.07 & 0.05 & 0.01 & 0.02 & 0.04 \\
\ion{Sc}{2} &\nodata& \nodata &\nodata  & \nodata  & \nodata & \nodata & \nodata & \nodata\\
\ion{Cr}{1} & ~~~~2 & ~~1.50 & --0.47 & 0.15 & 0.01 & 0.01 & 0.02 & 0.08 \\
\ion{Mn}{1} &\nodata& \nodata &\nodata  & \nodata  & \nodata & \nodata & \nodata & \nodata\\
\ion{Fe}{1} & ~~49 & ~~3.83 & ~~0.00 & 0.03 & 0.12 & 0.02 & 0.03 & 0.06 \\
\ion{Ni}{1} & ~~~~2 & ~~2.90 & ~~0.35 & 0.15 & 0.03 & 0.00 & 0.02 & 0.08 \\
\ion{Zn}{1} &\nodata& \nodata &\nodata  & \nodata  & \nodata & \nodata & \nodata & \nodata\\
\ion{Ba}{2} & ~~~~1 & $<$--2.89 & $<$--1.40 & 0.20 & \nodata & \nodata & \nodata & 0.20 \\
\hline
\multicolumn{9}{c}{J0925}\\
\hline
\ion{Li}{1} & ~~~~1 & ~~1.32 & ~~3.80 & 0.20 & \nodata & \nodata & \nodata & 0.20 \\
\ion{Mg}{1} & ~~~~2 & ~~4.42 & ~~0.35 & 0.09 & 0.02 & 0.02 & 0.05 & 0.05 \\
\ion{K}{1}  &\nodata& \nodata &\nodata  & \nodata  & \nodata & \nodata & \nodata & \nodata\\
\ion{Ca}{1} &\nodata& \nodata &\nodata  & \nodata  & \nodata & \nodata & \nodata & \nodata\\
\ion{Sc}{2} &\nodata& \nodata &\nodata  & \nodata  & \nodata & \nodata & \nodata & \nodata\\
\ion{Cr}{1} &\nodata& \nodata &\nodata  & \nodata  & \nodata & \nodata & \nodata & \nodata\\
\ion{Mn}{1} &\nodata& \nodata &\nodata  & \nodata  & \nodata & \nodata & \nodata & \nodata\\
\ion{Fe}{1} & ~~~~7 & ~~3.97 & ~~0.00 & 0.05 & 0.11 & 0.00 & 0.02 & 0.06 \\
\ion{Ni}{1} &\nodata& \nodata &\nodata  & \nodata  & \nodata & \nodata & \nodata & \nodata\\
\ion{Zn}{1} &\nodata& \nodata &\nodata  & \nodata  & \nodata & \nodata & \nodata & \nodata\\
\ion{Ba}{2} &\nodata & \nodata & \nodata & \nodata & \nodata & \nodata & \nodata & \nodata \\
\hline
\multicolumn{9}{c}{J1037}\\
\hline
\ion{Li}{1} & ~~~~1 & ~~2.21 & ~~3.66 & 0.20 & \nodata & \nodata & \nodata & 0.20 \\
\ion{Na}{1} & ~~~~2 & ~~3.33 & --0.41 & 0.12 & 0.01 & 0.01 & 0.01 & 0.06 \\
\ion{Mg}{1} & ~~~~4 & ~~5.40 & ~~0.30 & 0.05 & 0.02 & 0.03 & 0.01 & 0.03 \\
\ion{K}{1}  &\nodata& \nodata &\nodata  & \nodata  & \nodata & \nodata & \nodata & \nodata\\
\ion{Ca}{1} & ~~~~4 & ~~3.96 & ~~0.12 & 0.04 & 0.03 & 0.01 & 0.02 & 0.03 \\
\ion{Cr}{1} & ~~~~1 & ~~2.90 & --0.24 & 0.17 & 0.01 & 0.00 & 0.01 & 0.09 \\
\ion{Mn}{1} &\nodata& \nodata &\nodata  & \nodata  & \nodata & \nodata & \nodata & \nodata\\
\ion{Fe}{1} & ~~21 & ~~5.00 & ~~0.00 & 0.04 & 0.08 & 0.0 & 0.02 & 0.05 \\
\ion{Ni}{1} & ~~~~1 & ~~4.06 & ~~0.34 & 0.17 & 0.01 & 0.0 & 0.01 & 0.09 \\
\ion{Zn}{1} &\nodata& \nodata &\nodata  & \nodata  & \nodata & \nodata & \nodata & \nodata\\
\ion{Ba}{2} &\nodata & \nodata & \nodata & \nodata & \nodata & \nodata & \nodata & \nodata \\
\hline
\multicolumn{9}{c}{J1311}\\
\hline
\ion{Li}{1} & \nodata & \nodata & \nodata & \nodata & \nodata & \nodata & \nodata & \nodata\\
\ion{Na}{1} & ~~~~1 & ~~3.38 & --0.12 & 0.05 & 0.02 & 0.01 & 0.02 & 0.03 \\
\ion{Mg}{1} & ~~~~2 & ~~4.64 & --0.22 & 0.04 & 0.02 & 0.02 & 0.03 & 0.03 \\
\ion{K}{1}  &\nodata& \nodata &\nodata  & \nodata  & \nodata & \nodata & \nodata & \nodata\\
\ion{Ca}{1} &\nodata& \nodata &\nodata  & \nodata  & \nodata & \nodata & \nodata & \nodata\\
\ion{Sc}{2} &\nodata& \nodata &\nodata  & \nodata  & \nodata & \nodata & \nodata & \nodata\\
\ion{Cr}{1} & ~~~~2 & ~~2.55 & --0.35 & 0.04 & 0.00 & 0.00 & 0.00 & 0.02 \\
\ion{Mn}{1} &\nodata& \nodata &\nodata  & \nodata  & \nodata & \nodata & \nodata & \nodata\\
\ion{Fe}{1} & ~~~~2 & ~~4.76 & 0.00 & 0.04 & 0.08 & 0.00 & 0.01 & 0.04 \\
\ion{Ni}{1} &\nodata& \nodata &\nodata  & \nodata  & \nodata & \nodata & \nodata & \nodata\\
\ion{Zn}{1} &\nodata& \nodata &\nodata  & \nodata  & \nodata & \nodata & \nodata & \nodata\\
\ion{Ba}{2} & ~~~~1 & $<$--0.76 & $<$--0.2 & 0.20 & \nodata & \nodata & \nodata & 0.20 \\
\hline
\multicolumn{9}{c}{J1317}\\
\hline
\ion{Li}{1} & ~~~~1 & ~~2.41 & ~~3.73 & 0.20 & \nodata & \nodata & \nodata & 0.20 \\
\ion{Na}{1} & ~~~~2 & ~~3.79 & --0.08 & 0.15 & 0.02 & 0.01 & 0.01 & 0.08 \\
\ion{Mg}{1} & ~~~~3 & ~~5.53 & ~~0.30 & 0.12 & 0.01 & 0.02 & 0.04 & 0.06 \\
\ion{K}{1}  &\nodata& \nodata &\nodata  & \nodata  & \nodata & \nodata & \nodata & \nodata\\
\ion{Ca}{1} & ~~~~6 & ~~4.21 & ~~0.24 & 0.1 & 0.03 & 0.00 & 0.00 & 0.05 \\
\ion{Sc}{2} &\nodata& \nodata &\nodata  & \nodata  & \nodata & \nodata & \nodata & \nodata\\
\ion{Cr}{1} & ~~~~1 & ~~3.18 & --0.09 & 0.21 & 0.00 & 0.00 & 0.01 & 0.11 \\
\ion{Mn}{1} &\nodata& \nodata &\nodata  & \nodata  & \nodata & \nodata & \nodata & \nodata\\
\ion{Fe}{1} & ~~19 & ~~5.13 & ~~0.00 & 0.05 & 0.08 & 0.00 & 0.02 & 0.05 \\
\ion{Ni}{1} & ~~~~3 & ~~4.54 & ~~0.69 & 0.12 & 0.03 & 0.00 & 0.01 & 0.06 \\
\ion{Zn}{1} &\nodata& \nodata &\nodata  & \nodata  & \nodata & \nodata & \nodata & \nodata\\
\ion{Ba}{2} &\nodata & \nodata & \nodata & \nodata & \nodata & \nodata & \nodata & \nodata \\
\hline
\multicolumn{9}{c}{J1522}\\
\hline
\ion{Li}{1} & ~~~~1 & ~~1.65 & ~~4.30 & 0.20 & \nodata & \nodata & \nodata & 0.20 \\
\ion{Mg}{1} & ~~~~2 & ~~4.32 & ~~0.42 & 0.13 & 0.02 & 0.01 & 0.05 & 0.07 \\
\ion{K}{1}  &\nodata& \nodata &\nodata  & \nodata  & \nodata & \nodata & \nodata & \nodata\\
\ion{Ca}{1} &\nodata& \nodata &\nodata  & \nodata  & \nodata & \nodata & \nodata & \nodata\\
\ion{Sc}{2} &\nodata& \nodata &\nodata  & \nodata  & \nodata & \nodata & \nodata & \nodata\\
\ion{Cr}{1} &\nodata& \nodata &\nodata  & \nodata  & \nodata & \nodata & \nodata & \nodata\\
\ion{Mn}{1} &\nodata& \nodata &\nodata  & \nodata  & \nodata & \nodata & \nodata & \nodata\\
\ion{Fe}{1} & ~~~~6 & ~~3.80 & ~~0.00 & 0.07 & 0.10 & 0.00 & 0.01 & 0.06 \\
\ion{Ni}{1} &\nodata& \nodata &\nodata  & \nodata  & \nodata & \nodata & \nodata & \nodata\\
\ion{Zn}{1} &\nodata& \nodata &\nodata  & \nodata  & \nodata & \nodata & \nodata & \nodata\\
\ion{Ba}{2} &\nodata &\nodata & \nodata & \nodata & \nodata & \nodata & \nodata & \nodata \\
\hline
\multicolumn{9}{c}{J1650}\\
\hline
\ion{Li}{1} & ~~~~1 & ~~2.35 & ~~3.47 & 0.20 & \nodata & \nodata & \nodata & 0.20  \\
\ion{Na}{1} & ~~~~2 & ~~3.48 & --0.60 & 0.20 & 0.01 & 0.01 & 0.01 & 0.10 \\
\ion{Mg}{1} & ~~~~4 & ~~5.57 & ~~0.13 & 0.13 & 0.02 & 0.02 & 0.02 & 0.07 \\
\ion{K}{1}  & ~~~~1 & ~~3.56 & ~~0.69 & 0.28 & 0.01 & 0.00 & 0.02 & 0.14 \\
\ion{Ca}{1} & ~~~~6 & ~~4.53 & ~~0.35 & 0.07 & 0.02 & 0.00 & 0.01 & 0.04 \\
\ion{Sc}{2} &\nodata& \nodata &\nodata  & \nodata  & \nodata & \nodata & \nodata & \nodata\\
\ion{Cr}{1} & ~~~~2 & ~~3.26 & --0.22 & 0.20 & 0.01 & 0.01 & 0.02 & 0.10 \\
\ion{Mn}{1} &\nodata& \nodata &\nodata  & \nodata  & \nodata & \nodata & \nodata & \nodata\\
\ion{Fe}{1} & ~~25 & ~~5.34 & ~~0.00 & 0.06 & 0.07 & 0.00 & 0.02 & 0.05 \\
\ion{Ni}{1} & ~~~~2 & ~~4.79 & ~~0.73 & 0.20 & 0.02 & 0.01 & 0.02 & 0.10 \\
\ion{Zn}{1} &\nodata& \nodata &\nodata  & \nodata  & \nodata & \nodata & \nodata & \nodata\\
\ion{Ba}{2} & ~~~~1 & $<$--0.63 & $<$--0.64 & 0.20 & \nodata & \nodata & \nodata & 0.20 \\
\hline
\multicolumn{9}{c}{J1705}\\
\hline
\ion{Li}{1} & ~~~~1 & ~~1.96 & ~~3.55 & 0.20 & \nodata & \nodata & \nodata & 0.20  \\
\ion{Na}{1} & ~~~~2 & ~~3.16 & --0.44 & 0.11 & 0.02 & 0.00 & 0.01 & 0.06 \\
\ion{Mg}{1} & ~~~~4 & ~~5.21 & ~~0.25 & 0.06 & 0.02 & 0.02 & 0.02 & 0.03 \\
\ion{K}{1}  &\nodata& \nodata &\nodata  & \nodata  & \nodata & \nodata & \nodata & \nodata\\
\ion{Ca}{1} & ~~~~4 & ~~4.00 & ~~0.30 & 0.04 & 0.02 & 0.00 & 0.01 & 0.02 \\
\ion{Sc}{2} &\nodata& \nodata &\nodata  & \nodata  & \nodata & \nodata & \nodata & \nodata\\
\ion{Cr}{1} & ~~~~1 & ~~2.93 & --0.07 & 0.16 & 0.01 & 0.00 & 0.00 & 0.08 \\
\ion{Fe}{1} & ~~20 & ~~4.86 & ~~0.00 & 0.04 & 0.08 & 0.00 & 0.02 & 0.05 \\
\ion{Ni}{1} &\nodata& \nodata &\nodata  & \nodata  & \nodata & \nodata & \nodata & \nodata\\
\ion{Zn}{1} &\nodata& \nodata &\nodata  & \nodata  & \nodata & \nodata & \nodata & \nodata\\
\ion{Ba}{2} & ~~~~1 & $<$--0.96 & $<$--0.50 & 0.20 & \nodata & \nodata & \nodata & 0.20 \\
\hline
\multicolumn{9}{c}{J2241}\\
\hline
\ion{Li}{1} & \nodata & \nodata & \nodata & \nodata & \nodata & \nodata & \nodata & \nodata \\
\ion{Na}{1} & ~~~~2 & ~~2.93 & --0.60 & 0.29 & 0.03 & 0.00 & 0.01 & 0.15 \\
\ion{Mg}{1} & ~~~~2 & ~~4.28 & --0.61 & 0.29 & 0.04 & 0.01 & 0.01 & 0.15 \\
\ion{K}{1}  &\nodata& \nodata &\nodata  & \nodata  & \nodata & \nodata & \nodata & \nodata\\
\ion{Ca}{1} &\nodata& \nodata &\nodata  & \nodata  & \nodata & \nodata & \nodata & \nodata\\
\ion{Sc}{2} &\nodata& \nodata &\nodata  & \nodata  & \nodata & \nodata & \nodata & \nodata\\
\ion{Cr}{1} &\nodata& \nodata &\nodata  & \nodata  & \nodata & \nodata & \nodata & \nodata\\
\ion{Mn}{1} &\nodata& \nodata &\nodata  & \nodata  & \nodata & \nodata & \nodata & \nodata\\
\ion{Fe}{1} & ~~~~3 & ~~4.79 & ~~0.00 & 0.24 & 0.09 & 0.00 & 0.01 & 0.13 \\
\ion{Ni}{1} &\nodata& \nodata &\nodata  & \nodata  & \nodata & \nodata & \nodata & \nodata\\
\ion{Zn}{1} &\nodata& \nodata &\nodata  & \nodata  & \nodata & \nodata & \nodata & \nodata\\
\ion{Ba}{2} & ~~~~1 & $<$--1.03 & $<$--0.50 & 0.20 & \nodata & \nodata & \nodata & 0.20 \\
\hline
\multicolumn{9}{c}{J2242}\\
\hline
\ion{Li}{1} & \nodata & \nodata & \nodata & \nodata & \nodata & \nodata & \nodata & \nodata \\
\ion{Na}{1} & ~~~~2 & ~~1.82 & --1.02 & 0.51 & 0.02 & 0.00 & 0.02 & 0.26 \\
\ion{Mg}{1} & ~~~~4 & ~~4.44 & ~~0.24 & 0.06 & 0.05 & 0.00 & 0.04 & 0.04 \\
\ion{K}{1}  & ~~~~1 & ~~3.11 & ~~1.48 & 0.72 & 0.04 & 0.01 & 0.01 & 0.36 \\
\ion{Ca}{1} & ~~~~4 & ~~3.65 & ~~0.71 & 0.28 & 0.05 & 0.01 & 0.02 & 0.14 \\
\ion{Sc}{2} & ~~~~1 & --0.36 & --0.11 & 0.72 & 0.07 & 0.07 & 0.02 & 0.36 \\
\ion{Cr}{1} & ~~~~2 & ~~2.25 & ~~0.01 & 0.51 & 0.00 & 0.00 & 0.02 & 0.26 \\
\ion{Mn}{1} &\nodata& \nodata &\nodata  & \nodata  & \nodata & \nodata & \nodata & \nodata\\
\ion{Fe}{1} & ~~40 & ~~4.10 & ~~0.00 & 0.11 & 0.11 & 0.02 & 0.03 & 0.08 \\
\ion{Ni}{1} & ~~~~1 & ~~3.71 & ~~0.89 & 0.72 & 0.05 & 0.01 & 0.03 & 0.36 \\
\ion{Zn}{1} &\nodata& \nodata &\nodata  & \nodata  & \nodata & \nodata & \nodata & \nodata\\
\ion{Ba}{2} &\nodata & \nodata & \nodata & \nodata & \nodata & \nodata & \nodata & \nodata \\
\hline
\multicolumn{9}{c}{J2341}\\
\hline
\ion{Li}{1} & \nodata & \nodata & \nodata & \nodata & \nodata & \nodata & \nodata & \nodata \\
\ion{Na}{1} & ~~~~2 & ~~3.81 & ~~0.74 & 0.17 & 0.00 & 0.02 & 0.09 & 0.10 \\
\ion{Mg}{1} & ~~~~4 & ~~4.79 & ~~0.36 & 0.04 & 0.05 & 0.02 & 0.03 & 0.04 \\
\ion{K}{1}  & ~~~~1 & ~~2.45 & ~~0.59 & 0.24 & 0.03 & 0.02 & 0.03 & 0.12 \\
\ion{Ca}{1} & ~~10 & ~~3.42 & ~~0.25 & 0.05 & 0.05 & 0.01 & 0.03 & 0.04 \\
\ion{Sc}{2} & ~~~~4 & ~~0.07 & ~~0.09 & 0.09 & 0.08 & 0.08 & 0.03 & 0.07 \\
\ion{Cr}{1} & ~~~~5 & ~~2.09 & --0.38 & 0.08 & 0.01 & 0.00 & 0.02 & 0.04 \\
\ion{Mn}{1} &\nodata& \nodata &\nodata  & \nodata  & \nodata & \nodata & \nodata & \nodata\\
\ion{Fe}{1} & ~~88 & ~~4.33 & ~~0.00 & 0.03 & 0.12 & 0.03 & 0.04 & 0.07 \\
\ion{Ni}{1} & ~~~~2 & ~~3.12 & ~~0.07 & 0.17 & 0.02 & 0.00 & 0.03 & 0.09 \\
\ion{Zn}{1} &\nodata& \nodata &\nodata  & \nodata  & \nodata & \nodata & \nodata & \nodata\\
\ion{Ba}{2} & ~~~~2 & --2.60 & --1.61 & 0.06 & \nodata & \nodata & \nodata & 0.06 \\
\hline
\enddata
\tablecomments{SE stands for the standard error
of the mean, while TE indicates the total error
calculated by quadrature sum of SE and errors
caused by deviation of \teff, \logg, and \vt. The Li and Ba abundance ratios
are derived from the spectral synthesis, while the rest by the equivalent width analysis.
The $<$ symbol indicates an upper limit estimate.}
\end{deluxetable}  


\begin{thebibliography}{123456789}

\bibitem[Aguado et al.(2019)]{aguado2019} Aguado, D.~S., Ahumada, R., Almeida, A., et al.\ 2019, \apjs, 240, 23 
\bibitem[Allende Prieto et al.(2008)]{allende2008} Allende Prieto, C., Sivarani, T., Beers, T. C., et al. 2008, \aj, 136, 2070 
\bibitem[Alonso et al.(1999)]{alonso1999b} Alonso, A., Arribas, S., \& Mart{\'\i}nez-Roger, C.\ 1999, \aaps, 140, 261 
\bibitem[Amarsi et al.(2016)]{amarsi2016} Amarsi, A.~M., Lind, K., Asplund, M., et al.\ 2016, \mnras, 463, 1518 
\bibitem[Andrievsky et al.(2007)]{andrievsky2007} Andrievsky, S.~M., Spite, M., Korotin, S.~A., et al.\ 2007, \aap, 464, 1081 
\bibitem[Anthony-Twarog \& Twarog(1994)]{anthony1994} Anthony-Twarog, B.~J., \& Twarog, B.~A.\ 1994, \aj, 107, 1577
\bibitem[Arentsen et al.(2022)]{arentsen2022} Arentsen, A., Placco, V. M., Lee, Y. S., et al. 2022, \mnras, 515, 4082 
\bibitem[Arnett et al.(1971)]{arnett1971} Arnett, W.~D., Truran, J.~W., \& Woosley, S.~E.\
1971, \apj, 165, 87 

\bibitem[Aoki et al.(2005)]{aoki2005} Aoki, W., Honda, S., Beers, T.~C., et al.\ 2005, \apj, 632, 611 
\bibitem[Aoki et al.(2007)]{aoki2007} Aoki, W., Beers, T.~C., Christlieb, N., et al.\ 2007, \apj, 655, 492
\bibitem[Aoki et al.(2013)]{aoki2013} Aoki, W., Beers, T.~C., Lee, Y.~S., et al.\ 2013, \aj, 145, 13
\bibitem[Aoki et al.(2014)]{aoki2014} Aoki, W., Tominaga, N., Beers, T.~C., et al.\ 2014, Science, 345, 912 
\bibitem[Aoki et al.(2018)]{aoki2018} Aoki, W., Matsuno, T., Honda, S., et al.\ 2018, \pasj, 70, 94
\bibitem[Asplund et al.(2009)]{asplund2009} Asplund, M., Grevesse, N., Sauval, A. J., \& Scott, P. 2009, \araa, 47, 481 
\bibitem[Astropy Collaboration et al.(2018)]{astropy2} Astropy Collaboration, et al.\ 2018, \aj, 156, 123 
\bibitem[Astropy Collaboration et al.(2022)]{astropy3} Astropy Collaboration, et al.\ 2022, \apj, 935, 167 


\bibitem[Bastian \& Lardo(2018)]{bastian2018} Bastian, N. \& Lardo, C.\ 2018, \araa, 56, 83 
\bibitem[Beers et al.(1985)]{beers1985} Beers, T.~C., Preston, G.~W., \& Shectman, S.~A.\ 1985, \aj, 90, 2089 
\bibitem[Beers et al.(1992)]{beers1992} Beers, T.~C., Preston, G.~W., \& Shectman, S.~A.\ 1992, \aj, 103, 1987 
\bibitem[Beers et al.(2000)]{beers2000} Beers, T.~C., Chiba, M., Yoshii, Y., et al.\ 2000, \aj, 119, 2866 
\bibitem[Beers et al.(2012)]{beers2012} Beers, T.~C., Carollo, D., Ivezi{\'c}, {\v{Z}}., et al.\ 2012, \apj, 746, 34 
\bibitem[Beers \& Christlieb(2005)]{beers2005} Beers, T. C., \& Christlieb, N. 2005, \araa, 43, 531 
\bibitem[Blanton et al.(2017)]{blanton2017} Blanton, M.~R., Bershady, M.~A., Abolfathi, B., et al.\ 2017, \aj, 154, 28 
\bibitem[Bonifacio et al.(2010)]{bonifacio2010} Bonifacio, P., Monaco, L., Sbordone, L., et al.\ 2010, Light Elements in the Universe, 268, 269 
\bibitem[Bonifacio et al.(2018)]{bonifacio2018} Bonifacio, P., Caffau, E., Spite, M., et al.\ 2018, \aap, 612, A65 
\bibitem[Bromm \& Loeb(2003)]{bromm2003} Bromm, V. \& Loeb, A.\ 2003, \nat, 425, 812. 
\bibitem[Bromm \& Larson(2004)]{bromm2004} Bromm, V. \& Larson, R.~B.\ 2004, \araa, 42, 79 

\bibitem[Caffau et al.(2011)]{caffau2011} Caffau, E., Bonifacio, P., Fran{\c{c}}ois, P., et al.\ 2011, \aap, 534, A4
\bibitem[Caffau et al.(2013)]{caffau2013} Caffau, E., Bonifacio, P., Sbordone, L., et al.\ 2013, \aap, 560, A71 
\bibitem[Carretta et al.(2012)]{carretta2012} Carretta, E., D'Orazi, V., Gratton, R.~G., et al.\ 2012, \aap, 543, A117 
\bibitem[Casagrande et al.(2010)]{casagrande2010} Casagrande, L., Ram{\'\i}rez, I., Mel{\'e}ndez, J., et al.\ 2010, \aap, 512, A54
\bibitem[Castelli et al.(1997)]{castelli1997} Castelli, F., Gratton, R.~G., \& Kurucz, R.~L.\ 1997, \aap, 318, 841
\bibitem[Castelli \& Kurucz(2003)]{castelli2003} Castelli, F., \& Kurucz, R.~L.\ 2003, Modelling of Stellar Atmospheres, A20
\bibitem[Castelli \& Kurucz(2004)]{castelli2004} Castelli, F. \& Kurucz, R.~L.\ 2004, \aap, 419, 725 
\bibitem[Cayrel et al.(2004)]{cayrel2004} Cayrel, R., Depagne, E., Spite, M., et al.\ 2004, \aap, 416, 1117 
\bibitem[Cenarro et al.(2019)]{cenarro2019} Cenarro, A.~J., Moles, M., Crist{\'o}bal-Hornillos, D., et al.\ 2019, \aap, 622, A176 
\bibitem[Cescutti et al.(2006)]{cescutti2006} Cescutti, G., Fran{\c{c}}ois, P., Matteucci, F., et al.\ 2006, \aap, 448, 557 
\bibitem[Chen et al.(2014)]{chen2014} Chen, Y. Q., Zhao, G., Carrell, K., et al. 2014, \apj, 795, 52 
\bibitem[Chen{\'e} et al.(2014)]{chene2014} Chen{\'e}, A.-N., Padzer, J., Barrick, G., et al.\ 2014, \procspie, 915147
\bibitem[Chen{\'e} et al.(2021)]{chene2021} Chen{\'e}, A.-N., Mao, S., Lundquist, M., et al.\ 2021, \aj, 161, 109 
\bibitem[Chieffi \& Limongi(2002)]{chieffi2002} Chieffi, A. \& Limongi, M.\ 2002, \apj, 577, 281 
\bibitem[Choplin et al.(2018)]{choplin2018} Choplin, A., Hirschi, R., Meynet, G., et al.\ 2018, \aap, 618, A133 
\bibitem[Choplin et al.(2022)]{choplin2022} Choplin, A., Siess, L, \& Goriely, S.\ 2022, \aap, 667, A155 
\bibitem[Christlieb et al.(2002)]{christlieb2002} Christlieb, N., Bessell, M.~S., Beers, T.~C., et al.\ 2002, \nat, 419, 904 
\bibitem[Christlieb et al.(2008)]{christlieb2008} Christlieb, N., Sch{\"o}rck, T., Frebel, A., et al.\ 2008, \aap, 484, 721 
\bibitem[Coc \& Vangioni(2017)]{coc2017} Coc, A. \& Vangioni, E.\ 2017, International Journal of Modern Physics E, 26, 1741002
\bibitem[Cohen \& Kirby(2012)]{cohen2012} Cohen J. G. \& Kirby E. N.\ 2012, \apj, 760, 86
\bibitem[Conroy et al.(2019)]{conroy2019} Conroy, C., Bonaca, A., Cargile, P., et al.\ 2019, \apj, 883, 107 
\bibitem[C{\^o}t{\'e} et al.(2018)]{cote2018} C{\^o}t{\'e}, B., Denissenkov, P., Herwig, F., et al.\ 2018, \apj, 854, 105 

\bibitem[C{\^o}t{\'e} et al.(2019)]{cote2019} C{\^o}t{\'e}, B., Eichler, M., Arcones, A., et al. 2019, \apj, 875, 106

\bibitem[Cowan \& Rose(1977)]{cowan1977} Cowan, J.~J. \& Rose, W.~K.\ 1977, \apj, 212, 149 
\bibitem[Cowan et al.(2021)]{cowan2021} Cowan, J.~J., Sneden, C., Lawler, J.~E., et al.\ 2021, Rev. Mod. Phys., 93, 015002 
\bibitem[Cui et al.(2012)]{cui2012} Cui, X. Q., Zhao, Y. H., Chu, Y. Q, et al. 2012, RAA, 12, 1197
\bibitem[Cyburt et al.(2016)]{cyburt2016} Cyburt, R.~H., Fields, B.~D., Olive, K.~A., et al.\ 2016, Reviews of Modern Physics, 88, 015004 

\bibitem[Dawson et al.(2013)]{dawson2013} Dawson, K.~S., Schlegel, D.~J., Ahn, C.~P., et al.\ 2013, \aj, 145, 10 
\bibitem[De Silva et al.(2015)]{desilva2015} De Silva, G.~M., Freeman, K.~C., Bland-Hawthorn, J., et al.\ 2015, \mnras, 449, 2604 
\bibitem[Demarque et al.(2004)]{demarque2004} Demarque, P., Woo, J.-H., Kim, Y.-C., et al.\ 2004, \apjs, 155, 667

\bibitem[Denissenkov et al.(2017)]{denissenkov2017} Denissenkov, P. A., Herwig, F., Battino, U., et al.\ 2017, \apjl, 834, L10 
\bibitem[Denissenkov et al.(2019)]{denissenkov2019} Denissenkov, P. A., Herwig, F., Woodward, P., et al.\ 2019, \mnras, 488, 4258 
\bibitem[Drout et al.(2017)]{drout2017} Drout, M. R., Piro, A. L., Shappee, B. J., et al. 2017, Science, 358, 1570

\bibitem[Ezzeddine et al.(2019)]{ezzeddine2019} Ezzeddine, R., Frebel, A., Roederer, I.~U., et al.\ 2019, \apj, 876, 97 

\bibitem[Fern{\'a}ndez-Trincado et al.(2017)]{trincado2017} Fern{\'a}ndez-Trincado, J.~G., Zamora, O., Garc{\'\i}a-Hern{\'a}ndez, D.~A., et al.\ 2017, \apjl, 846, L2
\bibitem[Frebel et al.(2005)]{frebel2005} Frebel, A., Aoki, W., Christlieb, N., et al.\ 2005, \nat, 434, 871
\bibitem[Frebel et al.(2006)]{frebel2006} Frebel, A., Christlieb, N., Norris, J. E., et al. 2006, \apj, 652, 1585
\bibitem[Frebel et al.(2008)]{frebel2008} Frebel, A., Collet, R., Eriksson, K., et al.\ 2008, \apj, 684, 588 
\bibitem[Frebel et al.(2013)]{frebel2013} Frebel, A., Casey, A.~R., Jacobson, H.~R., et al.\ 2013, \apj, 769, 57
\bibitem[Frebel \& Norris(2015)]{frebel2015} Frebel, A., \& Norris, J. E. 2015, \araa, 53, 631
\bibitem[Frischknecht et al.(2012)]{frischknecht2012} Frischknecht, U., Hirschi, R., \& Thielemann, F.-K.\ 2012, \aap, 538, L2 
\bibitem[Frischknecht et al.(2016)]{frischknecht2016} Frischknecht, U., Hirschi, R., Pignatari, M., et al.\ 2016, \mnras, 456, 1803 

\bibitem[Gaia Collaboration et al.(2016)]{gaia2016} Gaia Collaboration, Brown, A. G. A., Vallenari, A., et al. 2016, \aap, 595, 2 
\bibitem[Gaia Collaboration et al.(2018)]{gaia2018} Gaia Collaboration, Brown, A. G. A., Vallenari, A., et al. 2018, \aap, 616, 1
\bibitem[Gaia Collaboration et al.(2021)]{gaia2021} Gaia Collaboration, Brown, A.~G.~A., Vallenari, A., et al.\ 2021, \aap, 649, A1 
\bibitem[Gaia Collaboration et al.(2022)]{gaia2022} Gaia Collaboration, Vallenari, A., Brown, A.~G.~A., et al.\ 2022, arXiv:2208.00211 
\bibitem[Gilmore et al.(2012)]{gilmore2012} Gilmore, G., Randich, S., Asplund, M., et al.\ 2012, The Messenger, 147, 25 
\bibitem[Gonz{\'a}lez Hern{\'a}ndez \& Bonifacio(2009)]{gonzales2009} Gonz{\'a}lez Hern{\'a}ndez, J.~I., \& Bonifacio, P.\ 2009, \aap, 497, 497 
\bibitem[Gratton et al.(2004)]{gratton2004} Gratton, R., Sneden, C., \& Carretta, E.\ 2004, \araa, 42, 385 

\bibitem[Hampel et al.(2016)]{hampel2016} Hampel, M., Stancliffe, R. J., Lugaro, M., \& Meyer, B. S. 2016, \apj, 831, 171
\bibitem[Hansen et al.(2016a)]{hansen2016a} Hansen, T. T., Andersen, J., Nordstr\"om, B., et al. 2016a, \aap, 586, 160
\bibitem[Hansen et al.(2016b)]{hansen2016b} Hansen, T. T., Andersen, J., Nordstr\"om, B., et al. 2016b, \aap, 588, 3
\bibitem[Hartwig et al.(2018)]{hartwig2018} Hartwig, T., Yoshida, N., Magg, M., et al. 2018, \mnras, 478, 1795
\bibitem[Hashimoto et al.(1989)]{hashimoto1989} Hashimoto, M., Nomoto, K., \& Shigeyama, T.\ 1989, \aap, 210, L5 
\bibitem[Heger \& Woosley(2010)]{heger2010} Heger, A., \& Woosley, S.~E.\ 2010, \apj, 724, 341 
\bibitem[Henden et al.(2016)]{henden2016} Henden, A.~A., Templeton, M., Terrell, D., et al.\ 2016, VizieR Online Data Catalog, II/336 
\bibitem[Herwig (2005)]{herwig2005} Herwig, F. 2005, \araa, 43, 435
\bibitem[Hirano et al.(2014)]{hirano2014} Hirano, S., Hosokawa, T., Yoshida, N., et al.\ 2014, \apj, 781, 60 
\bibitem[Hirschi(2007)]{hirschi2007} Hirschi, R.\ 2007, \aap, 461, 571 
\bibitem[Hoyle \& Fowler(1960)]{hoyle1960} Hoyle, F. \& Fowler, W.~A.\ 1960, \apj, 132, 565 

\bibitem[Hunter(2007)]{matplotlib} Hunter, J. D. 2007, Computing in Science and Engineering, 9, 90 

\bibitem[Ishigaki et al.(2014)]{ishigaki2014} Ishigaki, M.~N., Tominaga, N., Kobayashi, C., et al.\ 2014, \apjl, 792, L32 
\bibitem[Ishigaki et al.(2018)]{ishigaki2018} Ishigaki, M.~N., Tominaga, N., Kobayashi, C., et al.\ 2018, \apj, 857, 46
\bibitem[Ito et al.(2013)]{ito2013} Ito, H., Aoki, W., Beers, T.~C., et al.\ 2013, \apj, 773, 33 
\bibitem[Ivans et al.(2003)]{ivans2003} Ivans, I.~I., Sneden, C., James, C.~R., et al.\ 2003, \apj, 592, 906 
\bibitem[Iwamoto et al.(1999)]{iwamoto1999} Iwamoto, K., Brachwitz, F., Nomoto, K., et al.\ 1999, \apjs, 125, 439 

\bibitem[Jacobson et al.(2015)]{jacobson2015} Jacobson, H.~R., Keller, S., Frebel, A., et al.\ 2015, \apj, 807, 171 
\bibitem[Jorissen et al.(2016)]{jorissen2016} Jorissen, A., Van Eck, S., Van Winckel, H., et al. 2016, \aap, 586, A158

\bibitem[Keller et al.(2007)]{keller2007} Keller, S.~C., Schmidt, B.~P., Bessell, M.~S., et al.\ 2007, \pasa, 24, 1 
\bibitem[Keller et al.(2014)]{keller2014} Keller, S.~C., Bessell, M.~S., Frebel, A., et al.\ 2014, \nat, 506, 463 
\bibitem[Kemp et al.(2018)]{kemp2018} Kemp, A.~J., Casey, A.~R., Miles, M.~T., et al.\ 2018, \mnras, 480, 1384 
\bibitem[Kielty et al.(2021)]{kielty2021} Kielty, C.~L., Venn, K.~A., Sestito, F., et al.\ 2021, \mnras, 506, 1438
\bibitem[Kim et al.(2002)]{kim2002} Kim, Y.-C., Demarque, P., Yi, S.~K., et al.\ 2002, \apjs, 143, 499 
\bibitem[Kim et al.(2016)]{kim2016} Kim, B., An, D., Stauffer, J. R., et al.\ 2016, \apjs, 222,19
\bibitem[Kim et al.(2022)]{kim2022} Kim, C., Lee, Y.~S., Beers, T.~C., et al.\ 2022, JKAS, 55, 23 
\bibitem[Kirby et al.(2011)]{kirby2011} Kirby, E.~N., Martin, C.~L., \& Finlator, K.\ 2011, \apjl, 742, L25 
\bibitem[Kobayashi et al.(2006)]{kobayashi2006} Kobayashi, C., Umeda, H., Nomoto, K., et al.\ 2006, \apj, 653, 1145 
\bibitem[Kobayashi \& Nomoto(2009)]{kobayashi2009} Kobayashi, C. \& Nomoto, K.\ 2009, \apj, 707, 1466 
\bibitem[Kobayashi et al.(2011)]{kobayashi2011a} Kobayashi, C., Karakas, A.~I., \& Umeda, H.\ 2011, \mnras, 414, 3231 
\bibitem[Kobayashi et al.(2011)]{kobayashi2011b} Kobayashi, C., Izutani, N., Karakas, A.~I., et al.\ 2011, \apjl, 739, L57 
\bibitem[Kobayashi et al.(2020)]{kobayashi2020} Kobayashi, C., Karakas, A.~I., \& Lugaro, M.\ 2020, \apj, 900, 179 
\bibitem[Komiya et al.(2007)]{komiya2007} Komiya, Y., Suda, T., Minaguchi, H., et al. 2007, \apj, 658, 367
\bibitem[Koo et al.(2022)]{koo2022} Koo, J.-R., Lee, Y.~S., Park, H.-J., et al.\ 2022, \apj, 925, 35 
\bibitem[Kurucz(1993)]{kurucz1993} Kurucz, R.\ 1993, ATLAS9 Stellar Atmosphere Programs and 2 km/s grid. Kurucz CD-ROM No. 13. Cambridge, 13

\bibitem[Lee et al.(2008a)]{lee2008a} Lee, Y. S., Beers, T. C., Sivarani, T., et al. 2008a, \aj, 136, 2022
\bibitem[Lee et al.(2008b)]{lee2008b} Lee, Y. S., Beers, T. C., Sivarani, T., et al. 2008b, \aj, 136, 2050
\bibitem[Lee et al.(2011)]{lee2011} Lee, Y. S., Beers, T. C., Allende Prieto, C., et al. 2011, \aj, 141, 90
\bibitem[Lee et al.(2013)]{lee2013} Lee, Y. S., Beers, T. C., Masseron, T., et al. 2013, \aj, 146, 132
\bibitem[Lee et al.(2014)]{lee2014} Lee, Y.~S., Suda, T., Beers, T.~C., et al.\ 2014, \apj, 788, 131
\bibitem[Lee et al.(2017)]{lee2017} Lee, Y. S., Beers, T. C., Kim, Y. K., et al. 2017, \apj, 836, 91
\bibitem[Lee et al.(2019)]{lee2019} Lee, Y.~S., Beers, T.~C., \& Kim, Y.~K.\ 2019, \apj, 885, 102
\bibitem[Li et al.(2022)]{li2022} Li, H., Aoki, W., Matsuno, T., et al.\ 2022, \apj, 931, 147 
\bibitem[Lind et al.(2009)]{lind2009} Lind, K., Primas, F., Charbonnel, C., et al.\ 2009, \aap, 503, 545 
\bibitem[Lind et al.(2012)]{lind2012} Lind, K., Bergemann, M., \& Asplund, M.\ 2012, \mnras, 427, 50 
\bibitem[Lee et al.(2015)]{lee2015} Lee, Y.~S., Beers, T.~C., Carlin, J.~L., et al.\ 2015, \aj, 150, 187 
\bibitem[Lindegren et al.(2021)]{lindegren2021} Lindegren, L., Bastian, U., Biermann, M., et al.\ 2021, \aap, 649, A4
\bibitem[Lucatello et al.(2006)]{lucatello2006} Lucatello, S., Beers, T. C., Christlieb, N. C., et al. 2006, \apj, 652, L3
\bibitem[Lugaro et al.(2012)]{lugaro2012} Lugaro, M., Karakas, A. I., Stancliffe, R. J., \& Rijs, C. 2012, \apj, 742, 2
\bibitem[Majewski et al.(2017)]{majewski2017} Majewski, S.~R., Schiavon, R.~P., Frinchaboy, P.~M., et al.\ 2017, \aj, 154, 94 
\bibitem[Maeder et al.(2015)]{maeder2015} Maeder, A., Meynet, G., \& Chiappini, C.\ 2015, \aap, 576, A56 
\bibitem[Mardini et al.(2019)]{mardini2019} Mardini, M.~K., Placco, V.~M., Taani, A., et al.\ 2019, \apj, 882, 27 
\bibitem[Martell et al.(2011)]{martell2011} Martell, S.~L., Smolinski, J.~P., Beers, T.~C., et al.\ 2011, \aap, 534, A136 
\bibitem[Masseron et at.(2010)]{masseron2010} Masseron, T., Johnson, J. A., Plez, B., et al. 2010, \aap, 509, 93
\bibitem[Matas Pinto et al.(2021)]{mataspinto2021} Matas Pinto, A.~M., Spite, M., Caffau, E., et al.\ 2021, \aap, 654, A170 
\bibitem[Matsuno et al.(2017)]{matsuno2017} Matsuno, T., Aoki, W., Beers, T.~C., et al.\ 2017, \aj, 154, 52
\bibitem[Mel{\'e}ndez et al.(2010)]{melendez2010} Mel{\'e}ndez, J., Casagrande, L., Ram{\'\i}rez, I., et al.\ 2010, \aap, 515, L3 
\bibitem[Mendes de Oliveira et al.(2019)]{mendes2019} Mendes de Oliveira, C., Ribeiro, T., Schoenell, W., et al.\ 2019, \mnras, 489, 241 
\bibitem[Meynet et al.(2006)]{meynet2006} Meynet, G., Ekstr\"om, S., \& Maeder, A.. 2006, \aap, 447, 623
\bibitem[Mucciarelli et al.(2012)]{mucciarelli2012} Mucciarelli, A., Bellazzini, M., Ibata, R., et al.\ 2012, \mnras, 426, 2889
\bibitem[Mucciarelli et al.(2015)]{mucciarelli2015} Mucciarelli, A., Bellazzini, M., Merle, T., Plez, B., Dalessandro, E., \& Ibata R.\ 2015, \apj, 801, 68

\bibitem[Nishmura et al.(2015)]{nishimura2015} Nishimura, N., Takiwaki, T., \& Thielemann, F. 2015, \apj, 810, 2
\bibitem[Nomoto et al.(2013)]{nomoto2013} Nomoto, K., Kobayashi, C., \& Tominaga, N. 2013, \araa, 51, 457 
\bibitem[Norris et al.(2001)]{norris2001} Norris, J. E., Ryan, S. G., \& Beers, T. C. 2001, \apj, 561, 1034
\bibitem[Norris et al.(2002)]{norris2002} Norris, J.~E., Ryan, S.~G., Beers, T.~C., et al.\ 2002, \apjl, 569, L107 
\bibitem[Norris et al.(2013)]{norris2013} Norris, J.~E., Yong, D., Bessell, M.~S., et al.\ 2013, \apj, 762, 28
\bibitem[Norris \& Yong(2019)]{norris2019} Norris, J.~E., \& Yong, D.\ 2019, \apj, 879, 37
\bibitem[Nordlander et al.(2019)]{nordlander2019} Nordlander, T., Bessell, M.~S., Da Costa, G.~S., et al.\ 2019, \mnras, 488, L109 


\bibitem[Pancino et al.(2017)]{pancino2017} Pancino, E., Romano, D., Tang, B., et al.\ 2017, \aap, 601, A112 
\bibitem[Placco et al.(2014)]{placco2014} Placco, V. M., Frebel, A., Beers, T. C., Stancliffe, R. J. 2014 \apj, 797, 21
\bibitem[Placco et al.(2015a)]{placco2015a} Placco, V.~M., Frebel, A., Lee, Y.~S., et al.\ 2015, \apj, 809, 136
\bibitem[Placco et al.(2015b)]{placco2015b} Placco, V.~M., Beers, T. C., Ivans, I., et al.\ 2015, \apj, 812, 109
\bibitem[Placco et al.(2016)]{placco2016} Placco, V.~M., Frebel, A., Beers, T.~C., et al.\ 2016, \apj, 833, 21
\bibitem[Placco et al.(2020)]{placco2020} Placco, V.~M., Santucci, R.~M., Yuan, Z., et al.\ 2020, \apj, 897, 78 
\bibitem[Placco et al.(2021)]{placco2021a} Placco, V.~M., Sneden, C., Roederer, I.~U., et al.\ 2021, Research Notes of the American Astronomical Society, 5, 92 
\bibitem[Placco et al.(2021)]{placco2021b} Placco, V.~M., Roederer, I.~U., Lee, Y.~S., et al.\ 2021, \apjl, 912, L32
\bibitem[Placco et al.(2022)]{placco2022} Placco, V., Almeida-Fernandes, F., Arentsen, A., et al. 2022, \apjs, 262, 8


\bibitem[Rasmussen et al.(2020)]{rasmussen2020} Rasmussen, K.~C., Zepeda, J., Beers, T.~C., et al.\ 2020, \apj, 905, 20 
\bibitem[Reggiani et al.(2017)]{reggiani2017} Reggiani, H., Mel{\'e}ndez, J., Kobayashi, C., et al.\ 2017, \aap, 608, A46 
\bibitem[Robitaille et al.(2013)]{astropy1} Robitaille, T. P., Tollerud, E. J., Greenfield, P., et al.\ 2013, \aap, 558, A33

\bibitem[Roederer(2013)]{roederer2013} Roederer, I.~U.\ 2013, \aj, 145, 26. 
\bibitem[Roederer et al.(2014)]{roederer2014} Roederer, I.~U., Preston, G.~W., Thompson, I.~B., et al.\ 2014, \aj, 147, 136 
\bibitem[Rockosi et al.(2022)]{rockosi2022} Rockosi, C.~M., Lee, Y.~S., Morrison, H.~L., et al.\ 2022, \apjs, 259, 60 
\bibitem[Romano et al.(2010)]{romano2010} Romano, D., Karakas, A.~I., Tosi, M., et al.\ 2010, \aap, 522, A32 
\bibitem[Rossi et al.(1999)]{rossi1999} Rossi, S., Beers, T. C., \& Sneden, C. 1999, in ASP Conf. Ser. 165, Third Stromlo Symposium: The Galactic Halo, eds. B. Gibson, T. Axelrod, \& M. Putman (San Francisco: ASP), 264
\bibitem[Ryan et al.(1996)]{ryan1996} Ryan, S. G., Norris, J. E., \& Beers, T. C.\ 1996, \apj, 471, 254 
\bibitem[Ryan et al.(1999)]{ryan1999} Ryan, S. G., Norris, J. E., \& Beers, T. C.\ 1999, \apj, 523, 654 
\bibitem[Ryan et al.(2000)]{ryan2000} Ryan, S. G., Beers, T. C., Olive, K. A., Fields, B. D., \& Norris, J. E.\ 2000, apjl, 530, L57 


\bibitem[Sbordone et al.(2010)]{sbordone2010} Sbordone, L., Bonifacio, P., Caffau, E., et al.\ 2010, \aap, 522, A26 
\bibitem[Schlafly \& Finkbeiner(2011)]{sfd2011} Schlafly, E.~F. \& Finkbeiner, D.~P.\ 2011, \apj, 737, 103 
\bibitem[Schlegel et al.(1998)]{schlegel1998} Schlegel, D.~J., Finkbeiner, D.~P., \& Davis, M.\ 1998, \apj, 500, 525
\bibitem[Shetrone et al.(2003)]{shetrone2003} Shetrone, M., Venn, K.~A., Tolstoy, E., et al.\ 2003, \aj, 125, 684 
\bibitem[Shimansky et al.(2003)]{shimansky2003} Shimansky, V.~V., Bikmaev, I.~F., Galeev, A.~I., et al.\ 2003, Astronomy Reports, 47, 750 
\bibitem[Siegel et al.(2019)]{siegel2019} Siegel, D. M., Barnes, J., \& Metzger, B. D. 2019, \nat, 569, 241
\bibitem[Simpson et al.(2021)]{simpson2021} Simpson, J.~D., Martell, S.~L., Buder, S., et al.\ 2021, \mnras, 507, 43 
\bibitem[Skrutskie et al.(2006)]{skrutskie2006} Skrutskie, M.~F., Cutri, R.~M., Stiening, R., et al.\ 2006, \aj, 131, 1163
\bibitem[Smolinski et al.(2011)]{smolinski2011} Smolinski, J. P., Lee, Y. S., Beers, T. C., et al. 2011, \aj, 141, 89 
\bibitem[Sneden(1973)]{sneden1973} Sneden, C.~A.\ 1973, Ph.D. Thesis 
\bibitem[Sneden et al.(2008)]{sneden2008} Sneden, C., Cowan, J. J., \& Gallino, R. 2008, \araa, 46, 241
\bibitem[Sobeck et al.(2011)]{sobeck2011} Sobeck, J.~S., Kraft, R.~P., Sneden, C., et al.\ 2011, \aj, 141, 175 
\bibitem[Spergel et al.(2007)]{spergel2007} Spergel, D.~N., Bean, R., Dor{\'e}, O., et al.\ 2007, \apjs, 170, 377 
\bibitem[Spite \& Spite(1982)]{spite1982} Spite, F. \& Spite, M.\ 1982, \aap, 115, 357 
\bibitem[Spite et al.(2014)]{spite2014} Spite, M., Spite, F., Bonifacio, P., et al.\ 2014, \aap, 571, A40 
\bibitem[Spite et al.(2018)]{spite2018} Spite, M., Spite, F., Fran{\c{c}}ois, P., et al.\ 2018, \aap, 617, A56 
\bibitem[Starkenburg et al.(2014)]{starkenburg2014} Starkenburg, E., Shetrone, M. D., McConnanchie, A. W., \& Venn, K. A. 2014, \mnras, 441, 1217
\bibitem[Starkenburg et al.(2017)]{starkenburg2017} Starkenburg, E., Oman, K. A., Navarro, J. F., et al. 2017 \mnras, 465, 2212
\bibitem[Starkenburg et al.(2018)]{starkenburg2018} Starkenburg, E., Aguado, D.~S., Bonifacio, P., et al.\ 2018, \mnras, 481, 3838
\bibitem[Stacy \& Bromm(2013)]{stacy2013} Stacy, A. \& Bromm, V.\ 2013, \mnras, 433, 1094 
\bibitem[Stacy et al.(2016)]{stacy2016} Stacy, A., Bromm, V., \& Lee, A.~T.\ 2016, \mnras, 462, 1307


\bibitem[Suda et al.(2004)]{suda2004} Suda, T., Aikawa, M., Machida, M. N., \& Fujimoto, M. Y. 2004, \apj, 611, 476
\bibitem[Susa et al.(2014)]{susa2014} Susa, H., Hasegawa, K., \& Tominaga, N.\ 2014, \apj, 792, 32 
\bibitem[Thielemann \& Arnett(1985)]{thielemann1985} Thielemann, F.~K. \& Arnett, W.~D.\ 1985, \apj, 295, 604 
\bibitem[Thielemann et al.(1996)]{thielemann1996} Thielemann, F.-K., Nomoto, K., \& Hashimoto, M.-A.\ 1996, \apj, 460, 408 
\bibitem[Tinsley(1979)]{tinsley1979} Tinsley, B.~M.\ 1979, \apj, 229, 1046 
\bibitem[Tody(1986)]{tody1986} Tody, D.\ 1986, \procspie, 733 
\bibitem[Tody(1993)]{tody1993} Tody, D.\ 1993, Astronomical Data Analysis Software and Systems II, 173 
\bibitem[Tolstoy et al.(2009)]{tolstoy2009} Tolstoy, E., Hill, V., \& Tosi, M.\ 2009, \araa, 47, 371 
\bibitem[Tominaga et al.(2007)]{tominaga2007} Tominaga, N., Umeda, H., \& Nomoto, K.\ 2007, \apj, 660, 516 
\bibitem[Tominaga et al.(2014)]{tominaga2014} Tominaga, N., Iwamoto, N., \& Nomoto, K. 2014, \apj, 785, 98
\bibitem[Travaglio et al.(1999)]{travaglio1999} Travaglio, C., Galli, D., Gallino, R., et al.\ 1999, \apj, 521, 691 
\bibitem[Truran \& Arnett(1971)]{truran1971} Truran, J.~W. \& Arnett, W.~D.\ 1971, \apss, 11, 430 

\bibitem[Umeda \& Nomoto(2002)]{umeda2002} Umeda, H. \& Nomoto, K.\ 2002, \apj, 565, 385 
\bibitem[Umeda \& Nomoto(2003)]{umeda2003} Umeda, H., \& Nomoto, K. 2003, \nat, 422, 871


\bibitem[van der Walt et al.(2011)]{numpy} van der Walt, S., Colbert, S. C., \& Varoquaux, G. 2011, Computing in Science and Engineering, 13, 22
\bibitem[Venn et al.(2004)]{venn2004} Venn, K.~A., Irwin, M., Shetrone, M.~D., et al.\ 2004, \aj, 128, 1177
\bibitem[Virtanen et al.(2020)]{scipy} Virtanen, P., Gommers, R., Oliphant, T. E., et al. 2020, Nature Methods, 17, 261 

\bibitem[Wisotzki et al.(1996)]{wisotzki1996} Wisotzki, L., Koehler, T., Groote, D., et al.\ 1996, \aaps, 115, 227
\bibitem[Woosley et al.(1973)]{woosley1973} Woosley, S.~E., Arnett, W.~D., \& Clayton, D.~D.\ 1973, \apjs, 26, 231 
\bibitem[Woosley \& Weaver(1995)]{woosley1995} Woosley, S.~E. \& Weaver, T.~A.\ 1995, \apjs, 101, 181. 


\bibitem[Yanny et al.(2009)]{yanny2009} Yanny, B., Newberg, H. J., Johnson, J. A., et al. 2009, \aj, 137, 4377
\bibitem[Yong et al.(2013)]{yong2013} Yong, D., Norris, J.~E., Bessell, M.~S., et al.\ 2013, \apj, 762, 26
\bibitem[Yong et al.(2021)]{yong2021} Yong, D., Kobayashi, C., Da Costa, G.~S., et al.\ 2021, \nat, 595, 223 
\bibitem[Yoon et al.(2016)]{yoon2016} Yoon, J., Beers, T. C., Placco, V. M., et al. 2016, \apj,  833, 20
\bibitem[Yoon et al.(2018)]{yoon2018} Yoon, J., Beers, T. C., Dietz, S., et al. 2018, \apj, 861, 146
\bibitem[Yoon et al.(2019)]{yoon2019} Yoon, J., Beers, T. C., Tian, D., \& Whitten, D. D. 2019, \apj, 878, 97
\bibitem[York et al.(2000)]{york2000} York, D. G., Adelman, J., Anderson, J. E., Jr., et al. 2000, \aj, 120, 1579


\end{thebibliography}
\end{document}